\documentclass[aps,prb,twocolumn,superscriptaddress,unsortedadress,showpacs]{revtex4-1}

\usepackage{amsmath}
\usepackage{amssymb}
\usepackage{todonotes}
\usepackage{color}
\usepackage{bbm}
\usepackage{mathtools}

\newcommand{\ep}{\varepsilon}
\newcommand{\ee}{\text{e}}
\newcommand{\ii}{\text{i}}
\newcommand{\dd}{\text{d}}
\newcommand{\sgn}{\text{sign}}

\newlength{\graphwid}
\def\showgraph#1#2{
	\settowidth{\graphwid}{\includegraphics[#1,clip=true]{#2}}
	\parbox[c]{\graphwid}{\includegraphics[#1,clip=true]{#2}}
}

\begin{document}

\title{Quantum magnetooscillations in the ac conductivity of disordered graphene}

\author{U. Briskot}
\affiliation{Institut f\"ur Nanotechnologie, Karlsruhe Institute of Technology,  76021 Karlsruhe, Germany}
\affiliation{Institut f\"ur Theorie der Kondensierten Materie and Center for Functional Nanostructures, Karlsruhe Institute of Technology, 76128 Karlsruhe, Germany}
\author{I.~A. Dmitriev}
\affiliation{Institut f\"ur Nanotechnologie, Karlsruhe Institute of Technology,  76021 Karlsruhe, Germany}
\affiliation{Institut f\"ur Theorie der Kondensierten Materie and Center for Functional Nanostructures, Karlsruhe Institute of Technology, 76128 Karlsruhe, Germany}
\affiliation{Ioffe Physical Technical Institute, 194021 St.~Petersburg, Russia}
\author{A.~D. Mirlin}
\affiliation{Institut f\"ur Nanotechnologie, Karlsruhe Institute of Technology,  76021 Karlsruhe, Germany}
\affiliation{Institut f\"ur Theorie der Kondensierten Materie and Center for Functional Nanostructures, Karlsruhe Institute of Technology, 76128 Karlsruhe, Germany}
\affiliation{Petersburg Nuclear Physics Institute, 188300 St. Petersburg, Russia}

\date{\today}

\begin{abstract}
The dynamic conductivity $\sigma(\omega)$ of graphene in the presence of diagonal white noise disorder and quantizing magnetic field $B$ is calculated.
We obtain analytic expressions for $\sigma(\omega)$ in various parametric regimes ranging from the quasiclassical Drude limit corresponding to 
strongly overlapping Landau levels (LLs) to the extreme quantum limit where the conductivity is determined by the optical selection rules of the clean graphene.
The nonequidistant LL spectrum of graphene renders its transport characteristics quantitatively different from conventional 2D electron systems with parabolic spectrum. 
Since the magnetooscillations in the semiclassical density of states are anharmonic and are described by a quasi-continuum of cyclotron frequencies, 
both the ac Shubnikov-de Haas oscillations and the quantum corrections to $\sigma(\omega)$ that survive to higher temperatures manifest a slow beating 
on top of fast  oscillations with the local energy-dependent cyclotron frequency. Both types of quantum oscillations possess nodes whose index scales as $\omega^2$.
In the quantum regime of separated LLs, we study both the cyclotron resonance transitions, which have a rich spectrum due to the nonequidistant spectrum of LLs, and disorder-induced transitions
which violate the clean selection rules of graphene. We identify the strongest disorder-induced transitions in recent magnetotransmission experiments. We also compare the temperature-
and chemical potential-dependence of $\sigma(\omega)$ in various frequency ranges from the dc limit allowing intra-LL transition only to the universal high-frequency limit where the Landau quantization
provides a small $B$-dependent correction to the universal value of the interband conductivity $\sigma=e^2/4 \hbar$ of the clean graphene.
\end{abstract}

\pacs{72.80.Vp, 73.43.Qt, 78.67.Wj}
\keywords{graphene, optical properties of graphene, magnetotransport of graphene}

\maketitle

\section{Introduction}\noindent
Since its discovery in 2004,\cite{Nov04Graphene} the two dimensional carbon allotrope -- graphene -- is attracting outstanding interest in the condensed matter community. It has been verified that carriers in graphene show a linear dispersion relation with Fermi velocity $v_0\approx10^6 \text{m}/\text{s}$ and are governed by the massless Dirac equation.\cite{Sem84,CNeto09} 
Due to the carriers' Dirac nature, graphene shows remarkable properties. The clean density of states (DOS) is linear in energy and vanishes at the Dirac point, ${\nu_0(\ep)=|\ep|/2\pi v_0^2}$ ($\hbar=1$). 
The absence of scales at the Dirac point gives rise to a universal dc conductivity of the order of the conductance quantum. 
The universal high-frequency conductivity $\sigma=e^2/4 \hbar$ yields the constant absorption coefficient of $~2.3\%$ which makes graphene attractive for broadband optical applications.\cite{Nair08}
The nontrivial topology of graphene leads to the characteristic half-integer quantum Hall (QH) effect\cite{Ostr08AnQH} and to the nonequidistant Landau level (LL) spectrum
including the unusual zeroth LL at $\ep=0$.\cite{novoselov:2006} Due to the large cyclotron energy $\omega_c = v_0 \sqrt{2 e B/c}$ compared 
to conventional 2D electronic systems (2DES) the QH effect can be observed up to room temperature.\cite{Nov07QH}
Recent rapid developments demonstrate great potential of graphene in optoelectronics.\cite{FerrariGrapheneOptoRev10,AvGraphenePhotRev10,Koppens2011,Grigorenko2012,Engel2012} From this perspective, the application of a quantizing magnetic field creates a suitable environment for applications in e.g. laser physics.\cite{LLLaser09} The quantization introduces a tunable energy scale in the otherwise scale-free graphene, while the nonequidistant LL spectrum makes it highly selective in the frequency domain.

While experimental data on quantum magnetooscillations in graphene is limited (although growing),\cite{Tan11SdH,SadowskiGrLayers06,OrlitaPotMOptRev10,Pot11Scattering} there are comprehensive studies of related phenomena for semiconductor 2DES where the effects both near and far from equilibrium have been extensively studied.\cite{ando:1982a,IvanReview11} In the linear response regime, quantum magnetooscillations in the ac conductivity were theoretically predicted\cite{Ando74IVOscCond,Ando75CRShape} and observed\cite{Abstreiter76} already in the seventies. In Ref.~\onlinecite{Ivan03CRHarmonics} the theory was generalized to high mobility 2DES with smooth disorder potential; the findings are also confirmed in recent experiment.\cite{Stud10QO2DEG}

This work is (i) motivated by experimental and technical advances in the graphene research and  (ii) generalizes the theory on quantum magnetooscillations in the conventional 2DES mentioned above. We study quantum oscillations in Landau quantized graphene in the presence of disorder. Previous theoretical works on the magnetoconductivity in graphene accounted for the disorder in terms of a phenomenological broadening in the single-particle spectrum\cite{Gus07,Gus07_2,Gus07Anomalous} or focused on the dc magnetoconductivity.\cite{AndoJPSJ2DGraphite98,MagnetoRes2012Ioffe} The optical conductivity has also been studied for vacancies within the T-matrix approximation.\cite{PeresGuinea06} Here we perform a systematic calculation of the dynamic magnetoconductivity within the self-consistent Born approximation (SCBA).

The paper is organized as follows. In Sec.~\ref{sec:formalism} we outline the model of the SCBA in graphene and present results on the spectrum of disordered graphene in a magnetic field, in the semiclassical as well as in the quantum regime. Further details on the calculation of the density of states can be found in Appendix~\ref{sec:app_SCBA_overlapping}. Section~\ref{sec:conductivity} presents the formalism for the calculation of the dynamic conductivity. The following Secs.~\ref{sec:cond_overlapping}~and~\ref{sec:cond_separated} are devoted to the dynamic conductivity in the regimes of strongly overlapping and well-separated LLs respectively. In Sec.~\ref{sec:single_level} we calculate the high-frequency conductivity of a moderately disordered graphene. The summary of results is presented in Sec.~\ref{sec:summary}. Details on the calculation regarding Secs.~\ref{sec:cond_overlapping}~and~\ref{sec:cond_separated} can be found in Appendix~\ref{sec:app_vertex_corrections}.

\section{Landau level spectrum within SCBA\label{sec:formalism} }\noindent
In the following we assume that disorder does not mix the two valleys in graphene and therefore calculate all quantities per spin and valley. 
Correspondingly, to get the full conductivity of graphene one should multiply the results for conductivity below by the degeneracy factor of 4.
In the presence of a constant magnetic field in $z$-direction the electrons in a single valley of clean graphene are described by the 2D Dirac equation
\begin{equation}
	\hat{H} = v_0 \vec{\sigma} \cdot \left[\: \hat{\vec{p}} - e \vec{A}(\hat{\vec{r}}) \:\right]
	\: , \quad
	\vec{A}(\vec{r}) = -y B \hat{\vec{x}}
	\: .
	\label{sec:formalism:eq:hamiltonian}
\end{equation}
Here we have chosen the Landau gauge for the vector potential $\vec{A}$. The vectors $\vec{r}=(x,y)^T$, $\hat{\vec{x}}=(1,0)^T$, and $\vec{\sigma}=(\sigma_x,\sigma_y)$ denote the Pauli matrices. The positions of the Landau levels (LLs) in clean graphene are given by
\begin{equation}
	E_n=\sgn(n)\omega_c\sqrt{|n|} \: , \: n \in \mathbb{Z}
	\: .
	\label{sec:formalism:eq:Landau_levels}
\end{equation}
The optical selection rules of the clean graphene allow transitions from the LL $n$ to $m$ if
\begin{equation}
	|n|-|m|=\pm1 \: ,
	\label{sec:formalism:eq:selection_rules}
\end{equation}
therefore enabling both intra- [$\sgn(n)=\sgn(m)$] and interband [$\sgn(n)\neq\sgn(m)$] transitions. Since the LLs move closer at higher energy, it is convenient to introduce a local cyclotron frequency
\begin{equation}
	\omega_n
	= E_{|n|+1}-E_{|n|}
	\: ;
	\label{sec:formalism:eq:local_cyclotron_freq}
\end{equation}
that is, the distance between neighboring LLs. In high LLs, $n\gg1$ ($\ep\gg\omega_c$), it approaches the quasiclassical cyclotron frequency of a massless particle
\begin{equation}
	\omega_n\simeq\omega_c^{\text{loc}} = \frac{\omega_c^2}{2|\ep|}
	\: .
	\label{sec:formalism:eq:loc_cyclotron_freq_asymptotic}
\end{equation}

The disorder is included into the self-energy $\hat{\Sigma}$ which enters the impurity averaged electronic Green's function
\begin{equation}
	\hat{G}=(\ep-\hat{H}-\hat{\Sigma})^{-1}
	\: .
	\label{sec:formalism:eq:dressed_GF}
\end{equation}
Within the SCBA,\cite{Ando74ILevlBroad,AndoJPSJ2DGraphite98} the self-energy is given by
\begin{equation}
	\hat{\Sigma}_{\nu\mu} = \int \frac{\dd \vec{q}}{(2\pi)^2} \: W_{\nu\eta\sigma\mu}(\vec{q}) \: \ee^{+\ii\vec{q}\cdot\hat{\vec{r}}} \hat{G}_{\eta\sigma} \ee^{-\ii\vec{q}\cdot\hat{\vec{r}}}
	\: .
	\label{sec:formalism:eq:SCBA}
\end{equation}
Weak disorder is characterized by the disorder correlator $W(\vec{q})$ which in the case of graphene is generally a fourth rank tensor in valley and sublattice space. Here we consider short-range impurities which do scatter between the sublattices but do not produce any intervalley scattering. In a given valley, this gives diagonal white noise disorder with the correlator
\begin{equation}
	W_{\nu\eta\sigma\mu}(\vec{q}) = 2\pi v_0^2 \alpha \: \delta_{\nu\eta}\otimes\delta_{\sigma\mu}
	\: ,
	\label{sec:formalism:eq:dis_correlator}
\end{equation}
characterized by a single parameter $\alpha\ll1$. For the diagonal disorder, the self-energy is independent of the LL index and becomes diagonal in the sublattice space; still, it carries an asymmetry between the sublattices  $a$ and $b$,
\begin{equation}
	\hat\Sigma = \text{diag}(\Sigma_a,\Sigma_b)
	\: .
	\label{sec:formalism:eq:selfenergy_structure}
\end{equation}
The asymmetry is due to the fact that -- in a given valley of the clean graphene -- the wave function of the zeroth LL resides in one sublattice only. In what follows, we choose the valley such that the wave function of the clean zeroth LL resides in the sublattice $b$. The SCBA equation \eqref{sec:formalism:eq:SCBA} for disorder with the correlator \eqref{sec:formalism:eq:dis_correlator} acquires the form
\begin{equation}
	\Sigma_{a(b)} = \frac{\alpha \omega_c^2 }{2} \: \sum_{n=1(0)}^{+\infty}
	\frac{\varepsilon-\Sigma_{b(a)}}{(\varepsilon-\Sigma_{a})(\varepsilon-\Sigma_{b})-\omega_c^2n}
	\: .
	\label{sec:formalism:eq:selfenergy_1}
\end{equation}
Away from the zeroth LL the difference between the two self-energy components is negligible, $\Sigma_{a}\simeq\Sigma_{b}$, [see discussion under Eq.~\eqref{sec:formalism:eq:selfenergy_separated_2_2} and Fig.~\ref{sec:formalism:fig:DOS}] yielding
\begin{equation}
	\Sigma_{a,b}\simeq\Sigma = \frac{\alpha \omega_c^2}{2} \: \sum_{n=0}^{+\infty}
	\frac{(\ep-\Sigma)}{(\ep-\Sigma)^2-\omega_c^2 n}
	\: .
	\label{sec:formalism:eq:selfenergy_2}
\end{equation}
Apart from specifics of the zeroth LL due to its pronounced sublattice asymmetry, one can consider two limiting cases we address separately below: (i) clean, or quantum limit, when the disorder-broadened LLs remain well separated and (ii) dirty, or classical limit when LLs strongly overlap almost restoring the linear slope of the DOS at $B=0$. Unlike conventional 2D systems with parabolic spectrum, in graphene the two cases (i) and (ii) frequently coexist: LLs, well separated near the Dirac point, start to overlap at higher energies where the local cyclotron frequency~\eqref{sec:formalism:eq:loc_cyclotron_freq_asymptotic} strongly reduces compared to $\omega_c$.
\subsection{Separated Landau levels}\noindent
We start with the limit of well separated LLs. In this case, the main contribution to the self-energy at energy $\ep$ comes from the states in the nearest LL 
of the clean graphene to which we assign the integer number $N$ closest to $\ep^2/\omega_c^2$. 
LLs with index $n\neq N$ contribute to logarithmic energy renormalization as detailed below. 

For $N\neq 0$, the sublattice asymmetry can be neglected, and the solution
to Eq.~\eqref{sec:formalism:eq:selfenergy_2} for the retarded self-energy reads [the advanced self-energy $\displaystyle\Sigma^A=(\Sigma^R)^{*}$]
\begin{equation}
	\ep-\Sigma^R = \frac{\tilde\ep + E_N}{2} + \frac{\ii}{2} \sqrt{\Gamma_N^2-(\tilde\ep-E_N)^2}\,,\quad N\neq0
	\: ,
	\label{sec:formalism:eq:selfenergy_separated_1}
\end{equation}
where the width of the $N$th LL 
\begin{equation}
	\Gamma_N=\sqrt{\alpha}\omega_c/Z(E_N)\,,\quad N\neq0
	\: ,
	\label{sec:formalism:eq:gammaN}
\end{equation}
and the condition of applicability is $\Gamma_N\ll\omega_N$. Apart from the usual renormalization of energy by the factor of 2 inside the LLs,\cite{Ando74ILevlBroad,Ando75CRShape} in graphene the energy gets additionally renormalized according to
\begin{equation}
	\tilde\ep=\ep/Z(E_N).
	\label{sec:formalism:eq:renorm}
\end{equation}
Here the renormalization constant is
\begin{equation}
	Z(\ep)=1-\alpha\ln(\Delta_c/|\ep|)
	\: ,
	\label{sec:formalism:eq:renorm_const}
\end{equation}
and $\Delta_c$ is the high energy cut-off of the order of the band width. Within the SCBA, the additional logarithmic correction describes the influence of states in distant LLs [$n\neq N$ in Eq.~\eqref{sec:formalism:eq:selfenergy_1}], similar to renormalization group (RG) corrections.\cite{Ostr08AnQH} For the type of disorder we are investigating here, the SCBA corrections without magnetic field are known to be in quantitative accordance with RG calculations.\cite{Ostr08AnQH} Apart from the additional logarithmic renormalization and from the non-equidistant spectrum $E_N$ of clean LLs,  Eq.~\eqref{sec:formalism:eq:selfenergy_separated_1} reproduces the well-known semicircular law obtained by Ando for 2DES with parabolic spectrum.\cite{Ando74ILevlBroad,Ando75CRShape}
From the requirement ${\alpha\ln(\Delta_c/|\ep|)\ll1}$, we obtain that the results are justified for energies above the exponentially small energy scale $\Delta_c\ee^{-1/\alpha}$.

In the vicinity of the zeroth LL, $N=0$, one needs to take into account the explicit sublattice asymmetry. Equation \eqref{sec:formalism:eq:selfenergy_1} yields the self-energies\cite{Ostr08AnQH}
\begin{align}
	& \ep-\Sigma^R_b = \frac{\tilde\ep}{2} +\frac{\ii}{2} \sqrt{\Gamma_0^2-\tilde\ep^2} ,
	\label{sec:formalism:eq:selfenergy_separated_2_2} \\
    & \ep-\Sigma^R_a = \frac{1+Z(E_1)}{2}\tilde\ep +\ii \frac{1-Z(E_1)}{2}\sqrt{\Gamma_0^2-\tilde\ep^2}.
	\label{sec:formalism:eq:selfenergy_separated_2_1}
\end{align}
The width of the zeroth LL in both sublattices is
\begin{equation}
	\Gamma_0=\sqrt{2\alpha}\omega_c/Z(E_1)=\sqrt{2}\Gamma_1\,;
	\label{sec:formalism:eq:gamma0}
\end{equation}
the factor $\sqrt{2}$ comes from different degeneracy of the zeroth LL -- the total number of states ${\cal N}_{\rm tot}=L^2/2\pi l_B^2$ is the same in all LLs, but in the zeroth LL all these states reside in one sublattice. We observe that only a small amount of the spectral weight $\propto 1-Z(E_1)\propto \alpha\ll1$ is scattered into the afore empty sublattice $a$.
At the same time, the strong renormalization of energy by the factor of 2 related to the high degeneracy of clean LLs in Eqs.~\eqref{sec:formalism:eq:selfenergy_separated_1} and \eqref{sec:formalism:eq:selfenergy_separated_2_2} is absent in Eq.~\eqref{sec:formalism:eq:selfenergy_separated_2_1}. Only the logarithmic renormalization due to distant LLs with $n\neq 0$ remains.

\subsection{Overlapping Landau levels}\noindent
 In view of the condition $\alpha\ll1$ of weak disorder, the regime of strongly overlapping LLs is realized at 
high energies $\ep>\ep_{\rm ov}\simeq \omega_c/\sqrt{\alpha}\gg\omega_c$. Indeed, 
the LL number $N_{\rm ov}$ where LLs start to overlap is given by the condition $\Gamma_N\simeq\omega_N$
which gives [see Eqs.~(\ref{sec:formalism:eq:Landau_levels}), (\ref{sec:formalism:eq:local_cyclotron_freq}), and (\ref{sec:formalism:eq:gammaN})] 
\begin{equation}
	N_{\rm ov}\simeq\alpha^{-1}\gg 1.
	\label{sec:formalism:eq:crossover}
\end{equation}
Accordingly, we assume $\Sigma_{a,b}\simeq\Sigma$ and use the Poisson summation formula to rewrite Eq.~\eqref{sec:formalism:eq:selfenergy_2} into a rapidly convergent sum in the Fourier space [see Eq.~\eqref{sec:appendix:eq:SCBA_poisson}].
The small parameter that controls such an expansion is the Dingle or coherence factor
\begin{equation}
	\lambda = \ee^{-2\alpha\pi^2{\tilde\ep}^2/\omega_c^2}
	\: ,
	\label{sec:formalism:eq:Dingle_factor}
\end{equation}
which is the analog of the Dingle factor $\delta=\ee^{-\pi/\omega_c\tau_q}$ in 2DES with parabolic spectrum and describes the smearing of quantum oscillations in the disordered system. To zeroth order in $\lambda$, we obtain a broadening of the single-particle states at $B=0$,
\begin{equation}
	\left.\Sigma^R \right|_{\lambda\to 0} = - {\ii}/{2\tau_q(\tilde\ep)}.
	\label{sec:formalism:eq:selfenergy_0}
\end{equation}
with the energy-dependent quantum scattering time given by
\begin{equation}
	\tau_q(\tilde\ep) = \frac{Z(\ep)}{\alpha\pi|\tilde\ep|}
	\: .
	\label{sec:formalism:eq:quantum_time}
\end{equation}
The renormalization constant $Z(\ep)$ which defines the renormalized energy $\tilde\ep=\ep/Z(\ep)$ is given by Eq.~\eqref{sec:formalism:eq:renorm_const}. To first order in $\lambda\ll1$, the self-energy acquires the form
\begin{equation}
	\ep-\Sigma^R = \tilde\ep + \frac{\ii}{2\tau_q(\tilde\ep)} \:
	\left[ 2\lambda \: \ee^{\ii2\pi\tilde\ep|\tilde\ep|/\omega_c^2} + 1\right]
	\: .
	\label{sec:formalism:eq:selfenergy_overlapping}
\end{equation}

\subsection{Density of states}\noindent
The DOS per spin and valley ${\nu(\ep)=-{\rm Im\;tr}\: G^R(\vec{r},\vec{r},\ep)/\pi}$ in the case of white noise disorder is given by
\begin{equation}
	\nu(\ep) = -\frac{1}{2\pi^2v_0^2\alpha} \: \text{Im} \: \text{tr} \Sigma^R
	\: .
	\label{sec:formalism:eq:DOS_definition}
\end{equation}
\begin{figure}[ht]
	\centering
	\includegraphics[width=0.5\textwidth]{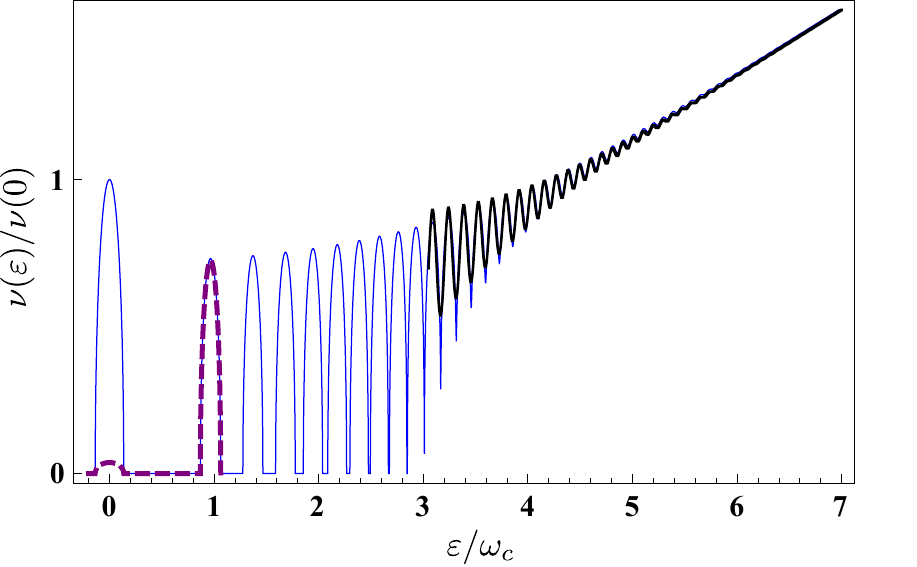}
	\caption{The DOS in disordered graphene. The thin line is the DOS in the sublattice $b$ obtained numerically from the SCBA equation~\eqref{sec:formalism:eq:selfenergy_1} for $\alpha=0.01$. For the first two levels we show the DOS in the sublattice $a$ as the dashed line. Already in the second LL there is no noticeable difference between the $a$ and $b$ sublattice. The thick curve shows our result for the DOS according to Eqs.~\eqref{sec:formalism:eq:selfenergy_overlapping} and \eqref{sec:formalism:eq:DOS_definition}. It is in good agreement with the numerical results for high energies.}
	\label{sec:formalism:fig:DOS}
\end{figure}
The DOS in the sublattice $b$ obtained numerically from the SCBA equation~\eqref{sec:formalism:eq:selfenergy_1} is plotted in Fig.~\ref{sec:formalism:fig:DOS} as the thin line. Separated LLs show a semicircle DOS, see Eqs.~\eqref{sec:formalism:eq:selfenergy_separated_1} and \eqref{sec:formalism:eq:selfenergy_separated_2_2}. The dashed line for the DOS in the sublattice $a$ shows that (i) in the zeroth LL the DOS transferred into the sublattice $a$ is small and vanishes if disorder is turned off, see Eq.~\eqref{sec:formalism:eq:selfenergy_separated_2_1} and (ii) at higher energy the DOS (self-energy) in both sublattices is approximately equal. Finally, the thick line, calculated according to Eqs.~\eqref{sec:formalism:eq:selfenergy_overlapping} and \eqref{sec:formalism:eq:DOS_definition}, illustrates the limit of strongly overlapping LLs corresponding to large $\ep\gg\ep_{\rm ov}$ (which gives $\lambda\ll 1$). 

\subsection{Semiclassical regime \label{sec:semiclassical}}\noindent
Here we address the semiclassical regime of high LLs in graphene, $N\gg1$, additionally assuming that these levels strongly overlap,
$\ep\gg\ep_\text{ov}\simeq\omega_c/\sqrt{\alpha}$. The aim is to provide a link to the the results in the correspondent parabolic band with effective 
energy-dependent mass such that the local characteristics of graphene and conventional 2DES with such parabolic spectrum at high LLs are identical.

In the limit $\lambda\to 0$, Eqs.~\eqref{sec:formalism:eq:selfenergy_0} and \eqref{sec:formalism:eq:DOS_definition} yield the $B=0$ renormalization of the clean DOS $\nu_0$ by disorder
\begin{equation}
	\nu_0=\frac{\ep}{2\pi v_0^2}\quad\rightarrow\quad\tilde{\nu}_0\equiv Z(\ep)\nu(\tilde{\ep})|_{\lambda\to 0} =\frac{\tilde{\ep}}{2\pi v_0^2}.
	\label{sec:formalism:eq:DOS_correction_0}
\end{equation}
According to Eq.~\eqref{sec:formalism:eq:selfenergy_overlapping}, the correction $\propto\lambda$ reads
\begin{equation}
	\tilde\nu_{\text{osc}}\equiv\tilde\nu(\tilde\ep)-\tilde{\nu}_0 = 2\tilde{\nu}_0\lambda \: \cos\frac{2\pi\tilde\ep|\tilde\ep|}{\omega_c^2}
	\: ,
	\label{sec:formalism:eq:DOS_correction}
\end{equation}
The energy renormalization induced by the Landau quantization shows similar (but phase-shifted by $\pi/2$) oscillations
\begin{equation}
	\Delta\text{Re}\Sigma(\tilde\ep)= \frac{\lambda}{\tau_q(\tilde\ep)}\:\sin\frac{2\pi\tilde\ep|\tilde\ep|}{\omega_c^2}
	\: .
	\label{sec:formalism:eq:energy_renorm}
\end{equation}
Apart from (i) the large-scale logarithmic renormalization \eqref{sec:formalism:eq:DOS_correction_0}  which is specific for the linear spectrum of graphene
and (ii) a phase shift related to a nontrivial Berry's phase of graphene
the local characteristics of graphene and 2DES with parabolic spectrum at high LLs are identical as expected. Namely, quasiparticles in graphene at energy $\tilde{\ep}$ close to $\tilde{\ep}^*\gg\omega_c$ behave equivalent to massive electrons with mass, energy, and cyclotron energy given by
\begin{align}
	& m_{\text{eff}}=\tilde{\ep}^*/v_0^2 ,
	\label{sec:formalism:eq:equiv1} \\
	& \ep_{\text{eff}}=\tilde\ep/2-\omega^{(\text{eff})}_c/2,
	\label{sec:formalism:eq:equiv2}\\
	&\omega^{(\text{eff})}_c=e B/m_{\text{eff}} c,
	\label{sec:formalism:eq:equiv3}
\end{align}
such that the local velocity in the parabolic band ${v_{\text{eff}}=\sqrt{2 \ep_{\text{eff}}/m_{\text{eff}}}}$ coincides with $v_0$.
The term $-\omega^{(\text{eff})}_c/2$ accounts for the shift of LLs due to vacuum fluctuations in the parabolic band which is absent due to the nontrivial Berry's phase in graphene, see also Eq.~\eqref{sec:formalism:eq:equiv6} below.
The effective cyclotron frequency $\omega^{(\text{eff})}_c$ coincides with the renormalized local cyclotron frequency in graphene, $\tilde{\omega}_c^{\text{loc}}(\tilde{\ep}^*)= \omega_c^2/2|\tilde{\ep}^*|=Z(\ep^*)\omega_c^{\text{loc}}(\ep^*)$.
If one additionally introduces the renormalized $\tilde{\tau}_q(\tilde{\ep})=\tau_q(\ep)/Z(\ep)=1/\pi\alpha|\tilde{\ep}|$, equation \eqref{sec:formalism:eq:DOS_correction} transforms into conventional expression for parabolic band, specifically,
\begin{align}
	& \tilde{\nu}_0=\frac{\tilde{\ep}^*}{2\pi v_0^2}\quad\to\quad \frac{m_{\text{eff}}}{2\pi} ,
	\label{sec:formalism:eq:equiv4} \\
	& \lambda=\ee^{-2\alpha\pi^2{\tilde\ep}^2/\omega_c^2}\quad\to\quad\ee^{-\pi/\tilde{\omega}_c^{\text{(eff)}}\tilde{\tau}_q(\tilde{\ep})},
	\label{sec:formalism:eq:equiv5}\\
	&\frac{2\pi\tilde\ep|\tilde\ep|}{\omega_c^2}\quad\to\quad \frac{2\pi\ep_{\text{eff}}}{\omega_c^{\text{(eff)}}} +\pi.
	\label{sec:formalism:eq:equiv6}
\end{align}
\section{Dynamic Conductivity \label{sec:conductivity}}\noindent
Below we calculate the dynamic conductivity of disordered graphene in the presence of a quantizing magnetic field. We use the Kubo formula for the real part of the diagonal conductivity
\begin{equation}
	\sigma(\omega) = \int \frac{\dd\ep}{4\pi} \frac{f_{\ep}-f_{\ep+\omega}}{\omega} \: K(\ep,\ep+\omega)
	\: ,
	\label{sec:formalism:eq:Kubo_formula}
\end{equation}
to calculate the magnetoconductivity in the linear response. Here $f_{\ep}$ is the equilibrium Fermi-Dirac distribution function, and the conductivity kernel is given by
\begin{equation}
	K(\ep_1,\ep_2) = - \text{tr}\left[\left(\hat{G}^{R}_{\ep}-\hat{G}^{A}_{\ep_1}\right)\hat{j}_x\left(\hat{G}^{R}_{\ep_2}-\hat{G}^{A}_{\ep_2}\right)\hat{j}_x\right]
	\: .
	\label{sec:formalism:eq:kernel_definition}
\end{equation}
From the Hamiltonian~\eqref{sec:formalism:eq:hamiltonian}, the current density operator
\begin{equation}
	\hat{\vec{j}} = \frac{e}{L} \frac{\partial\hat{H}}{\partial{\vec{k}}} = \frac{e v_0}{L} \: \vec{\sigma}\,,
	\label{sec:formalism:eq:current_density}
\end{equation}
where $L$ is the length of the system. Due to the linear spectrum the current operator does not depend on the magnetic field. Since the two valleys are decoupled, we calculate the conductivity per spin and valley.

The effect of the disorder averaging in Eq.~\eqref{sec:formalism:eq:kernel_definition} is twofold: (i) the bare Green's function is replaced by the impurity averaged Green's function~\eqref{sec:formalism:eq:dressed_GF}, (ii) the summation of the diagrams~\ref{sec:app_vertex_corrections:fig:combined}(c) leads to vertex corrections to the current operator~\eqref{sec:formalism:eq:current_density}. In 2DES with parabolic spectrum and white noise disorder, the vertex corrections are absent. By contrast, in graphene the vertex corrections are present for the diagonal white noise disorder as well. They originate from the nontrivial Berry's phase of Dirac fermions. We present details on the calculation of the conductivity including vertex corrections in App.~\ref{sec:app_vertex_corrections}. As expected, we find that in the quasiclassical regime of strongly overlapping LLs the vertex corrections give rise to the replacement of $\tau_q$ by the transport scattering time $\tau_\text{tr}=2\tau_q$ in the Drude part of the conductivity, see Eq.~\eqref{sec:conductivity:eq:Drude} below, while the quantum time appears only in the quantum corrections related to the Landau quantization. In particular, $\tau_q$ enters the Dingle factor $\lambda$, see Eq.~\eqref{sec:formalism:eq:Dingle_factor}.

\section{Overlapping Landau levels \label{sec:cond_overlapping}}\noindent
In this subsection we consider the case of highly doped graphene with the Fermi energy $\ep_F \gg\omega_c,\omega$. Therefore, only intraband processes are possible and one can neglect the sublattice asymmetry in the self-energy. It follows that
the conductivity in this regime should not change in the presence of intervalley scattering. Using the continuity equation, the conductivity kernel~\eqref{sec:formalism:eq:kernel_definition} is expressed in terms of density-density correlators $\Pi$. Their general form is given in Eq.~\eqref{sec:app_vertex_corrections:eq:Pi_correlators_def}, which simplifies to
\begin{equation}
	\begin{split}
		\Pi^{RR(RA)}_{\ep_1,\ep_2} =
		& \frac{\omega_c^2}{2\pi} \sum_{n=0}^{\infty} \:
			\bigg[G^{R}_{n+1,-}(\ep_1)+G^{R}_{n+1,+}(\ep_1)\bigg]\\
			& \times \bigg[G^{R(A)}_{n,-}(\ep_2)+G^{R(A)}_{n,+}(\ep_2)\bigg]
		\label{sec:conductivity:eq:Pi_correlators_def}
	\end{split}
\end{equation}
if the sublattice asymmetry is absent. Here we introduced the chiral Green's functions
\begin{equation}
	G_{n,\pm}^{R(A)}(\tilde\ep) = \frac{1}{\tilde\ep-\text{Im}\Sigma^{R(A)}\mp|E_n|}	
	\: .
	\label{sec:conductivity:eq:chiral_GF}
\end{equation}
In the following we write $\ep$ for $\tilde\ep$ for brevity. In terms of the $\Pi$-correlators, the kernel~\eqref{sec:formalism:eq:kernel_definition} acquires the form
\begin{equation}
	\begin{split}
		K(\ep_1,\ep_2) = e^2 \text{Re}\Big[ \Big( \big[\Pi^{RA}_{\ep_1,\ep_2}\big]^{-1}-\alpha\pi \Big)^{-1} \\
			- \Big( \big[\Pi^{RR}_{\ep_1,\ep_2}\big]^{-1}-\alpha\pi \Big)^{-1}  + \{\ep_1\leftrightarrow\ep_2\}\:\Big]
		\: .
	\end{split}
	\label{sec:conductivity:eq:kernel_Pi}
\end{equation}
As previously, we use the Poisson formula to rewrite the sums occurring in Eq.~\eqref{sec:conductivity:eq:Pi_correlators_def} as rapidly convergent sums in the Fourier space. With the self-energies for overlapping LLs from Eq.~\eqref{sec:formalism:eq:selfenergy_overlapping}, we obtain
\begin{equation}
	\begin{split}
		& \Pi^{RR(RA)}_{\ep_1,\ep_2}
		= \frac{2\big(\ep_1-\Sigma_{\ep_1}^{R}\big)\big(\ep_2-\Sigma_{\ep_2}^{R(A)}\big)}{\big(\ep_1-\Sigma_{\ep_1}^{R}\big)^2-\big(\ep_2-\Sigma_{\ep_2}^{R(A)}\big)^2-\omega_c^2} \\
		& \times \bigg\{
			\sgn(\ep_1) \: \tau_{q,\ep_1}\Sigma^{R}_{\ep_1}
			\:-\:
			\sgn(\ep_2) \: \tau_{q,\ep_2}\Sigma^{R(A)}_{\ep_2}
		\bigg\}
		\: .
	\end{split}
	\label{sec:conductivity:eq:Pi_correlator_2}
\end{equation}
We organize the following analysis in orders of $\lambda$. Details of the calculation are presented in Appendix \ref{sec:app_vertex_corrections}.
	
It follows from Eq.~\eqref{sec:formalism:eq:selfenergy_overlapping} that to zeroth order in $\lambda$ only the RA-sector of Eq.~\eqref{sec:conductivity:eq:Pi_correlator_2} contributes to the conductivity for the intraband processes.
For $\omega=|\ep_1-\ep_2|\ll\ep_F$ the kernel \eqref{sec:conductivity:eq:kernel_Pi} has a Drude form with the energy-dependent broadening \eqref{sec:formalism:eq:quantum_time} and local cyclotron frequency \eqref{sec:formalism:eq:loc_cyclotron_freq_asymptotic}. If the temperature $T\ll\ep_F$, these parameters can be evaluated on-shell. The conductivity $\sigma_D=\sigma_{D,+}+\sigma_{D,-}$ in this regime reads
\begin{equation}
	\sigma_{D,\pm}(\omega)
	= \frac{1}{4\pi} \frac{\mathcal{D}/\tau_\text{tr}}{(\omega\pm\omega_c^{\text{loc}})^2+\tau_\text{tr}^{-2}}
	\: .
	\label{sec:conductivity:eq:Drude}
\end{equation}
Due to the vertex corrections the transport scattering time $\tau_\text{tr}=2\tau_q$ replaces the quantum scattering time as explained in App.~\ref{sec:app_vertex_corrections}. The Drude weight $\mathcal{D}$ for Dirac fermions is
\begin{equation}
	\mathcal{D} = \frac{e^2 |\ep_F|}{2}
	\: ,
	\label{sec:conductivity:eq:Drude_weight}
\end{equation}
which produces the standard Drude weight $\mathcal{D}=e^2 \ep_\text{eff}$ in the correspondent parabolic band, see Eq.~\eqref{sec:formalism:eq:equiv2}. The Drude part of the conductivity in graphene was also obtained in Refs.~\onlinecite{Peres07Drude,Gus06HallOptCond}. Without magnetic field the Drude form has been confirmed experimentally\cite{Horng11DrudeGraphene} though deviations (possibly due to interactions) were also observed.

On top of the semiclassical Drude conductivity quantum oscillations are superimposed. There are two major and competing damping mechanisms present. Finite $T$ leads to the thermal damping of the Shubnikov-de Haas (SdH) oscillations in the ac- and dc-response. Scattering off disorder also smears quantum oscillations and this is captured by the coherence factor $\lambda$. The latter mechanism is dominant for $T$ below the Dingle temperature $T_D=1/2\pi\tau_q$.
\footnote{\label{footnote_interactions} Interactions may contribute a third damping mechanism. For systems with a parabolic spectrum, electron-electron interactions are known to have no direct influence on the damping of SdH oscillations. Only if one considers the combined effect of disorder and interactions, the damping of SdH oscillations is influenced by a renormalization of the effective mass and scattering time.\cite{Martin03QMO_FL,AdamovGornyiMirlin06Interactions} By contrast, interactions directly influence\cite{dmitriev:2009b} the higher-order quantum corrections $\propto\lambda^2$ similar to these in Eq.~(\ref{sec:conductivity:eq:highT_QO_1}).}

The leading order quantum corrections which are strongly damped by finite temperature describe SdH oscillations in the dynamic and dc conductivity. The corresponding contribution to the kernel $K=K_++K_-$, Eq.~\eqref{sec:formalism:eq:kernel_definition}, reads
\begin{equation}
	\begin{split}
		K^{\hspace{-0.2mm}(1)}_\pm(\ep_1,\ep_2)
		=
		& \: 4\pi\:\sigma_{D,\pm} \: \\
		& 	\hspace{-20mm}
			\times \bigg\{
			\frac{2a_\pm^2}{a_\pm^2+1} \: \left[\frac{\tilde\nu_{\text{osc}}(\ep_2)}{\tilde\nu_0(\ep_2)}+\frac{\tilde\nu_{\text{osc}}(\ep_1)}{\tilde\nu_0(\ep_1)}\right] \\
			& \hspace{-20mm} \:+\:
			\frac{a_\pm}{a_\pm^2+\omega^2/\ep_1^2} \: \left[\frac{\Delta\text{Re}\Sigma(\ep_2)}{1/\tau_q(\ep_2)}-\frac{\Delta\text{Re}\Sigma(\ep_1)}{1/\tau_q(\ep_1)}\right]
		\bigg\}
		\: ,
	\end{split}
	\label{sec:conductivity:eq:kernel_1st_order}
\end{equation}
where we used the abbreviation ${a_\pm=\tau_\text{tr}(\omega\pm\omega_c^{\text{loc}})}$.

\begin{figure}[ht]
	\centering
	\includegraphics[width=0.45\textwidth]{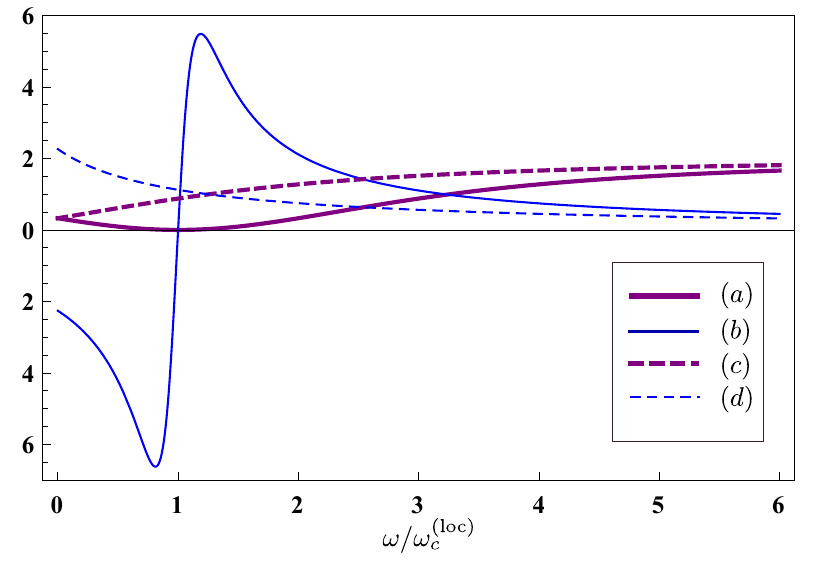}
	\caption{The prefactors from Eq.~\eqref{sec:conductivity:eq:kernel_1st_order}: $(a)$ $2a_-^2/(a_-^2+1)$, $(b)$ $a_-/(a_-^2+\omega^2/\ep_F^2)$, $(c)$ $2a_+^2/(a_+^2+1)$, $(d)$ $a_+/(a_+^2+\omega^2/\ep_F^2)$, where $a_\pm=\tau_\text{tr}(\omega\pm\omega_c^\text{loc})$. All curves are calculated for $\ep_F/\omega_c=6$ and $\alpha=0.01$. Away from the cyclotron resonance, $a_\pm\gg1$, the subleading effects due to the energy renormalization (thin lines) can be ignored.}
	\label{sec:conductivity:fig:prefactors_SdH}
\end{figure}
The first term in Eq.~\eqref{sec:conductivity:eq:kernel_1st_order} is produced by the quantum correction to the DOS, Eq.~\eqref{sec:formalism:eq:DOS_correction}, while the second term accounts for the energy renormalization, Eq.~\eqref{sec:formalism:eq:energy_renorm}. The first term is dominant away from the cyclotron resonance, see Fig.~\ref{sec:conductivity:fig:prefactors_SdH}. In this case $a_\pm\gg1$, and the kernel \eqref{sec:conductivity:eq:kernel_1st_order} acquires the form
\begin{equation}
	K^{\hspace{-0.2mm}(1)}_\pm(\ep_1,\ep_2) \simeq
	8\pi\:\sigma_{D,\pm} \: \left[\frac{\tilde\nu_{\text{osc}}(\ep_2)}{\tilde\nu_0(\ep_2)}+\frac{\tilde\nu_{\text{osc}}(\ep_1)}{\tilde\nu_0(\ep_1)}\right]
	\: .
	\label{sec:conductivity:eq:kernel_asymptotics}
\end{equation}
This form of the kernel is expected from a golden rule consideration where the conductivity (in the classically strong field, $a_\pm\gg1$) is determined by the product of initial and final DOS $\nu(\ep)\nu(\ep+\omega)$.
To first order in the coherence factor $\lambda$, this yields ${\nu(\ep)\nu(\ep+\omega)-\nu_0^2\simeq\nu_0[\nu_{\text{osc}}(\ep+\omega)+\nu_{\text{osc}}(\ep)]\propto\lambda}$.

\begin{figure}[ht]
	\centering
	\includegraphics[width=0.48\textwidth]{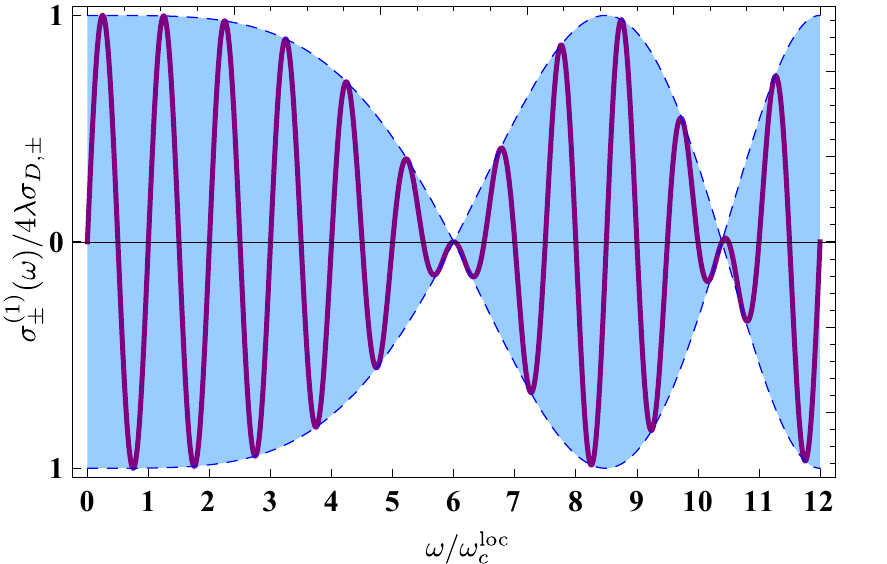}
	\caption{The SdH oscillations in the dynamic conductivity (solid line) and the envelope function showing slow modulation due to nonequidistant spectrum of LLs (dashed line) calculated according to Eq.~\eqref{sec:conductivity:eq:conductivity_1st_order} for $\ep_F/\omega_c=6$.}
	\label{sec:conductivity:fig:SdH_beating}
\end{figure}

For small $T,\omega\ll\ep_F$ we evaluate the smooth functions $\sigma_{D,\pm}$ and $\lambda$ on-shell 
but keep the energy dependence in the rapidly oscillating parts while calculating the conductivity \eqref{sec:formalism:eq:Kubo_formula}. The asymptotics of the resulting Fresnel integrals provides a reasonably simple expression for the correction to the Drude conductivity away from the cyclotron resonance at $T=0$,
\begin{equation}
	\sigma^{\hspace{-0.2mm}(1)}_\pm(\omega) = 4\sigma_{D,\pm} \lambda \cos\left(\frac{2\pi(\ep_F^2+\omega^2)}{\omega_c^2}\right)\frac{\sin\frac{2\pi\omega}{\omega_c^{\text{loc}}}}{2\pi\omega/\omega_c^{\text{loc}}}
	\: .
	\label{sec:conductivity:eq:conductivity_1st_order}
\end{equation}
Equation \eqref{sec:conductivity:eq:conductivity_1st_order} describes SdH oscillations in the dynamic conductivity illustrated in Fig.~\ref{sec:conductivity:fig:SdH_beating}.

The fast harmonic oscillations with the local cyclotron frequency $\omega_c^{(\text{loc})}$ are similar to those known for systems with a parabolic spectrum except for the absence of the vacuum shift due to the Berry's phase of Dirac fermions and hence a shift in the zero of energy to ${\ep_F/2-\omega_c^{(\text{loc})}}$.

However, the magnetooscillations in Fig.~\ref{sec:conductivity:fig:SdH_beating} further show a slow modulation on the scale of the cyclotron frequency $\omega_c$ at the Dirac point. This beating in the quantum oscillations is due to the difference of the cyclotron frequencies at the initial and final state of the optical transition. Remarkably, the quantum oscillations show nodes stemming from destructive interference of the density of states in the initial and final state. If the Fermi energy $\ep_F$ is situated in the center of a LL, the node occurs at
\begin{equation}
	\omega = \sqrt{2k+1} \: \frac{\omega_c}{2} \: , \quad k \in \mathbb{N}
	\: .
	\label{sec:conductivity:eq:nodes}
\end{equation}
Detuning of the Fermi energy from the center of the LL shifts the frequency nodes such that ${\ep_F^2+\omega^2=\omega_c^2(n/2+1/4)}$, $n \in \mathbb{N}$ holds. Note that although we assumed ${\omega\ll\ep_F}$ the frequencies in Eq.~\eqref{sec:conductivity:eq:nodes} are well within the range of our approximation for not too large $k$ in view of $\ep_F\gg\omega_c$.
For the values of $\omega$ in Eq.~\eqref{sec:conductivity:eq:nodes} the DOS oscillations acquire a relative phase shift of approximately $(2 k+1)\pi$ between $\ep$ and $\ep+\omega$ due to the $\ep$-variation of the period which leads to the destructive interference.

At $T>T_D$, the temperature smearing dominates over the quantum mechanical smearing. However, there are additional quantum oscillations $\propto\lambda^2$ that survive higher temperatures and become exponentially larger than SdH oscillations at $T\gg T_D$. They are well known for systems with a parabolic spectrum\cite{Ando75CRShape,Ivan03CRHarmonics} and originate from the terms in the $RA$-sector in Eq.~\eqref{sec:conductivity:eq:Pi_correlator_2} which do not oscillate with mean energy $\ep_1+\ep_2$ but do oscillate with the energy difference $\omega\equiv|\ep_1-\ep_2|$. Such $\omega$-oscillations are insensitive to the position of the chemical potential with respect to LLs. Therefore, averaging over the temperature window according to Eq.~\eqref{sec:formalism:eq:Kubo_formula} does not lead to additional damping in contrast to SdH oscillations.
The relevant contribution at $T\gg T_D$ is
\begin{equation}
	\begin{split}
		\sigma^{(q)}_\pm(\omega)
		\:=\:
		2 \sigma_{D,\pm} \:
		\lambda^2
		\bigg( \: &
			\frac{a_\pm^2-1}{a_\pm^2+1} \: \cos \frac{2\pi\omega}{\omega_c^{\text{loc}}} \\
			& \hspace{-25mm} \:+\:
			\frac{2a_\pm}{a_\pm^2+1} \:  \sin \frac{2\pi\omega}{\omega_c^{\text{loc}}}
		\bigg)
		\:
		\frac{
			\sin\frac{2\pi \omega^2}{\omega_c^2}
		}
		{
			2\pi\omega^2/\omega_c^2
		}
		\: .
		\label{sec:conductivity:eq:highT_QO_1}
	\end{split}
\end{equation}
Away from the cyclotron resonance, $a_\pm\gg1$, the quantum correction \eqref{sec:conductivity:eq:highT_QO_1} acquires a simpler form,
\begin{equation}
	\sigma^{(q)}_{\pm}(\omega)
	\:\simeq\:
	2 \: \sigma_{D,\pm} \:
	\lambda^2  \:
	\frac{
		\sin\frac{2\pi \omega^2}{\omega_c^2}
	}
	{
		2\pi\omega^2/\omega_c^2
	}
	\:\cos \frac{2\pi\omega}{\omega_c^{\text{loc}}}
	\: .
	\label{sec:conductivity:eq:highT_QO_2}
\end{equation}
\begin{figure}[ht]
	\centering
	\includegraphics[width=0.45\textwidth]{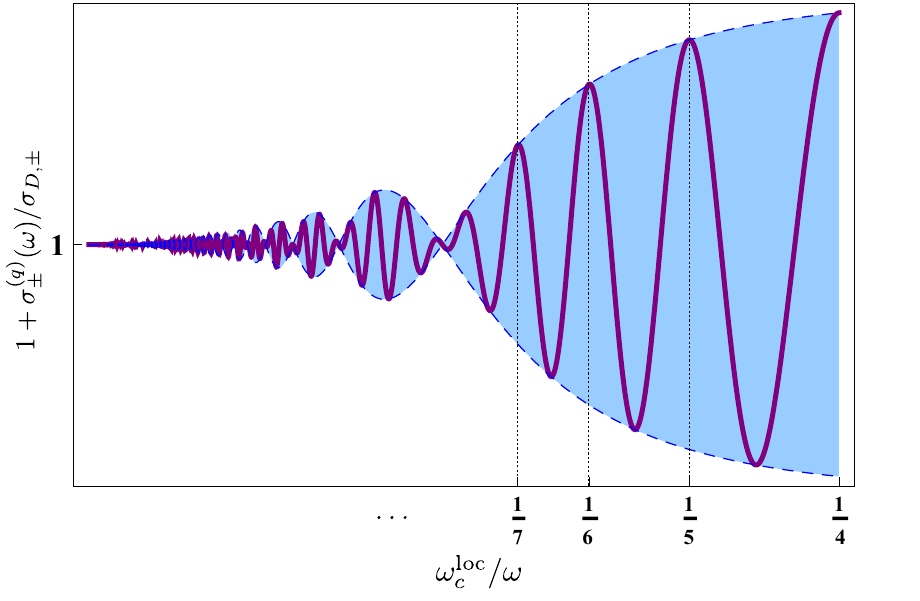}
	\caption{Magnetoconductivity $\sigma_{D,\pm}(\omega)+\sigma^{(q)}_{\pm}(\omega)$ of graphene at high temperatures [normalized to the Drude value \eqref{sec:conductivity:eq:Drude}], calculated according to \eqref{sec:conductivity:eq:highT_QO_2} for $\ep_F/\omega_c=6$. Vertical lines mark the position of integer harmonics of the cyclotron resonance.}
	\label{sec:conductivity:fig:2nd_order_QO}
\end{figure}

The conductivity at high temperature $T\gg T_D$ is plotted in Fig.~\ref{sec:conductivity:fig:2nd_order_QO}. Apart from the integer cyclotron resonance harmonics $\omega=n \omega_c^\text{loc}$, $n\in\mathbb{N}$, that have an analog in systems with a parabolic spectrum, we encounter again an additional modulation of the quantum oscillations due to the presence of multiple cyclotron frequencies. Since the temperature-stable quantum corrections $\propto\lambda^2$ are insensitive to the position of the Fermi energy with respect to LLs, the positions of the nodes, $\omega/\omega_c=\sqrt{n/2}$, $n\in \mathbb{N}$, are therefore solely determined by the probing frequency $\omega$.

\section{Separated Landau levels \label{sec:cond_separated}}\noindent
We now address the quantum regime of well resolved LLs, which involves LLs with numbers $|n|<N_\text{ov}$, see Eq.~\eqref{sec:formalism:eq:crossover}. In the following, $L_K$ denotes the K-th LL.

\begin{figure}[ht]
	\centering
	\includegraphics[width=0.5\textwidth]{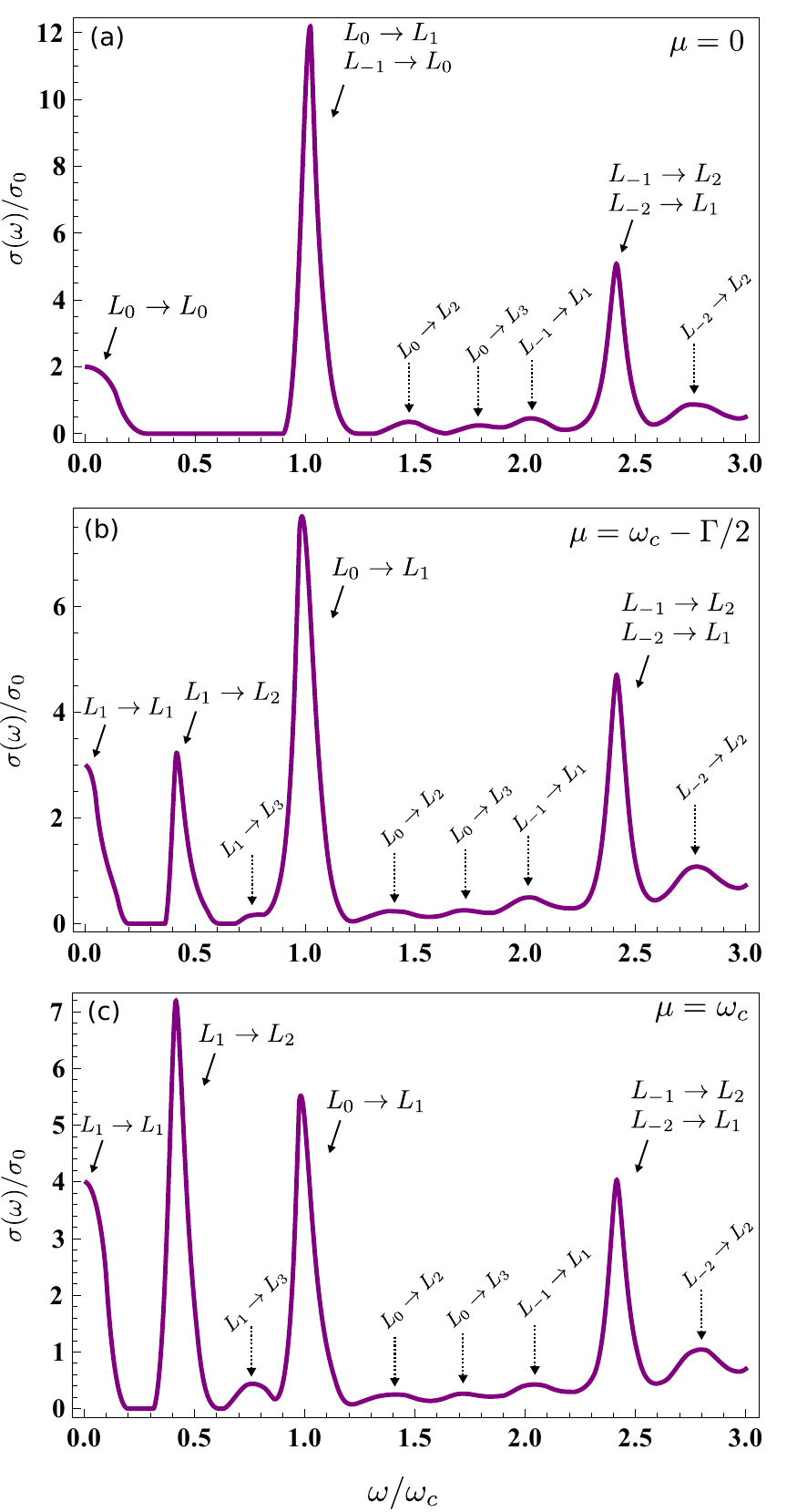}
	\caption{The dynamic conductivity according to Eq.~\eqref{sec:conductivity:eq:conductivity_sep_LL} for different chemical potentials and $T=0$. The cyclotron resonances are indicated by the solid arrows and the participating LLs, where $L_n$ denotes the n-th LL. The dashed arrows mark the disorder-induced transitions. The spectrum is calculated for a disorder strength $\alpha=0.01$.}
	\label{sec:conductivity:fig:cond_spectrum}
\end{figure}

Our results for the dynamic conductivity at $T=0$ are illustrated in Fig.~\ref{sec:conductivity:fig:cond_spectrum}. As long as LLs are separated, the conductivity is the sum of contributions from individual transitions between $L_{K}$ and $L_{M}$. Despite LLs are separated, the corresponding peaks in $\sigma(\omega)$ overlap. Indeed, some transitions are degenerate (for instance $L_0\rightarrow L_1$ and $L_{-1}\rightarrow L_0$ in Fig.~\ref{sec:conductivity:fig:cond_spectrum}). In other cases, the excitation energies for different transitions may become close to each other due to the nonequidistant spectrum of LLs.

From the Kubo formula~\eqref{sec:formalism:eq:Kubo_formula} we find that to leading order in $\Gamma/\omega_c$ the conductivity kernel is proportional to the density of the initial and final states. 
In this section, we neglect the renormalization of the energy described by $Z(\ep)$, Eq.~\eqref{sec:formalism:eq:renorm_const}, assuming that the contribution of states with energies $|\ep|<\Delta_c\ee^{-1/\alpha}$ is negligible. We thus put $Z(\ep)=1$ in Eqs.~\eqref{sec:formalism:eq:selfenergy_separated_1}-\eqref{sec:formalism:eq:gamma0}, which gives $\text{Im}\Sigma_a^R=0$ in Eq.~\eqref{sec:formalism:eq:selfenergy_separated_2_1}. Using Eq.~\eqref{sec:formalism:eq:DOS_definition}, we obtain the partial DOS
\begin{equation}
 	\nu_{K\neq0}(\ep) = \frac{1}{l_B^2\pi^2\Gamma^2} \: \text{Re}\sqrt{\Gamma^2-(\ep-E_K)^2} \: , \quad \Gamma=\sqrt{\alpha}\omega_c \: ,
	\label{sec:conductivity:eq:dos_sep_1}
\end{equation}
in either sublattice $a$ or $b$ for $K\neq0$, and
\begin{equation}
 	\nu_{K=0}(\ep) = \frac{1}{l_B^2\pi^2\Gamma^2} \: \text{Re}\sqrt{2\Gamma^2-(\ep)^2} \: ,
	\label{sec:conductivity:eq:dos_sep_2}
\end{equation}
residing in sublattice $b$ for $K=0$.
The total DOS, including both valleys, spin components and sublattices is given by $8\nu_{K\neq0}$ for $K\neq0$ and by $4\nu_{K=0}$ for $K=0$.
The width of the individual resonances is then $2(\Gamma_N+\Gamma_K)$, determined by the width of the DOS in $L_{K}$ and $L_{M}$. In the following we distinguish the cyclotron resonance peaks, $||K|-|M||=1$, from the disorder-induced peaks, $||K|-|M||\neq 1$. The latter vanishes if the disorder is switched off, while the former survive as they respect the clean selection rules in graphene.

Using $\omega_c\gg\Gamma$ for the $\Pi$-correlators~\eqref{sec:conductivity:eq:Pi_correlator_2}, 
we cast the conductivity~\eqref{sec:formalism:eq:Kubo_formula} in the form
\begin{equation}
	\begin{split}
		\sigma(\omega) = &  \sigma_0 \: \sum_{K,M} \: \mathcal{P}(K,M) \: \frac{\omega_c^2 c_K c_M}{\Gamma_K\Gamma_M} \: \tilde{\mathcal{F}}_{KM}(\omega,\mu,T) \: ,
	\end{split}
	\label{sec:conductivity:eq:conductivity_sep_LL}
\end{equation}
where $\sigma_0=e^2/4\pi^2$ and $c_K=1+\delta_{K,0}$, see App.~\ref{sec:app_vertex_corrections} and \ref{sec:app_Fintegrals} for details. The coefficients
\begin{equation}
	\mathcal{P}(K,M) = 1,\qquad ||K|-|M||=1,
	\label{sec:conductivity:eq:trans_prob_off_res_0}
\end{equation}
for the cyclotron resonance transitions;
for the disorder-induced transitions not involving the zeroth LL ($||K|-|M||\neq1$, ${K\neq0}$ and ${M\neq0}$), 
\begin{equation}
	\mathcal{P}(K,M) = \frac{\Gamma^2(E_M^2+E_K^2)}{2(E_K^2-E_M^2+\omega_c^2)^2} + \{M\leftrightarrow K\} \: ;
	\label{sec:conductivity:eq:trans_prob_off_res_1}
\end{equation}
finally, for the disorder-induced transitions involving the zeroth LL ($||K|-|M||\neq1$, ${K=0}$ or ${M=0}$)
\begin{equation}
	\mathcal{P}(K,M) = \frac{\Gamma^2(E_K^2+\omega_c^2)}{2(E_M^2-E_K^2+\omega_c^2)^2} + \{M\leftrightarrow K\} \: .
	\label{sec:conductivity:eq:trans_prob_off_res_2}
\end{equation}

\begin{figure}[ht]
	\centering
	\includegraphics[width=0.45\textwidth]{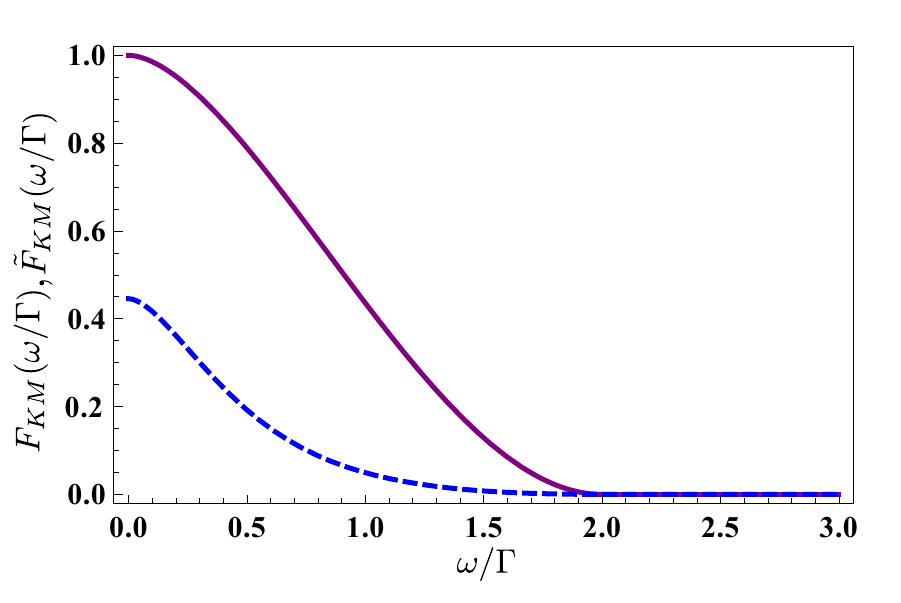}
	\caption{Solid line: The function $F_{KM}$, Eq.~\eqref{sec:app_Fintegrals:eq:fctFKM}, describing the shape of the disorder-induced conductivity peaks ($||K|-|M||\neq1$) in the regime $T,\omega_c\gg\Gamma$, see Eqs.~\eqref{sec:conductivity:eq:fctF_high_T_1} and \eqref{sec:conductivity:eq:conductivity_sep_LL}. Dashed line: The function $\tilde{F}_{KM}$, Eq.~\eqref{sec:app_Fintegrals:eq:fctTildeF_vertex}, replacing $F_{KM}$ in the case of the cyclotron resonance ($||K|-|M||=1$). The strength and the width of the cyclotron resonance is reduced due to the vertex corrections as compared to $F_{KM}$ that would result from the calculation with the bare polarization bubble.
	Here we assume $K,M \neq 0$, in which case $F_{KM}$ and $\tilde{F}_{KM}$ do not depend on $K$ and $M$.}
	\label{sec:conductivity:fig:functionF}
\end{figure}

The function $\tilde{\mathcal{F}}_{KM}$ in Eq.~\eqref{sec:conductivity:eq:conductivity_sep_LL} describes the shape of the peaks in the conductivity; its general form is given in App.~\ref{sec:app_Fintegrals}, see Eq.~\eqref{sec:app_Fintegrals:eq:fctTildecalF_vertex}. In the case of the disorder-induced transitions $||K|-|M||\neq1$, the vertex corrections are negligible: the conductivity kernel can be calculated using the bare polarization bubble and depends on energy only via the product $\nu_K(\ep)\nu_M(\ep+\omega)$. Correspondingly, for $||K|-|M||\neq1$ the function $\tilde{\mathcal{F}}_{KM}$ in Eq.~\eqref{sec:conductivity:eq:conductivity_sep_LL} is reduced to the bare $\mathcal{F}_{KM}\simeq\tilde{\mathcal{F}}_{KM}$, given by
\begin{equation}
	\begin{split}
		& \mathcal{F}_{KM}(\omega,\mu,T) = \frac{l_B^4\pi^4\Gamma_K\Gamma_M}{c_K c_M} \\
		& \times \int\dd\ep \: \frac{f_{\ep}-f_{\ep+\omega}}{\omega} \: \nu_{K}(\ep)\nu_{M}(\ep+\omega) \: ,
	\end{split}
	\label{sec:conductivity:eq:fctF_def}
\end{equation}
see Eq.~\eqref{sec:app_Fintegrals:eq:fctCalFKM}. At high $T\gg\Gamma$ and $K\neq M$, the distribution function can be considered smooth on the scale $\Gamma$; 
Eq.~\eqref{sec:conductivity:eq:fctF_def} reduces to
\begin{equation}
	\mathcal{F}_{KM}(\omega,\mu,T) = \frac{4\Gamma_K}{3\omega} \: (f_K-f_M) \: F_{KM}(\delta\omega/\Gamma) \: ,
	\label{sec:conductivity:eq:fctF_high_T_1}
\end{equation}
where we denote $f_K=f(E_K)$ and ${\delta\omega = \omega-E_M+E_K}$. The function $F_{KM}$ is given by Eq.~\eqref{sec:app_Fintegrals:eq:fctFKM}, see also Fig.~\ref{sec:conductivity:fig:functionF}. For $T\gg\Gamma$ and $K=M$ (intra-LL transitions),
\begin{equation}
	\mathcal{F}_{KK}(\omega,\mu,T) = \frac{\Gamma_K \: F_{KK}(\delta\omega/\Gamma_K)}{3T\cosh^2[(E_K-E_N)/2T]} \: ,
	\label{sec:conductivity:eq:fctF_high_T_2}
\end{equation}
where $N$ is the LL closest to the chemical potential $\mu$. % The function $F=F_{KK}$ is given by Eq.~\eqref{sec:app_Fintegrals:eq:fctF} 
 In the case of disorder-induced transitions, the conductivity vanishes in the clean limit $\Gamma\rightarrow 0$.

For the cyclotron resonance ${||N|-|M||=1}$, the vertex corrections are important leading to a more complicated $\ep$-dependence of the kernel; the corresponding functions $\tilde{\mathcal{F}}$ and $\tilde{F}$ entering Eq.~\eqref{sec:conductivity:eq:conductivity_sep_LL} are given in App.~\ref{sec:app_Fintegrals}.
The behavior of the functions $F$ and $\tilde{F}$ is illustrated in Fig.~\ref{sec:conductivity:fig:functionF}. 
The figure shows that the vertex corrections reduce the strength and width of the cyclotron resonances.

We now discuss several limiting cases which describe the rich pattern of resonances in Fig.~\ref{sec:conductivity:fig:cond_spectrum}.

\subsection{Intra-Landau level transitions} \noindent
We start with the case $\omega\lesssim\Gamma$ enabling only intra-LL transitions. This includes the dc limit $\omega\rightarrow 0$ and two distinct temperature regimes: a) $T\ll\Gamma$ and b) $T\gg\Gamma$. The contribution of the intra-LL transitions $L_{0}\rightarrow L_{0}$ and $L_{1}\rightarrow L_{1}$ is illustrated in Fig.~\ref{sec:conductivity:fig:cond_spectrum}.

\textbf{a)} When the temperature is the smallest scale, $T\ll\Gamma$, only the level $L_N$, determined by the position of the chemical potential, contributes. The conductivity~\eqref{sec:conductivity:eq:conductivity_sep_LL} becomes ($\sigma_0=e^2/4\pi^2$)
\begin{equation}
	\sigma = \sigma_0 \: \frac{c_N\omega_c^2}{\omega_N^2} \: \mathcal{F}_{NN}(\omega,\mu,T) \: .
	\label{sec:conductivity:eq:intra_lowT}
\end{equation}
In the dc limit $\omega\rightarrow 0$, Eq.~\eqref{sec:conductivity:eq:intra_lowT} acquires the form\cite{AndoJPSJ2DGraphite98}
\begin{equation}
	\sigma =  \sigma_0 \: \frac{c_N\omega_c^2}{\omega_N^2} \: \left(1-\frac{(\mu-E_N)^2}{\Gamma_N^2}\right) \: .
	\label{sec:conductivity:eq:dc_cond_lowT}
\end{equation}
Note that $\mathcal{F}$ is close to unity for small frequencies, hence the dependence on the disorder strength drops out for $\mu=E_N$. In this regime, the dependence $\sigma\propto\tau_\text{tr}^{-1}\propto\alpha$ characteristic for classically strong magnetic fields, $\omega_N\tau_\text{tr}\gg1$, is exactly compensated by the increased DOS inside the LL: the average of $\nu(\ep)^2$ over $L_N$ is proportional to $\omega_c^2/\Gamma^2\propto\alpha^{-1}$.

\textbf{b)} $T\gg\Gamma$, hence $T\gg\omega$: With the help of the high T expression~\eqref{sec:conductivity:eq:fctF_high_T_2}, the conductivity~\eqref{sec:conductivity:eq:conductivity_sep_LL} reads
\begin{equation}
	\sigma = \sigma_0 \sum_{n=-\infty}^\infty \: \frac{c_n\omega_c^2\Gamma_n \: F_{nn}(\omega/\Gamma_n)}{3\omega_n^2 T \cosh^2[(E_n-E_N)/2T]} \: .
	\label{sec:conductivity:eq:intra_highT}
\end{equation}
The summation limits are sent to infinity since the contribution of large energies where LLs overlap is exponentially small.

In Eq.~\eqref{sec:conductivity:eq:intra_highT} the zeroth LL is special since its width is bigger by a factor of $\sqrt{2}$ and its oscillator strength is enhanced by a factor of two. However, its contribution is significant only for $N=0$ and $T\lesssim\omega_c$. Note that for all other levels ($n\neq0$) the function $F_{nn}(\omega/\Gamma_n)$ does not depend on the LL index $n$. We obtain three high $T$ regimes:

\textbf{b.1)} For ${\Gamma\ll T\ll\omega_N}$ only $L_N$ contributes, while the contribution from the levels farther away from the chemical potential is exponentially suppressed,
\begin{equation}
	\sigma = \sigma_0 \: \frac{c_N \omega_c^2 \Gamma_N}{3\omega_N T}\: F_{NN}(\omega/\Gamma_N) \: .
	\label{sec:conductivity:eq:intra_highT_1}
\end{equation}
The conductivity $\sigma\propto T^{-1}$ is proportional to the slope of the Fermi function in $L_N$, $f\simeq1/2-(\ep-E_N)/4T$. Note that the width of the peak in Eq.~\eqref{sec:conductivity:eq:intra_highT_1} for $N=0$ is bigger by a factor $\sqrt{2}$ in view of $\Gamma_0=\sqrt{2}\Gamma$.

\textbf{b.2)} ${\omega_N\ll T\ll\mu<\ep_\text{ov}}$: As $\mu\gg T$, the influence of zeroth LL can be neglected. Furthermore, at $T\gg\omega_N$ the sum in Eq.~\eqref{sec:conductivity:eq:intra_highT} can be converted into an integral, which gives 
\begin{equation}
	\sigma = \sigma_0 \: \frac{64\Gamma|\mu|^3}{3\omega_c^4}\: F_{NN}(\omega/\Gamma) \: .
	\label{sec:conductivity:eq:intra_highT_2}
\end{equation}
The conductivity in this regime is $T$-independent.

\textbf{b.3)} For ${\omega_N,\mu\ll T<\ep_\text{ov}}$ we obtain
\begin{equation}
	\sigma = \sigma_0 \: \frac{96\zeta(3)\Gamma T^3}{\omega_c^4} \: F_{NN}(\omega/\Gamma) \: ,
	\label{sec:conductivity:eq:intra_highT_3}
\end{equation}
where $\zeta(z)$ is the Riemann $\zeta$-function. We observe that the temperature takes the role of the chemical potential in Eq.~\eqref{sec:conductivity:eq:intra_highT_2}.

The dc-limit $\omega\rightarrow 0$ of Eqs.~\eqref{sec:conductivity:eq:intra_highT_1},~\eqref{sec:conductivity:eq:intra_highT_2} and \eqref{sec:conductivity:eq:intra_highT_3} is obtained using $F_{NN}(0)=1$, and hence shows the same dependence on $\mu$ and $T$.

\begin{figure}[ht]
	\centering
	\includegraphics[width=0.45\textwidth]{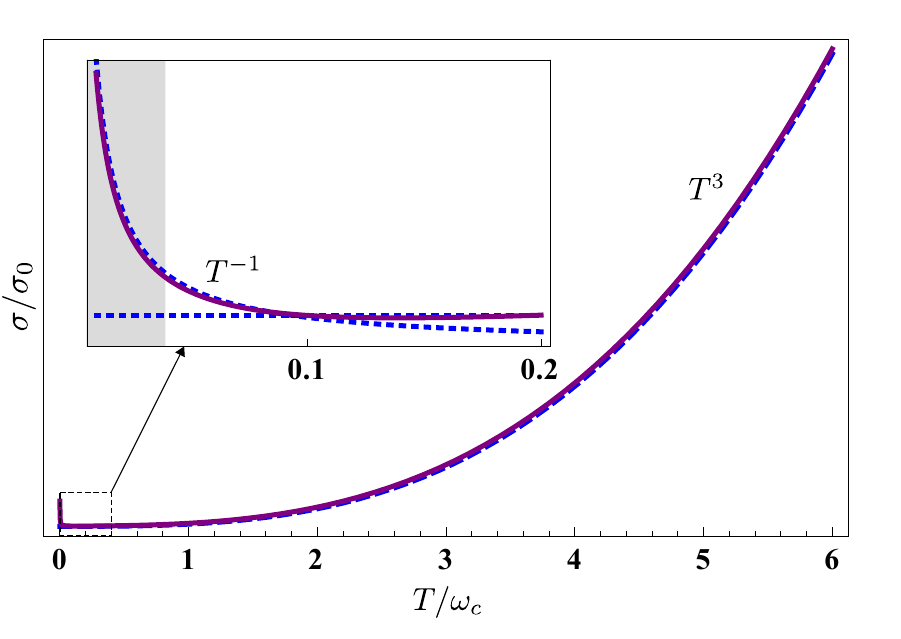}
	\caption{The temperature dependence of the conductivity for $T\gg\Gamma_N>\omega$ and $\mu=\omega_c$. The inset shows the small temperature behavior according to Eqs.~\eqref{sec:conductivity:eq:intra_highT_1} and \eqref{sec:conductivity:eq:intra_highT_2} (dashed lines), describing the $1/T$ decrease followed by a saturation into a T-independent regime. The thick line is calculated according to Eq.~\eqref{sec:conductivity:eq:intra_highT}. The shaded area indicates the regime $T\lesssim\Gamma$, where the conductivity saturates at a $\mu$-dependent value, see Eq.~\eqref{sec:conductivity:eq:intra_lowT}.}
	\label{sec:conductivity:fig:smallMu_T_dep_1st_level}
\end{figure}
\begin{figure}[ht]
	\centering
	\includegraphics[width=0.45\textwidth]{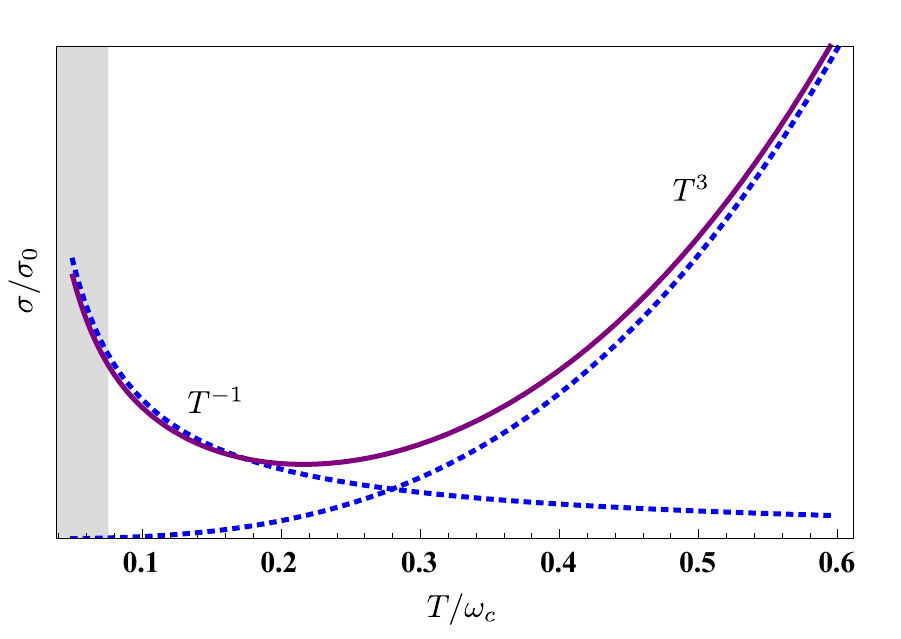}
	\caption{The temperature dependence of the conductivity for $\omega<\Gamma$ and $|\mu|<\Gamma_0\ll T$. The dashed lines are the contribution from the zeroth LL [Eq.~\eqref{sec:conductivity:eq:intra_highT_1}] and all other levels [Eq.~\eqref{sec:conductivity:eq:intra_highT_3}]. The thick line is calculated according to Eq.~\eqref{sec:conductivity:eq:intra_highT}. The shaded area indicates the regime $T\lesssim\Gamma$, where the conductivity saturates at a $\mu$-dependent value, see Eq.~\eqref{sec:conductivity:eq:intra_lowT}.}
	\label{sec:conductivity:fig:smallMu_T_dep_dp}
\end{figure}

The overall $T$-dependence of the dc conductivity (${\omega\rightarrow 0}$) for $T\gg\Gamma$ is shown in Fig.~\ref{sec:conductivity:fig:smallMu_T_dep_1st_level} (for $\mu\simeq \omega_c$) and in Fig.~\ref{sec:conductivity:fig:smallMu_T_dep_dp} (for $\mu\simeq\Gamma$). In both cases it shows a nonmonotonous temperature dependence. 
In Fig.~\ref{sec:conductivity:fig:smallMu_T_dep_dp} the $T$-independent regime \eqref{sec:conductivity:eq:intra_highT_2} is not present since the corresponding conditions cannot be met. At small $T$ the contribution from $L_0$ dominates. It decreases due to thermal smearing within the zeroth LL. With increasing $T$ the influence of $L_0$ decreases and the other LLs take over, which leads to an enhancement of the conductivity due to thermal activation of higher energy states. In Figs.~\ref{sec:conductivity:fig:smallMu_T_dep_1st_level}~and~\ref{sec:conductivity:fig:smallMu_T_dep_dp} the shaded areas indicate the crossover to the regime~\eqref{sec:conductivity:eq:intra_lowT}, where the conductivity saturates at a $\mu$-dependent value. The unusual $T^3$ and $\mu^3$ dependence originates from the interplay between the energy dependence of the transition rates~\eqref{sec:conductivity:eq:trans_prob_off_res_1}~and~\eqref{sec:conductivity:eq:trans_prob_off_res_2} and the level spacing~\eqref{sec:formalism:eq:loc_cyclotron_freq_asymptotic}. It is therefore special for Dirac fermions and hence graphene.

In the dc limit $\omega\rightarrow0$, Eqs.~\eqref{sec:conductivity:eq:intra_highT_3}~and~\eqref{sec:conductivity:eq:intra_highT_2} reproduce the results of Ref.~\onlinecite{MagnetoRes2012Ioffe}, where the regime $T\gg\omega_N$ was studied for the dc conductivity. In addition, we find the $1/T$ behavior, Eq.~\eqref{sec:conductivity:eq:intra_highT_1}, in the low-T range of $\Gamma\ll T\ll\omega_N$.

\subsection{Inter-Landau level transitions} \noindent
Figure~\ref{sec:conductivity:fig:cond_spectrum} demonstrates that the nonequidistant LL spectrum of graphene leads to a rich spectrum of resonances. Their strength depends strongly on the chemical potential. In what follows we discuss separately a) the cyclotron resonance transitions and b) disorder-induced transitions.

\begin{figure}[ht]
	\centering
	\includegraphics[width=0.45\textwidth]{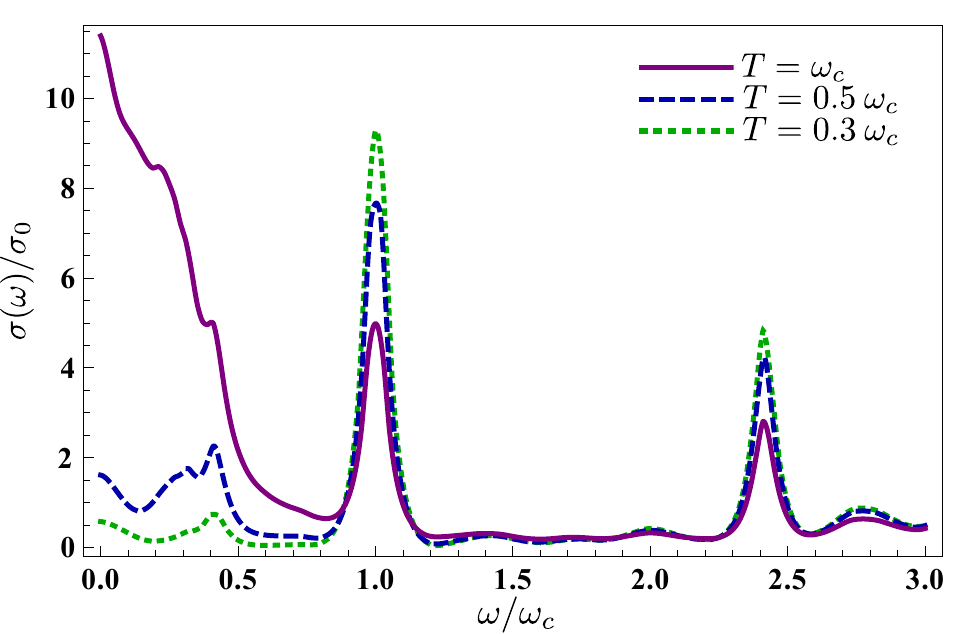}
	\caption{The dynamic conductivity, Eq.~\eqref{sec:conductivity:eq:conductivity_sep_LL}, for $\mu=0$ and three different temperatures. The spectrum is calculated for a disorder strength $\alpha=0.01$.}
	\label{sec:conductivity:fig:cond_highT}
\end{figure}

\textbf{a)} We start with the cyclotron resonances, $||K|-|M||=1$. Apart from the transition $L_K\rightarrow L_M$ also the transition $L_{-M}\rightarrow L_{-K}$ needs to be taken into account, since it has the same transition energy.

\textbf{a.1)} For $T\ll\Gamma$, Eq.~\eqref{sec:conductivity:eq:conductivity_sep_LL} reduces to
\begin{equation}
	\sigma(\omega) = \frac{c_K c_M \sigma_0\omega_c^2}{\Gamma_K\Gamma_M} \:[ \tilde{\mathcal{F}}_{KM}(\omega,\mu,T) + \tilde{\mathcal{F}}_{-M,-K}(\omega,\mu,T) ] \: .
	\label{sec:conductivity:eq:cyc_res_lowT}
\end{equation}

\textbf{a.2)} For $T\gg\Gamma$, Eq.~\eqref{sec:conductivity:eq:conductivity_sep_LL} yields
\begin{equation}
	\sigma(\omega) = \sigma_0 \: \frac{c_K c_M 4\omega_c^2}{3\Gamma\omega} \: (f_{K}-f_{M}+f_{-M}-f_{-K}) \: \tilde{F}_{KM}\left(\frac{\delta\omega}{\Gamma}\right) \: .
	\label{sec:conductivity:eq:cyc_res_highT}
\end{equation}
Here $f_{K}$ are introduced below Eq.~\eqref{sec:conductivity:eq:fctF_high_T_1}, and the functions $\tilde{\mathcal{F}}$ and $\tilde{F}$ are given by Eqs.~\eqref{sec:app_Fintegrals:eq:fctTildecalF_vertex}~and~\eqref{sec:app_Fintegrals:eq:fctTildeF_vertex}. For $M=0$, $K=-1$ the occupation of $L_0$ drops out,
\begin{equation}
	\sigma(\omega) = \sigma_0 \: \frac{8\omega_c^2}{3\Gamma\omega} \: (f_{-1}-f_{1}) \: \tilde{F}_{-1,0}(\delta\omega/\Gamma) \: .
	\label{sec:conductivity:eq:cyc_res_0th_highT}
\end{equation}
In Eqs.~\eqref{sec:conductivity:eq:cyc_res_highT} and \eqref{sec:conductivity:eq:cyc_res_0th_highT}, $\mu$ and $T$ enter via $f_{K,M}$ only. The corresponding expression in brackets takes values between zero and one.

\textbf{b)} $||K|-|M||\neq 1$: According to Eqs.~\eqref{sec:conductivity:eq:conductivity_sep_LL}~and~\eqref{sec:conductivity:eq:trans_prob_off_res_1}, the partial contribution of the disorder-induced transitions $L_K\rightarrow L_M$ in the case $K,M\neq 0$ is
\begin{equation}
	\sigma_{KM} = \sigma_0 \: \frac{2(|K|+|M|)[1+(|K|-|M|)^2]}{[(|K|-|M|)^2-1]^2} \: \mathcal{F}_{KM}(\omega,\mu,T) \: .
	\label{sec:conductivity:eq:dis_ind_lowT}
\end{equation} 
For $M=0$ or $K=0$,
\begin{equation}
	\begin{split}
		& \sigma_{KM} = \sigma_0 \: \frac{c_K c_M \Gamma^2}{2\Gamma_K\Gamma_M} \: \mathcal{F}_{KM}(\omega,\mu,T) \\
		& \times \frac{(K+1)[M-(K+1)]^2+(M+1)[K-(M+1)]^2}{[(M-K)^2-1]^2}  \: .
	\end{split}
	\label{sec:conductivity:eq:dis_ind_lowT_0thLL}
\end{equation}
where the function $\mathcal{F}_{KM}$ is given by Eq.~\eqref{sec:app_Fintegrals:eq:fctF_intra}.

To make the result more transparent, we rewrite Eq.~\eqref{sec:conductivity:eq:dis_ind_lowT} for a particular set $L_{-M}\rightarrow L_M$ ($M>0$) of mirror transitions and for $T\gg\Gamma$,
\begin{equation}
	\sigma(\omega) = \sigma_0 \: \frac{4\omega_c^2\Gamma}{3\omega_M^2\omega} \: (f_{-M}-f_{M}) \: F_{MM}(\delta\omega/\Gamma) \: .
	\label{sec:conductivity:eq:mirror_transition_1}
\end{equation}
The function $F_{MM}$ is illustrated in Fig.~\eqref{sec:conductivity:fig:functionF}. Since $\omega\simeq 2E_M$, it holds
\begin{equation}
	\frac{\omega_c^2\Gamma}{\omega_M^2\omega} = \frac{\omega\Gamma}{\omega_c^2}\: .
	\label{sec:conductivity:eq:mirror_transition_rel}
\end{equation}
We see that the strength of the disorder-induced transition is enhanced with increasing frequency ${\omega\propto\sqrt{M}}$.

Our results are illustrated in Figs.~\ref{sec:conductivity:fig:cond_spectrum} and \ref{sec:conductivity:fig:cond_highT}. The strongest response corresponds to the cyclotron resonances. Indeed, Eqs.~\eqref{sec:conductivity:eq:cyc_res_lowT} - \eqref{sec:conductivity:eq:cyc_res_0th_highT} show that the cyclotron peaks are a factor $\omega_c^2/\Gamma^2$ stronger than the disorder-induced peaks \eqref{sec:conductivity:eq:dis_ind_lowT} - \eqref{sec:conductivity:eq:mirror_transition_1}. Additionally, we observe that the interband cyclotron resonance is suppressed by a factor $\omega_c/\omega<1$ in comparison to the intraband cyclotron resonances, for which $\omega_c/\omega\ge 1$.

Among the disorder-induced transitions, the intra-LL and the mirror transitions $L_{-M}\rightarrow L_{M}$ are the strongest. The latter become more pronounced with increasing frequency.
The presence of the $L_{1}\rightarrow L_{3}$ and the $L_{1}\rightarrow L_{2}$ peaks in Figs.~\ref{sec:conductivity:fig:cond_spectrum}(b)~and~(c) indicates that the chemical potential lies in the first LL. Furthermore, the intensity ratio of the $L_{1}\rightarrow L_{2}$ and the $L_{0}\rightarrow L_{1}$ peaks provides information on the occupation of $L_1$. For instance, the comparison of the calculated spectra in Fig.~\ref{sec:conductivity:fig:cond_spectrum} to the measurements reported in Ref.~\onlinecite{SadowskiGrLayers06} shows that in this particular experiment the chemical potential was lying in $L_1$; since the $L_0\rightarrow L_1$ resonance was stronger than the $L_1\rightarrow L_2$ resonance, we conclude that $L_1$ was less than half-filled.
No measurements so far have reported the disorder-induced transitions. However, a closer inspection of Fig.~$1$(a)~and~(b) from Ref.~\onlinecite{Pot11Scattering} reveals a peak at roughly $140\:\text{meV}$ which should be attributed the disorder-induced transition $L_{-1}\rightarrow L_1$ according to its position with respect to the cyclotron resonances $L_0\rightarrow L_1$ and $L_1 \rightarrow L_2$.

In Fig.~\ref{sec:conductivity:fig:cond_highT} we illustrate the $T$-dependence of the conductivity for $\mu=0$. The strength of the cyclotron resonance decreases to accommodate the increase of resonances at lower frequencies coming from thermal occupation of higher LLs with smaller local cyclotron frequency. In particular, at moderate $T=0.3\:\omega_c$ the $L_{1}\rightarrow L_{2}$ transition becomes visible, while for higher $T$ it is absorbed into the background formed by other transitions.

\section{Single resolved Landau level \label{sec:single_level}}
\begin{figure}[h]
        \centering
        \includegraphics[width=0.5\textwidth]{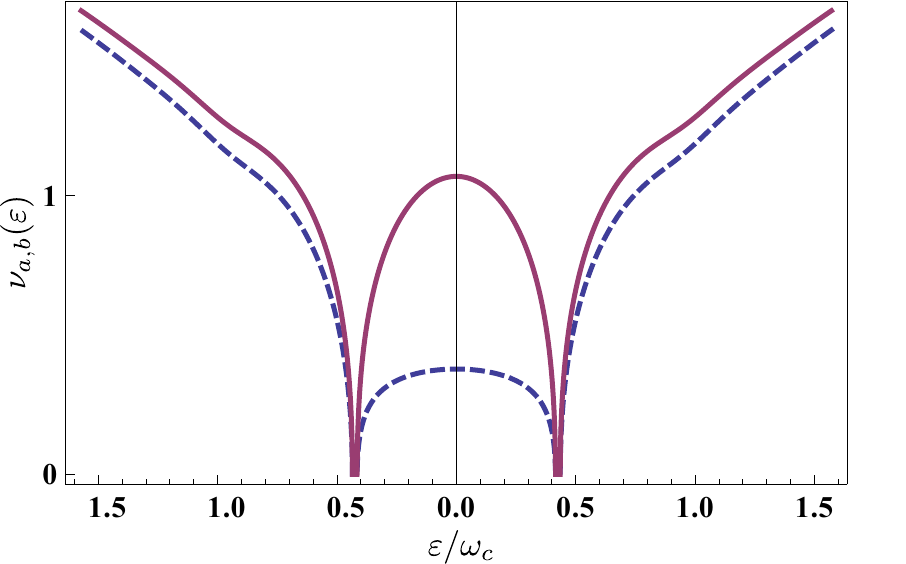}
        \caption{The DOS near the Dirac point in the sublattice $b$ (solid line) and $a$ (dashed line) obtained numerically from
	Eqs.~\eqref{sec:formalism:eq:selfenergy_1} and \eqref{sec:formalism:eq:DOS_definition} for $\alpha=0.2$.
	}
        \label{sec:conductivity:fig:DOS_single_level}
\end{figure}

\noindent
We finally consider the specific case of relatively dirty graphene with
only one resolved LL at zero energy $L_0$ within a quasi-continuum
$\mathcal{C}$. Due to the nonequidistant LL spectrum of graphene, this
situation can still be realized for a moderately weak disorder strength,
for instance, for $\alpha=0.2$, see
Fig.~\ref{sec:conductivity:fig:DOS_single_level}. In the
quasi-continuum, LLs are strongly broadened by disorder such that the
quantization is effectively absent. Additionally, we consider probing
frequencies much larger than the cyclotron frequency,
$\omega\gg\omega_c$, and assume $\omega\gg T$, hence the temperature is
effectively zero.

We then distinguish two types of processes: (i) transitions
$\mathcal{C}\rightarrow\mathcal{C}$ within the continuum leading to a
contribution to the conductivity $\sigma_{\mathcal{C}\mathcal{C}}$ and
(ii) transitions  $\mathcal{C}\rightarrow L_0$ and $L_0
\rightarrow\mathcal{C}$ giving the contribution $\sigma_{\mathcal{C}0}$.
In the case of $\omega>|\ep_F|+\Gamma_0$, which means that we indeed
probe $L_0$, the total conductivity
$\sigma(\omega)=\sigma_{\mathcal{C}0}+\sigma_{\mathcal{C}\mathcal{C}}$
is given by
\begin{align}
        \sigma_{\mathcal{C}0}(\omega) = & \int_{-\Gamma_0}^{\Gamma_0}
K_{\mathcal{C}0}(\ep,\ep+\omega) \frac{\dd\ep}{\omega} \: ,
\label{sec:conductivity:eq:0thLL_cond_basic} \\
        \sigma_{\mathcal{C}\mathcal{C}}(\omega) = &
\int_{\Gamma_0-\omega}^{-\Gamma_0}
K_{\mathcal{C}\mathcal{C}}\big(\ep,\ep+\sgn(|\ep|-|\ep_F|)\omega\big)
\frac{\dd\ep}{\omega} \: . \label{sec:conductivity:eq:cont_cond_basic}
\end{align}
Here the kernel for the transitions $\mathcal{C}\rightleftarrows L_0$ reads
\begin{equation}
        \begin{split}
                K_{\mathcal{C}0}(\ep,\ep+\omega)
                =
                \frac{e^2 \nu_0(\omega) \nu_{K=0}(\ep)}{\tau_q(\omega) |\omega^3| l_B^2}
                \: ,
        \end{split}
        \label{sec:conductivity:cont_landau_kernel}
\end{equation}
where $\nu_{K=0}(\ep)$ is given in Eq.~\eqref{sec:conductivity:eq:dos_sep_2}
and $\nu_0(\omega)=\omega/2\pi v_F^2$. The kernel describing the
transitions $\mathcal{C}\rightarrow\mathcal{C}$ is given by
\begin{equation}
        K_{\mathcal{C}\mathcal{C}}(\ep_1,\ep_2) = \frac{\alpha\pi
e^2|\ep_1\ep_2|(\ep_1^2+\ep_2^2-|\ep_1\ep_2|)}{(\ep_1^2-\ep_2^2\pm\omega_c^2)^2+\alpha^2\pi^2(\ep_1^2+\ep_2^2-|\ep_1\ep_2|)^2}
        \: ,
        \label{sec:conductivity:eq:cont_kernel}
\end{equation}
which leads, up to small disorder-induced corrections, to the well-known
universal value of the high-frequency conductivity in graphene,
\begin{equation}
        \sigma_{\mathcal{C}\mathcal{C}}(\omega)
        \:=\:
        \frac{e^2}{16}\left(1-\frac{5\pi^2}{16}\alpha^2\right)
\Theta(\omega-2|\ep_F|) + \mathcal{O}(\alpha^3)
        \: .
        \label{sec:conductivity:eq:cont_contribution}
\end{equation}
Multiplying the leading term in Eq.~\eqref{sec:conductivity:eq:cont_contribution} by the degeneracy factor of 4, we obtain the clean
conductivity $\sigma=e^2/4\hbar$ at $B=0$, where we restored Planck's constant
for convenience. Without vertex corrections, the conductivity $\sigma_{\mathcal{C}\mathcal{C}}$ obtains a
contribution of the order $\alpha$, which is canceled by the vertex
corrections. The resulting disorder-induced correction to the universal 
conductivity of the clean graphene is of the order $\alpha^2$.

The kernel~\eqref{sec:conductivity:cont_landau_kernel}
leads to the correction
\begin{equation}
        \sigma_{\mathcal{C}0}(\omega)
        =
        \frac{\alpha e^2}{4}\:\frac{\omega_c^2}{\omega^2}
        \: .
        \label{sec:conductivity:eq:cont_landau_contribution}
\end{equation}
The total conductivity for $\omega>2|\ep_F|$ is then given by
\begin{equation}
        \sigma(\omega)\big|_{\omega\gg\omega_c}
        = \frac{e^2}{16} \left( \: 1 + 4\alpha \:\frac{\omega_c^2}{\omega^2}
%-\frac{5\pi^2}{16}\alpha^2
\:\right) + \mathcal{O}(\alpha^2)
        \: .
        \label{sec:conductivity:eq:cont_cond}
\end{equation}
It is dominated by the universal background with a small quantum correction 
$\propto B$ stemming from the resolved zeroth LL.
Despite its smallness, the correction $\propto B$ should be detectable
in magnetooptical experiments since the background is $B$-independent.
The differential signal
\begin{equation}
        \omega^2\frac{\dd \sigma(\omega)}{\dd \omega_c^2}\bigg|_{\omega\gg\omega_c}
        \:=\:
        \: \frac{e^2\Gamma_0^2}{8\omega_c^2}
        \: ,
        \label{sec:conductivity:eq:derivative_omega}
\end{equation}
should provide an experimental access to the width of zeroth LL.

\section{Summary and conclusions\label{sec:summary}}\noindent
We have studied the linear transport properties of a single-layer
graphene in the presence of diagonal white noise disorder and a
moderately strong perpendicular magnetic field. Specifically, we
obtained analytic results for the ac conductivity both in the
semiclassical limit of high Landau levels (LLs) and in the quantum
limit of well separated LLs. In both cases, several transport regimes
are identified for different relations between temperature, the
external frequency, the LL separation, and the position of the
chemical potential.
Additionally, we studied a specific case of a single resolved LL
within a quasi-continuum
of states nearly not affected by the magnetic field.  Here, we show
that corrections to the universal value of
the background interband optical conductivity in graphene should
provide an experimental access to the effects of Landau quantization
even in the limit of very weak magnetic field.

The nontrivial topology of graphene, leading to the nonequidistant LL
spectrum and to the
unusual zeroth LL with strong sublattice asymmetry, makes its magnetotransport
properties very different from conventional 2D electronic systems with a
parabolic spectrum.

In the spectrum of graphene, both the quasiclassical regime of
strongly overlapping LLs and the quantum regime of well-separated LLs
can be realized simultaneously.
Which states dominate the transport properties is determined by the
position of the chemical potential, the disorder strength and the
external frequency.
Two extreme limits are (i) the classical limit of strongly doped (or
dirty) graphene where the effects of Landau quantization are
negligible and the conductivity
is given by the Drude expression and (ii) the quantum limit where the
dynamic conductivity is given by the set of delta-peaks with positions
governed by
the optical selection rules of the clean graphene.

In the disorder-dominated quasiclassical regime the Landau
quantization leads to quantum corrections to the semiclassical Drude
conductivity.
Since in high LLs $N\gg 1$ the DOS is almost periodic with the period
$\omega_c^\text{loc}\sim \omega_c/\sqrt{N}\ll \omega_c$,
the Shubnikov-de Haas (SdH) oscillations in the ac conductivity
$\sigma(\omega)$ are nearly harmonic at
$\omega\gtrsim\omega_c^\text{loc}$.
At larger frequencies $\omega\sim\omega_c$, the nonequidistant
spectrum of LLs leads to the beating of the SdH oscillations which
exhibit nodes for $\omega/\omega_c=\sqrt{2k+1}/2$ ($k\in \mathbb{N}$).
Apart from the SdH oscillations, we also studied quantum corrections
that survive at high temperatures ($T\gg T_D$) where the SdH
oscillations are strongly damped. These high-$T$ oscillations
also show a slow beating at $\omega\gg\omega_c^\text{loc}$. The nodes
in the high-$T$ quantum oscillations are found to occur at
$\omega/\omega_c=\sqrt{k/2}$ ($k\in \mathbb{N}$).

In the quantum limit of well-separated LLs, the conductivity is
dominated by the cyclotron resonance (CR) transitions between LLs with
indices $N$ and $M$ obeying $|N|-|M|=\pm 1$. For $T\ll\omega_c^\text{loc}$, only a single transition $L_N\rightarrow L_M$ determines the conductivity. The spectrum of resonance peaks then strongly depends on the position of the chemical potential as seen in Fig.~\ref{sec:conductivity:fig:cond_spectrum}. Here, Fig.~\ref{sec:conductivity:fig:cond_spectrum}(b) corresponds to the measurement reported in Ref.~\onlinecite{SadowskiGrLayers06}. In particular, we can deduce the position of the chemical potential from the intensity ratio of the CR peaks.
As for systems with a parabolic spectrum, the width of the individual resonance peaks for transitions between well-separated LLs, is given by $2(\Gamma_M+\Gamma_N)$, ($T\gg\Gamma$), while the width of the Drude peak in the semiclassical regime is given by $1/\tau_\text{tr}$. However, in graphene the ratio $\Gamma\tau_\text{tr}=\omega_c/\sqrt{\alpha}\pi|\ep|$ becomes energy dependent. The linear energy dependence of $1/\tau_\text{tr}$ has been confirmed in measurements of the cyclotron resonance line width in graphene stacks.\cite{Pot11Scattering}
Apart from the cyclotron resonance, disorder enables transitions not respecting the clean selection rules in graphene. Their strength in comparison to the CR is a direct measure for the disorder strength. The most prominent disorder-induced transitions are those between mirror symmetric LLs for which we find $\sigma\propto\omega\Gamma$. The magnetoconductivity measurements reported in Ref.~\onlinecite{Pot11Scattering}, contain a resonance peak that we attribute to such a disorder-induced mirror transition, ${L_{-1}\rightarrow L_1}$ (see Fig.~\ref{sec:conductivity:fig:cond_spectrum}).
For high $T$, multiple transitions can be addressed simultaneously. We studied the dependence of the low-frequency conductivity on the chemical potential and temperature and found that $\sigma\propto\mu^3$ in the regime $\mu\gg T$ and $T\gg\omega_c^\text{loc}$, whereas $\sigma\propto T^3$ for $T\gg\mu,\omega_c$. In contrast, the conductivity scales with $1/T$ for $T\ll\omega_c^\text{loc}$. The cubic dependence is a consequence of the Dirac LL spectrum. Specifically, it is a consequence of the energy-dependent transition rate which scales with $\omega_c^2/[\omega_c^\text{loc}]^2$ in combination with the level spacing $\omega_c^\text{loc}$.

Before closing the paper, we briefly discuss some of prospective research directions related to this work:\\
{(i)}
It would be very interesting to extend the experimental studies\cite{SadowskiGrLayers06,Pot11Scattering} to
the regime of overlapping LLs, where the beating of quantum oscillations should be observed, and to study
systematically the dependence on temperature and doping discussed in Sec.~\ref{sec:cond_overlapping}.
The disorder-induced optical transitions in the regime of separated LLs also deserve a detailed
experimental study.\\
{(ii)} Theoretically, the effects of electron-electron interactions on 
transport and optical properties of graphene in moderate magnetic field, in particular,
the interaction-induced damping of the magnetooscillations, is an interesting issue 
that has not been explored yet. Unlike conventional 2DES, in graphene interactions
are generally strong and can directly affect the transport. Another peculiar property is
that graphene supports fast directional thermalization\cite{MuellerSchmalian09} due to a forward 
scattering resonance in the electron collision integral\cite{MichaRates11,MichaelCoulDrag12} 
giving rise to effective photon energy conversion via carrier multiplication.\cite{KoppensCascade2012,KoppensCascade2}
This has tremendous consequences for potential applications of graphene in rapidly developing
field of graphene plasmonics and optoelectronics.\cite{FerrariGrapheneOptoRev10,AvGraphenePhotRev10,Koppens2011,Grigorenko2012,Engel2012}\\
{(iii)} The results of the present work, which studies the linear transport response
of graphene to ac and dc perturbations, serve a good starting point for studies of the nonequilibrium 
magnetotransport in strong ac and dc fields. First experimental work in this direction has reported the effects 
of heating of carriers by a strong dc bias on the SdH oscillations in graphene.\cite{Tan11SdH}
Just as different cyclotron frequencies lead to a modulation of the linear-response dynamic conductivity, 
they are expected to manifest themselves out of equilibrium as well. Especially near the Dirac point in graphene, 
the possibility to create nontrivial nonequilibrium steady states at moderate $B$ and high $T$, 
including population inversion\cite{PopInv12} leading to optical gain,\cite{RyzhiiNegCond07}  is tempting.
From previous studies on semiconductor 2DES we know that nonequilibrium
phenomena in moderate $B$ are very sensitive to the details of disorder
and interactions\cite{dmitriev:2009b} which are not accessible from standard measurements.\cite{IvanReview11} 
Thus experimental studies of nonequilibrium magnetotransport phenomena in graphene combined with their adequate 
theoretical description should provide, in particular, valuable information about the nature of disorder
and the role of interactions in electronic transport and relaxation which are still under debate.

\appendix
\section{SCBA equation \label{sec:app_SCBA_overlapping}}\noindent
With the help of the Poisson summation formula the SCBA equation~\eqref{sec:formalism:eq:selfenergy_2} is rewritten as the following equation for $\ep-\Sigma_\ep$:
\begin{widetext}
	\begin{equation}
		\begin{split}
			& (\varepsilon - \Sigma^R_\ep)^2
			\left\{
				1 - \frac{\alpha}{4}
					\left[
						2 \: \ln {
							\frac{-\Delta_c^2}{ (\varepsilon-\Sigma_\ep^R)^2 }
						}
						+ 4\pi \ii \:
						\sgn([\Theta^R_\ep]'') \cdot
						\frac{
							\lambda \cdot \ee^{\ii 2\pi |[\Theta^R_\ep]'| \sgn([\Theta^R_\ep]'') / \omega_c^2 }
						}
						{
							1 - \lambda \cdot \ee^{\ii 2\pi |[\Theta^R_\ep]'| \sgn([\Theta^R_\ep]'') / \omega_c^2 }
						}
					\right]
			\right\}
			- \varepsilon ( \varepsilon - \Sigma^R_\ep )
			+ \frac{ \alpha \omega_c^2 }{4}
			\:=\: 0
			\: .
		\end{split}
		\label{sec:appendix:eq:SCBA_poisson}
	\end{equation}
\end{widetext}
The coefficients of this quadratic equation also depend on the self-energies via
\begin{equation}
	\Theta^\beta_\ep=[\ep-\Sigma_a^\beta(\ep)][\ep-\Sigma^\beta_b(\ep)] \: ,
	\label{sec:appendix:eq:Theta_def}
\end{equation}
where $\beta\in\{R,A\}$. The real part of $\Theta_\ep$ is denoted as $\Theta'_\ep$ and the imaginary part as $\Theta''_\ep$.
We solve Eq.~\eqref{sec:appendix:eq:SCBA_poisson} by iteration in $\lambda$. Furthermore, for high energies we make the replacement
\begin{equation}
	\ln {\frac{-\Delta_c^2}{ (\varepsilon-\Sigma_\ep)^2 }}
	\rightarrow
	2\ln {\frac{\Delta_c}{|\ep|}}\pm\ii\:\sgn(\ep)\pi
	\: .
	\label{sec:appendix:eq:log_replacement}
\end{equation}
Here $\pm$ refers to the retarded or advanced self-energy. To leading order in $\alpha$ the result is given in Eq.~\eqref{sec:formalism:eq:selfenergy_overlapping}.

First we set all terms proportional to $\lambda$ equal to zero in Eq.~\eqref{sec:appendix:eq:SCBA_poisson} and solve
\begin{equation}
	(\varepsilon - \Sigma_\ep)^2 \left(Z(\ep)\mp\ii\frac{\alpha\pi\sgn(\ep)}{2}\right) - \varepsilon (\varepsilon - \Sigma_\ep) + \frac{\alpha\omega_c^2}{4} = 0 \: ,
	\label{sec:appendix:eq:SCBA_leading_order}
\end{equation}
where the renormalization constant $Z$ from Eq.~\eqref{sec:formalism:eq:renorm_const} has been used. Equation~\eqref{sec:appendix:eq:SCBA_leading_order} yields
\begin{equation}
	\ep-\Sigma = \tilde\ep + \frac{\ii}{2\tau_q} \: .
\end{equation}
This solution is reinserted for $\Theta$ into the oscillating part of Eq.~\eqref{sec:appendix:eq:SCBA_poisson} and one obtains
\begin{equation}
	\ep-\Sigma = \tilde\ep + \frac{\ii}{2\tau_q}\left(1+2\sum_{k=1}\lambda^k\ee^{\ii2\pi k\ep^2/\omega_c^2}\right) \: .
	\label{sec:appendix:eq:sigma}
\end{equation}
The leading terms in Eq.~\eqref{sec:appendix:eq:sigma} yield Eq.~\eqref{sec:formalism:eq:selfenergy_overlapping}.

\section{Conductivity with vertex corrections \label{sec:app_vertex_corrections}}\noindent
\subsection{Vertex corrections}
\begin{figure}[ht]
	\centering
	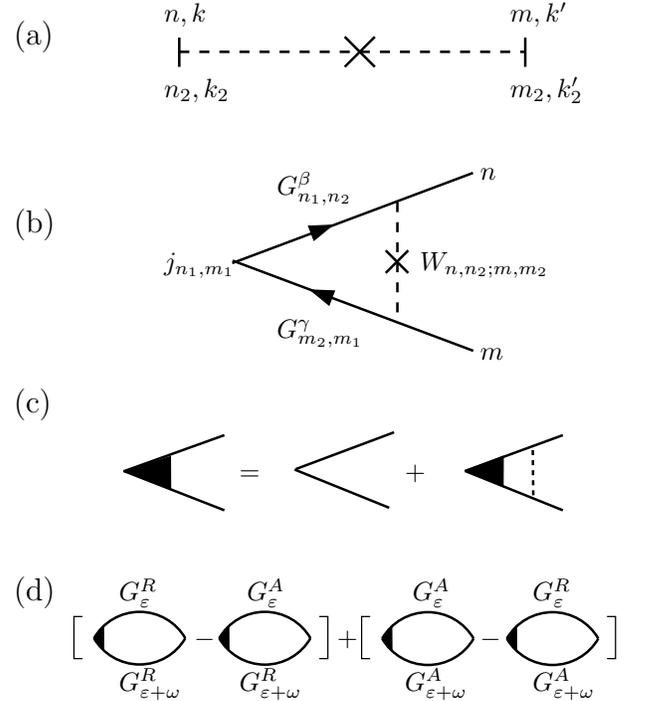
	\caption{(a) The disorder correlator in the Landau level basis, Eq.~\eqref{sec:app_vertex_corrections:eq:correlator_LL_basis_1}, as a two particle operator. (b) The vertex correction $\Gamma^{(1)}$ from Eq.~\eqref{sec:app_vertex_corrections:eq:vertex_corr_1}. For brevity we omitted the momenta here. (c) The equation for the dressed vertex $\Gamma$. (d) The contributions to the conductivity kernel Eq.~\eqref{sec:formalism:eq:kernel_definition} including the vertex corrections denoted by the black triangle according to (c).}
	\label{sec:app_vertex_corrections:fig:combined}
\end{figure}
In this chapter we describe the SCBA in graphene
and provide a detailed calculation of the conductivity in disordered LLs of graphene including vertex corrections.
 
We use the basis of eigenstates of the clean Hamiltonian \eqref{sec:formalism:eq:hamiltonian},
\begin{align}
	&\psi_{n\neq0,k}=\frac{\ee^{\ii k \hat{y}}}{\sqrt{2}}\:\begin{pmatrix}||n|-1\rangle\\ \text{sgn}(n)||n|\rangle\end{pmatrix} \: , \label{sec:app_vertex_corrections:eq:eigenstate_1} \\
	&\psi_{n=0,k}=\ee^{\ii k \hat{y}}\:\begin{pmatrix}0\\ \text{sgn}(n)|0\rangle\end{pmatrix} \: . \label{sec:app_vertex_corrections:eq:eigenstate_2}
\end{align}
Here $|n,k\rangle$ are the harmonic oscillator eigenfunctions with origin $l_B^2 k$. The matrix elements of the current operator $\hat{j}_x$ in the basis \eqref{sec:app_vertex_corrections:eq:eigenstate_1}, \eqref{sec:app_vertex_corrections:eq:eigenstate_2} are
\begin{equation}
	(\hat{j}_x)_{nk,mk'}=\frac{v_0 e}{2L}\:\delta_{kk'}\left[c_m\delta_{|m|,|n|-1}+c_n\delta_{|n|,|m|-1}\right]\: ,
	\label{sec:app_vertex_corrections:eq:current_me}
\end{equation}
where the coefficients $c_{n\neq0}=\text{sign}(n)$ and $c_{n=0}=\sqrt{2}$. With the self-energy from Eq.~\eqref{sec:formalism:eq:selfenergy_structure} we obtain for the Green's functions in the case $n,m\neq0$
\begin{equation}
	G_{n,k;m,k'} = \delta_{k,k'}\Big\{\delta_{n,m}[G_+]_{n,k}(\ep)+\delta_{n,-m}[G_-]_{n,k}(\ep)\Big\} \: ,
	\label{sec:app_vertex_corrections:eq:gf_me}
\end{equation}
where
\begin{align}
	& [G_+]_{n,k}(\ep)=\frac{\ep-(\Sigma_a+\Sigma_b)/2+E_n}{(\ep-\Sigma_a)(\ep-\Sigma_b)-E_n^2} \: , \label{sec:app_vertex_corrections:eq:gf_me_+} \\
	& [G_-]_{n,k}(\ep)=\frac{(\Sigma_a-\Sigma_b)/2}{(\ep-\Sigma_a)(\ep-\Sigma_b)-E_n^2} \: . \label{sec:app_vertex_corrections:eq:gf_me_-}
\end{align}
For $n=0$, the matrix element of the Green's function is
\begin{equation}
	G_{n=0}(\ep) = \frac{1}{\ep-\Sigma_b(\ep)} \: .
\end{equation}
The disorder correlator in graphene is defined as
\begin{equation}
	W(\vec{r}-\vec{r}') = \langle V(\vec{r}) \otimes V(\vec{r}') \rangle_\text{\text{dis}} \: ,
\end{equation}
where $V$ is the matrix of the impurity potential in the sublattice space. For diagonal white noise disorder we obtain for its Fourier transform the expression~\eqref{sec:formalism:eq:dis_correlator}. The matrix elements of the disorder correlator in LL basis are formally obtained from the operator expression for the SCBA self-energy \eqref{sec:formalism:eq:SCBA},
\begin{equation}
	\begin{split}
		& W_{nk,n_2k_2;mk',m_2k_2'} = \int \frac{\dd\vec{q}}{(2\pi)^2} \: W_{\nu\eta\sigma\mu}(\vec{q}) \\
		& \times \: \prescript{}{\nu}{\langle} \psi_{n,k}| \ee^{+\ii\vec{q}\cdot\hat{\vec{r}}}{|\psi_{n_2,k_2}\rangle}_{\eta} \:
		 \prescript{}{\sigma}{\langle} \psi_{m_2,k_2'}| \ee^{-\ii\vec{q}\cdot\hat{\vec{r}}}{|\psi_{m,k'}\rangle}_{\mu} \: .
	\end{split}
	\label{sec:app_vertex_corrections:eq:correlator_LL_basis_1}
\end{equation}
This equation is valid for a generic disorder in graphene. Here $|\psi_{nk}\rangle_{\nu}$ denotes the $\nu$-th spinor component of the eigenstates~\eqref{sec:app_vertex_corrections:eq:eigenstate_1}~and~\eqref{sec:app_vertex_corrections:eq:eigenstate_2}. For the vertex diagram in Fig.~\ref{sec:app_vertex_corrections:fig:combined}(b) we exploit the relations between the LL indices in this diagram to evaluate the product of the phase factors $\langle nk|\exp(\ii\vec{q}\cdot\hat{\vec{r}})|n'k'\rangle$ in the case of white noise disorder, Eq.~\eqref{sec:formalism:eq:dis_correlator}. Using the solutions of the LL states in the Landau gauge for $|n|>|n_2|$, we find
\begin{equation}
	\begin{split}
		& \int\frac{\dd\vec{q}}{(2\pi)^2}{\langle} \psi_{n,k}| \ee^{+\ii\vec{q}\cdot\hat{\vec{r}}}{|\psi_{n_2,k_2}\rangle} \:
		{\langle} \psi_{m_2,k_2'}| \ee^{-\ii\vec{q}\cdot\hat{\vec{r}}}{|\psi_{m,k'}\rangle} \\
		& = \frac{\delta_{k,k_2}}{8\pi l^2} \int\frac{\dd\varphi_q}{2\pi} \: \ee^{\ii\varphi_q(|n|-|m|+|m_2|-|n_2|)} \int\dd u \: \ee^{-u} u^{|n|-|n_2|} \\
		& \times \bigg[ \left(\frac{(|n_2|-1)!}{(|n|-1)!}\right)^2 \left(\frac{(|m_2|)!}{(|m|)!}\right)^2 L_{|n_2|-1}^{|n|-|n_2|}(u) L_{|m_2|}^{|n|-|n_2|}(u) \\
		& + \left(\frac{(|n_2|)!}{(|n|)!}\right)^2 \left(\frac{(|m_2|-1)!}{(|m|-1)!}\right)^2 L_{|n_2|}^{|n|-|n_2|}(u) L_{|m_2|-1}^{|n|-|n_2|}(u) \bigg]\: ,
	\end{split}
\end{equation}
where $L_{\alpha}^{n}(u)$ are the generalized Laguerre polynomials. Using the orthogonality of $L_{\alpha}^{n}(u)$, we obtain the matrix elements for the diagram in Fig.~\ref{sec:app_vertex_corrections:fig:combined}(b),
\begin{equation}
	\begin{split}
		& W_{nk,n_2k_2;mk',m_2k_2'} = \frac{\alpha\omega_c^2}{8} \: \delta_{k,k_2} \: \delta_{|n|-|m|,|n_2|-|m_2|} \\
		& \times \bigg\{ c_{n} c_{n_2}\delta_{|n_2|+1,|m_2|} + c_{m} c_{m_2}\delta_{|m_2|+1,|n_2|} \bigg\} \: .
	\end{split}
	\label{sec:app_vertex_corrections:eq:correlator_LL_basis_2}
\end{equation}
The vertex function [Fig.~\ref{sec:app_vertex_corrections:fig:combined}(b)], to first order in the disorder correlator, can be expressed as ($\beta,\gamma\in\{R,A\}$)
\begin{equation}
	\begin{split}
		& \Gamma^{(1),\beta\gamma}_{n,k;m,k'} = \sum_{\stackrel{n_i,m_i,}{k_i,k_i'}} \: W_{nk,n_2k_2;mk',m_2k_2'} \\
		& \times G^\gamma_{m_2 k_2';m_1k_1'}(\ep+\omega) \:  j_{m_1k_1';n_1k_1} \: G^\beta_{n_1k_1;n_2k_2}(\ep) \: ,
	\end{split}
	\label{sec:app_vertex_corrections:eq:vertex_corr_1}
\end{equation}
which gives 
\begin{equation}
	\begin{split}
		& \Gamma^{(1),\beta\gamma}_{nk,mk'} = \alpha\pi \delta_{k,k'} \: \bigg\{ \: c_n \delta_{|n|+1,|m|} \: \Pi^{\beta\gamma}(\ep,\ep+\omega) \\
		& + \: c_m \delta_{|m|+1,|n|} \: \Pi^{\gamma\beta}(\ep+\omega,\ep) \: \bigg\} \: .
	\end{split}
	\label{sec:app_vertex_corrections:eq:vertex_corr_2}
\end{equation}
The $\Pi$-correlators are already for the case of overlapping LLs are given by Eq.~\eqref{sec:conductivity:eq:Pi_correlators_def}. Their general definition is
\begin{equation}
	\Pi^{\beta\gamma}(\ep_1,\ep_2) = \frac{\omega_c^2}{2\pi} \: \frac{1}{4}\sum_{\substack{n,m\\n_2,m_2}} c_m c_{m_2}G^{\beta}_{n,n_2}(\ep_1)G^{\gamma}_{m_2,m}(\ep_2) \: ,
	\label{sec:app_vertex_corrections:eq:Pi_correlators_def}
\end{equation}
which can be written explicitly as
\begin{equation}
	\begin{split}
		& \Pi^{\beta\gamma}(\ep_1,\ep_2)
		= \frac{\omega_c^2}{2\pi} \sum_{n=0}^\infty \: \frac{\ep_1-\Sigma_b^\beta(\ep_1)}{[\ep_1-\Sigma_a^\beta(\ep_1)][\ep_1-\Sigma_b^\beta(\ep_1)]-\ep_{n+1}^2} \\
		& \times \: \frac{\ep_2-\Sigma_a^\gamma(\ep_2)}{[\ep_2-\Sigma_a^\gamma(\ep_2)][\ep_2-\Sigma_b^\gamma(\ep_2)]-\ep_{n}^2} \: .
	\end{split}
	\label{sec:app_vertex_corrections:eq:Pi_correlators_explicit}
\end{equation}
For $\Sigma_a=\Sigma_b$ we recover Eq.~\eqref{sec:conductivity:eq:Pi_correlators_def}. A closed expression for Eq.~\eqref{sec:app_vertex_corrections:eq:Pi_correlators_explicit} can be obtained in terms of the Digamma function $\Psi(z)$, which has been used in a similar context in Ref.~\onlinecite{Gus06HallOptCond},
\begin{equation}
	\begin{split}
		& \Pi^{\beta\gamma}(\ep_1,\ep_2)
		= \frac{\omega_c^2}{2\pi} \frac{[\ep_1-\Sigma_b^\beta(\ep_1)][\ep_2-\Sigma_a^\gamma(\ep_2)]}{\Theta^\gamma(\ep_2)-\Theta^\beta(\ep_1)+\omega_c^2} \\
		& \times \: \left\{ \Psi\left(-\Theta^\beta(\ep_1)/\omega_c^2\right)-\Psi\left(-\Theta^\gamma(\ep_2)/\omega_c^2\right) - \omega_c^2/\Theta^\beta(\ep_1) \right\} \: .
	\end{split}
	\label{sec:app_vertex_corrections:eq:Pi_correlators_closed_exp}
\end{equation}
Here $\Theta^\beta(\ep)$ is defined in Eq.~\eqref{sec:appendix:eq:Theta_def}. Summing up all ladder diagrams, we obtain the dressed vertex
\begin{equation}
	\begin{split}
		& \Gamma^{\beta\gamma}_{nk,mk'} = \delta_{k,k'}\bigg\{ \: \frac{c_n \: [\Pi^{\beta\gamma}(\ep,\ep+\omega)]^{-1}}{[\Pi^{\beta\gamma}(\ep,\ep+\omega)]^{-1}-\alpha\pi} \: \delta_{|n|+1,|m|} \\
		& + \: \frac{c_m \: [\Pi^{\gamma\beta}(\ep+\omega,\ep)]^{-1}}{[\Pi^{\gamma\beta}(\ep+\omega,\ep)]^{-1}-\alpha\pi} \: \delta_{|m|+1,|n|} \: \bigg\} \: .
	\end{split}
	\label{sec:app_vertex_corrections:eq:vertex_corr_summed}
\end{equation}
\subsection{Conductivity bubble}
The conductivity kernel $K(\ep,\ep+\omega)$ from Eq.~\eqref{sec:formalism:eq:kernel_definition} is the sum of all conductivity bubbles $K^{\beta\gamma}$ [see Fig.~\ref{sec:app_vertex_corrections:fig:combined}(d)], where $\beta,\gamma\in\{R,A\}$. At $B=0$, the dominant contribution away from the Dirac point and for intraband processes comes from the RA and AR bubbles while the RR and AA bubble give corrections of the order of the conductance quantum that are usually neglected.\cite{Ostr06Dis} At $B\neq0$, all four contributions are important.% and contribute equally to the SdH and quantum oscillations.

Without vertex corrections each bubble in Fig.~\ref{sec:app_vertex_corrections:fig:combined}(d) is given by the expression
\begin{equation}
	\begin{split}
		\showgraph{width=1cm}{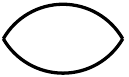} = & \sum_{n,m,n_i,m_i} \sum_{k,k',k_i,k_i'} \: j_{mk',nk} \: G^\beta_{nk,n_2k_2}(\ep) \\
		& \times j_{n_2k_2,m_2k_2'} \: G^\gamma_{m_2k_2',mk'}(\ep+\omega)
		\: .
	\end{split}
	\label{sec:app_vertex_corrections:eq:conductivity_no_vertex_1}
\end{equation}
It is straightforward to rewrite this in terms of the $\Pi$-correlators.
\begin{equation}
	\begin{split}
		\showgraph{width=1cm}{conductivity_bubble_small_2.pdf} = e^2 \left\{ \Pi^{\beta\gamma}(\ep,\ep+\omega) + \Pi^{\gamma\beta}(\ep+\omega,\ep) \right\}
		\: ,
	\end{split}
	\label{sec:app_vertex_corrections:eq:conductivity_no_vertex_2}
\end{equation}
which in the presence of vertex corrections turns into
\begin{equation}
	\begin{split}
		\showgraph{width=1cm}{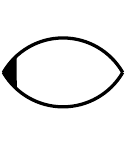} = e^2
		\bigg\{ & \frac{1}{[\Pi^{\beta\gamma}(\ep,\ep+\omega)]^{-1}-\alpha\pi} \\
		& + \frac{1}{[\Pi^{\gamma\beta}(\ep+\omega,\ep)]^{-1}-\alpha\pi} \bigg\}
		\: .
	\end{split}
	\label{sec:app_vertex_corrections:eq:conductivity_vertex}
\end{equation}
Taking into account all diagrams from Fig.~\ref{sec:app_vertex_corrections:fig:combined}(d) we obtain for the conductivity kernel,
\begin{widetext}
	\begin{equation}
		\begin{split}
			K(\ep_1,\ep_2) = e^2 \: \text{Re}\bigg\{ \frac{1}{[\Pi^{RA}(\ep_1,\ep_2)]^{-1}-\alpha\pi}
			- \frac{1}{[\Pi^{RR}(\ep_1,\ep_2)]^{-1}-\alpha\pi} + \{ \ep_1 \leftrightarrow \ep_2 \} \bigg\}
			\: .
		\end{split}
		\label{sec:app_vertex_corrections:eq:kernel_pi}
	\end{equation}
\end{widetext}

\subsubsection{Overlapping Landau levels}

In the regime of overlapping LL, the density-density correlators \eqref{sec:app_vertex_corrections:eq:Pi_correlators_explicit} are calculated using the Poisson formula. Using the abbreviation ${\Theta^\beta_\ep=(\ep-\Sigma^\beta_\ep)^2}$, we obtain 
\begin{widetext}
	\begin{equation}
		\begin{split}
			\Pi^{\beta\gamma}(\ep_1,\ep_2) = & \frac{1}{4\pi} \: \big[\ep_1-\Sigma_b^\beta(\ep_1)\big]\big[\ep_2-\Sigma_a^\gamma(\ep_2)\big]
			\: \sum_{k=-\infty}^{+\infty} \int\dd x \: \frac{\exp(\ii 2\pi k x)}{(\Theta^\beta_{\ep_1}-|x|-1)(\Theta^\gamma_{\ep_2}-|x|)} \\
			& + \frac{\omega_c^2}{2\pi} \: \frac{\ep_2-\Sigma^\gamma(\ep_2)}{\big[\ep_1-\Sigma^\beta(\ep_1)\big]\big[(\ep_2-\Sigma^\gamma(\ep_2))^2-\omega_c^2\big]} \: .
		\end{split}
		\label{sec:app_vertex_corrections:eq:Pi_correlators_Poisson}
	\end{equation}
\end{widetext}
for the correlators~\eqref{sec:app_vertex_corrections:eq:Pi_correlators_explicit}. Next, we neglect the second term in Eq.~\eqref{sec:app_vertex_corrections:eq:Pi_correlators_Poisson} since it is small for energies $\ep\gg\omega_c$. The sum is dominated by the terms with $k=0$ and $k=\pm1$ due to the presence of the coherence factor $\lambda^k$, Eq.~\eqref{sec:formalism:eq:Dingle_factor}, in each term in the sum of Eq.~\eqref{sec:app_vertex_corrections:eq:Pi_correlators_Poisson}. The corresponding three integrals yield
\begin{widetext}
	\begin{equation}
		\Pi^{RR(RA)}_{\ep_1,\ep_2}
		= \frac{2\big(\ep_1-\Sigma_{\ep_1}^{R}\big)\big(\ep_2-\Sigma_{\ep_2}^{R(A)}\big)}{\big(\ep_1-\Sigma_{\ep_1}^{R}\big)^2-\big(\ep_2-\Sigma_{\ep_2}^{R(A)}\big)^2-\omega_c^2} \:
		\bigg\{
			\sgn\Big[\text{Im}\big(\ep_1-\Sigma_{\ep_1}^{R}\big)^2\Big] \: \tau_{q,\ep_1}\Sigma^{R}_{\ep_1}
			\:-\:
			\sgn\Big[\text{Im}\big(\ep_2-\Sigma_{\ep_2}^{R(A)}\big)^2\Big] \: \tau_{q,\ep_2}\Sigma^{R(A)}_{\ep_2}
		\bigg\}
		\: .
		\label{sec:app_vertex_corrections:eq:Pi_correlators_final}
	\end{equation}
\end{widetext}
From the self-energy \eqref{sec:formalism:eq:selfenergy_overlapping} we deduce that one may replace $\sgn\Big[\text{Im}\big(\ep-\Sigma_{\ep}^{R}\big)^2\Big] \rightarrow \sgn(\ep)$ and we obtain the result~\eqref{sec:conductivity:eq:Pi_correlator_2} from the main text.

Let us turn to the conductivity including the vertex corrections. Again, using the explicit form of the self-energies, Eq.~\eqref{sec:formalism:eq:selfenergy_overlapping}, and expanding the correlators to leading order in the coherence factor $\lambda$, we obtain the Drude part
\begin{equation}
	\text{Re}\:\Pi^{RA}(\ep_1,\ep_2) = \frac{\alpha\pi|\ep_1\ep_2|(\ep_1^2+\ep_2^2)}{(\ep_1^2-\ep_2^2\pm\omega_c^2)^2+\alpha^2\pi^2(\ep_1^2+\ep_2^2)^2} \: .
\end{equation}
The corresponding kernel \eqref{sec:conductivity:eq:kernel_Pi} without vertex corrections reads
\begin{equation}
	K^{\hspace{-0.2mm}(0)}(\ep_1,\ep_2) = \frac{\alpha\pi e^2|\ep_1\ep_2|(\ep_1^2+\ep_2^2)}{(\ep_1^2-\ep_2^2\pm\omega_c^2)^2+\alpha^2\pi^2(\ep_1^2+\ep_2^2)^2}
	\: .
	\label{sec:conductivity:eq:pre_Drude_kernel}
\end{equation}
If we include the vertex corrections we obtain
\begin{equation}
	\begin{split}
		& \text{Re}\left[\frac{1}{[\Pi^{RA}(\ep_1,\ep_2)]^{-1}-\alpha\pi}\right] \\
		& = \frac{\alpha\pi|\ep_1\ep_2|(\ep_1^2+\ep_2^2-|\ep_1\ep_2|)}{(\ep_1^2-\ep_2^2\pm\omega_c^2)^2+\alpha^2\pi^2(\ep_1^2+\ep_2^2-|\ep_1\ep_2|)^2} \\
		& = \frac{\ep^* \tau_\text{tr}^{-1}/2}{(\omega\pm\omega_c^\text{loc})^2+1/\tau_\text{tr}^2}
		\: ,
	\end{split}
	\label{sec:app_vertex_corrections:eq:RePi_with_transport}
\end{equation}
where we introduced the transport time $\tau_\text{tr}=2\tau_q$. The factor $2$ in the transport time can be understood within the Boltzmann theory from the Dirac factors $(1+\cos\phi_{k,k'})$ coming from the scattering matrix elements in addition to the transport factor $(1-\cos\phi_{k,k'})$ in the definition of the transport and quantum scattering time,
\begin{equation}
	\begin{matrix}
		1/\tau_\text{tr} \\
		1/\tau_q
	\end{matrix}
	\bigg\}
	= 2\pi \sum_{\vec{k}'}
	\bigg\{
	\begin{matrix}
		1-\cos\phi_{k,k'}  \\
		1
	\end{matrix}
	\bigg\}
	\: |\langle \vec{k} |V(\vec{q})|\vec{k}'\rangle|^2 \: .
	\label{sec:app_vertex_corrections:eq:q_vs_tr1}
\end{equation}
Here $\phi_{k,k'}$ is the angle between the initial and final direction of momentum.
In the case of diagonal disorder, Eq.~\eqref{sec:app_vertex_corrections:eq:q_vs_tr1} gives
\begin{equation}
	\begin{split}
		\begin{matrix}
			1/\tau_\text{tr} \\
			1/\tau_q
		\end{matrix}
		\bigg\}
		= & \pi \sum_{\vec{k}'}
		\bigg\{
		\begin{matrix}
			1-\cos\phi_{k,k'}  \\
			1
		\end{matrix}
		\bigg\}
		W(\vec{q}) \: (1+\cos\phi_{k,k'}) \: .
	\end{split}
	\label{sec:app_vertex_corrections:eq:q_vs_tr2}
\end{equation}
From Eq.~\eqref{sec:app_vertex_corrections:eq:RePi_with_transport}, we obtain the result from Eq.~\eqref{sec:conductivity:eq:Drude} for the conductivity, which in the dc limit turns into
\begin{equation}
	\sigma(0) = \frac{1}{2\pi} \frac{\mathcal{D}\:\tau_\text{tr}^{-1}}{\left(\omega_c^\text{loc}\right)^2+1/\tau_\text{tr}^2}
	\stackrel{(B\rightarrow0)}{\rightarrow}
	\frac{\mathcal{D}\:\tau_\text{tr}}{2\pi} \: .
	\label{sec:app_vertex_corrections:eq:conductivity_with_transport}
\end{equation}
The higher order corrections in $\lambda$ yield results for the SdH oscillations and quantum corrections presented in the main text.

\subsubsection{Separated Landau levels}

Since LLs are separated and the kernel~\eqref{sec:app_vertex_corrections:eq:kernel_pi} is proportional to the product of the DOS at energies $\ep$ and $\ep+\omega$, the conductivity~\eqref{sec:formalism:eq:Kubo_formula} can be written as
\begin{equation}
	\begin{split}
		& \sigma(\omega) = \sum_{K,M}\int\frac{\dd\ep}{4\pi}\frac{f_\ep-f_{\ep+\omega}}{\omega} K(\ep,\ep+\omega)\\
		& \times \Theta(\Gamma_M-|\ep+\omega-E_M|)\Theta(\Gamma_K-|\ep-E_K|) \: ,
	\end{split}
	\label{sec:app_vertex_corrections:eq:cond_decomp_separated}
\end{equation}
Comparing Eq.~\eqref{sec:app_vertex_corrections:eq:cond_decomp_separated} with Eq.~\eqref{sec:conductivity:eq:conductivity_sep_LL}, we identify
\begin{equation}
	\begin{split}
		& \tilde{\mathcal{F}}_{KM} = \frac{\pi\Gamma_K\Gamma_M}{\omega_c^2c_Kc_M} \int\dd\ep\:\frac{f_\ep-f_{\ep+\omega}}{\omega} \\
		& \times \Theta(\Gamma_M-|\ep+\omega-E_M|)\Theta(\Gamma_K-|\ep-E_K|)\\
		& \times \frac{1}{P(K,M)}\text{Re}\bigg\{ \frac{1}{[\Pi^{RA}(\ep,\ep+\omega)]^{-1}-\alpha\pi} \\ 
		& - \frac{1}{[\Pi^{RR}(\ep,\ep+\omega)]^{-1}-\alpha\pi} + \{ \ep \leftrightarrow \ep+\omega \} \bigg\} \: ,
	\end{split}
	\label{sec:app_vertex_corrections:eq:def_F_tilde}
\end{equation}
where the factors $P(K,M)$ are given by Eqs.~\eqref{sec:conductivity:eq:trans_prob_off_res_0}-\eqref{sec:conductivity:eq:trans_prob_off_res_2}. 
Without vertex corrections taken into account, Eq.~\eqref{sec:app_vertex_corrections:eq:def_F_tilde} reduces to 
\begin{equation}
	\begin{split}
		& \mathcal{F}_{KM} = \frac{\pi\Gamma_K\Gamma_M}{\omega_c^2c_Kc_M} \int\dd\ep\:\frac{f_\ep-f_{\ep+\omega}}{\omega} \\
		& \times \Theta(\Gamma_M-|\ep+\omega-E_M|)\Theta(\Gamma_K-|\ep-E_K|)\\
		& \times \frac{\text{Re}\left\{\Pi^{RA}(\ep,\ep+\omega) - \Pi^{RR}(\ep,\ep+\omega) + \{ \ep \leftrightarrow \ep+\omega \} \right\}}{P(K,M)} \: ,
	\end{split}
	\label{sec:app_vertex_corrections:eq:def_F}
\end{equation}
leading to Eq.~\eqref{sec:conductivity:eq:fctF_def} in the main text.

With vertex corrections included, for the cyclotron resonance $||K|-|M||=1$ we obtain
\begin{equation}
	\begin{split}
		& \text{Re}\left(\frac{1}{[\Pi^{RA}_{\ep,\ep+\omega}]^{-1}-\alpha\pi}-\frac{1}{[\Pi^{RR}_{\ep,\ep+\omega}]^{-1}-\alpha\pi}\right) \\
		& = \frac{\omega_c^2(l_B \pi)^4}{\pi} \frac{3\Gamma^2 \nu_K(\ep)\nu_M(\ep+\omega)}{9\Gamma^2 - 8 (\delta_1-2\delta_2)(\delta_2-2\delta_1)} \: .
	\end{split}
	\label{sec:app_vertex_corrections:eq:sep_Pi_cycl_res_vertex}
\end{equation}
where we introduced the detunings from the center of the LLs $\delta_1=\ep-\ep_K$ and $\delta_2=\ep+\omega-\ep_M$ and used $\Gamma/\omega_c\ll 1$. The density of states $\nu_K$ is given by Eqs.~\eqref{sec:conductivity:eq:dos_sep_1}~and~\eqref{sec:conductivity:eq:dos_sep_2}.
The term Eq.~\eqref{sec:app_vertex_corrections:eq:def_F_tilde} with $\ep$ and $\ep+\omega$ interchanged does not contribute. For the cyclotron resonance $|K|-|M|=-1$ the situation is reversed.
The vertex corrections lead to an additional energy dependence in the denominator of Eq.~\eqref{sec:app_vertex_corrections:eq:sep_Pi_cycl_res_vertex} in comparison to
the expression 
\begin{equation}
	\text{Re}\left(\Pi^{RA}_{\ep,\ep+\omega}-\Pi^{RR}_{\ep,\ep+\omega}\right) = \frac{\omega_c^2(l_B \pi)^4}{\pi} \: \nu_K(\ep)\nu_M(\ep+\omega)
	\label{sec:app_vertex_corrections:eq:sep_Pi_cycl_res_1}
\end{equation}
that would result from the bare bubbles in Fig.~\ref{sec:app_vertex_corrections:fig:combined}(d).  

Vertex corrections are important only at the cyclotron resonances ($||K|-|M||=1$). In the presence of disorder the selection rules of clean graphene are relaxed and the disorder-induced transitions ($||K|-|M||\neq 1$) become possible. For disorder-induced transitions, one needs to calculate the correlator $\text{Re}\left(\Pi^{RA}_{\ep,\ep+\omega}-\Pi^{RR}_{\ep,\ep+\omega}\right)$ entering Eq.~\eqref{sec:app_vertex_corrections:eq:def_F}; the term with $\ep$ and $\ep+\omega$ interchanged is obtained by exchanging $K$ and $M$.

In the case $||K|-|M||\neq1$, $K,M\neq0$, calculation using $\Gamma/\omega_c\ll 1$ gives
\begin{equation}
	\begin{split}
		& \text{Re}\left(\Pi^{RA}_{\ep,\ep+\omega}-\Pi^{RR}_{\ep,\ep+\omega}\right) = \frac{\omega_c^2(l_B \pi)^4}{\pi} \: \nu_K(\ep)\nu_M(\ep+\omega) \\
			& \quad \times \frac{\Gamma^2(\ep_M^2+\ep_K^2)}{(\ep_K^2-\ep_M^2+\omega_c^2)^2} \: .
	\end{split}
	\label{sec:app_vertex_corrections:eq:sep_Pi_off_res}
\end{equation}
 In the case $K=M$ of intra-LL transitions,  Eq.~\eqref{sec:app_vertex_corrections:eq:sep_Pi_off_res} yields
\begin{equation}
	\begin{split}
		\text{Re}\left(\Pi^{RA}_{\ep,\ep+\omega}-\Pi^{RR}_{\ep,\ep+\omega}\right) = & \frac{\omega_c^2(l_B \pi)^4\Gamma^2}{2\pi\omega_K^2} \: \nu_K(\ep)\nu_K(\ep+\omega) \: .
	\end{split}
	\label{sec:app_vertex_corrections:eq:sep_Pi_res_vertex}
\end{equation}
For $K=0$ or $M=0$ and $||K|-|M||\neq1$, 
\begin{equation}
	\begin{split}
		& \text{Re}\left(\Pi^{RA}_{\ep,\ep+\omega}-\Pi^{RR}_{\ep,\ep+\omega}\right) = \frac{\omega_c^2(l_B \pi)^4}{\pi} \: \nu_K(\ep)\nu_M(\ep+\omega) \\
			& \quad \times \frac{\Gamma^2(\ep_K^2+\omega_c^2)}{2(\ep_M^2-\ep_K^2+\omega_c^2)^2} \: .
	\end{split}
	\label{sec:app_vertex_corrections:eq:sep_Pi_off_res_0thLL}
\end{equation}

\section{Shape of the conductivity peaks in the regime of separated LLs \label{sec:app_Fintegrals}}\noindent

\begin{figure}[h]
	\centering
	\includegraphics[width=0.45\textwidth]{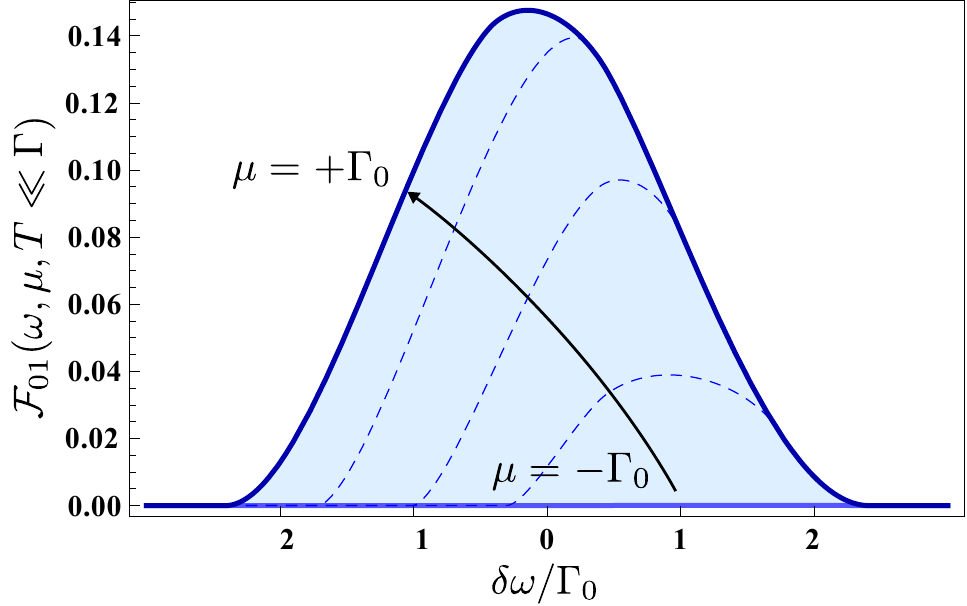}
	\caption{The evolution of the function $\mathcal{F}_{01}(\omega,\mu,T)$, Eq.~\eqref{sec:app_Fintegrals:eq:fctCalFKM}, for fixed $T\ll\Gamma$ and for $\mu$ varying from $\mu=-\Gamma_0$ to $\mu=\Gamma_0$ inside the zeroth LL.}
	\label{sec:app_Fintegrals:fig:F01_2}
\end{figure}
\begin{figure}[ht]
	\centering
	\includegraphics[width=0.45\textwidth]{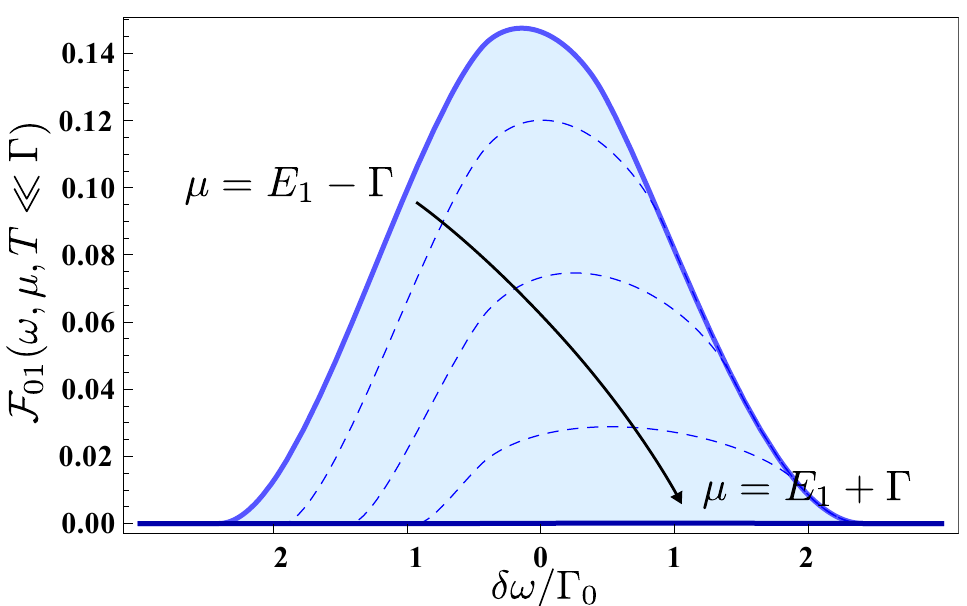}
	\caption{The evolution of the function $\mathcal{F}_{01}(\omega,\mu,T)$, Eq.~\eqref{sec:app_Fintegrals:eq:fctCalFKM}, for fixed $T\ll\Gamma$ and for $\mu$ varying from $\mu=E_1-\Gamma_0$ to $\mu=E_1+\Gamma_0$ inside the first LL.}
	\label{sec:app_Fintegrals:fig:F01_3}
\end{figure}

\noindent
Using the expressions~\eqref{sec:app_vertex_corrections:eq:sep_Pi_cycl_res_vertex}-\eqref{sec:app_vertex_corrections:eq:sep_Pi_off_res_0thLL} for the $\Pi$-correlators in Eqs.~\eqref{sec:app_vertex_corrections:eq:def_F_tilde} and \eqref{sec:app_vertex_corrections:eq:def_F}, here we derive the explicit expressions for the functions $\mathcal{F}$, $F$ and $\tilde{\mathcal{F}}$, $\tilde{F}$ describing the shape of the conductivity peaks in the regime of separated LLs.

For $K\neq M$ (inter-LL transitions) and $T\ll\Gamma$, we obtain for Eq.~\eqref{sec:app_vertex_corrections:eq:def_F},
\begin{equation}
	\begin{split}
		& \mathcal{F}_{KM}(\omega,\mu) = \frac{\Gamma}{\omega} \: \Theta(\Gamma_K+\Gamma_M-|\delta\omega|) \\
		& \times \Theta(\delta\mu+\text{min}\{\Gamma_K,\delta\omega+\Gamma_M\})\\
		& \times \Theta(\text{min}\{\Gamma_K-\delta\omega,\Gamma_M\}-\delta\mu) \\
		& \times \int_{\text{max}\{-\Gamma_K,-\Gamma_M-\delta\omega,\delta\mu-\omega\}/\Gamma}^{\text{min}\{\Gamma_K,\Gamma_M-\delta\omega,\delta\mu\}/\Gamma}\dd\ep \: \sqrt{1-\frac{\ep^2}{c_K}} \\
		& \times \sqrt{1-\frac{(\ep+\delta\omega/\Gamma)^2}{c_M}} \: ,
	\end{split}
	\label{sec:app_Fintegrals:eq:fctCalFKM}
\end{equation}
for the function~\eqref{sec:conductivity:eq:fctF_def}. Here $\delta\omega=\omega-E_M+E_K$ and ${\delta\mu=\mu-E_K}$.

In the case $K=M$ and for $T\ll\Gamma$, Eq.~\eqref{sec:app_vertex_corrections:eq:def_F} yields
\begin{equation}
	\begin{split}
		& \mathcal{F}_{KK}(\omega,\mu) = \: \frac{\Gamma_K}{\omega} \: \Theta\left(2\Gamma_K-\left|\delta\omega\right|\right) \\
		& \times \Theta(\delta\mu+\text{min}\{\Gamma_K,\Gamma_K+\delta\omega\}) \\
		& \times \Theta(\text{min}\{\Gamma_K-\delta\omega,\Gamma_K\}-\delta\mu) \\
		& \times \int_{\text{max}\{-1,-1-\delta\omega/\Gamma_K,\delta\mu/\Gamma_K\}}^{\text{min}\{1,1-\delta\omega/\Gamma_K,\delta\mu/\Gamma_K\}}\dd\ep \: \sqrt{1-\ep^2} \\
		& \times \sqrt{1-(\ep+\delta\omega/\Gamma_K)^2} \: .
	\end{split}
	\label{sec:app_Fintegrals:eq:fctF_intra}
\end{equation}

\begin{figure}[h]
	\centering
	\includegraphics[width=0.45\textwidth]{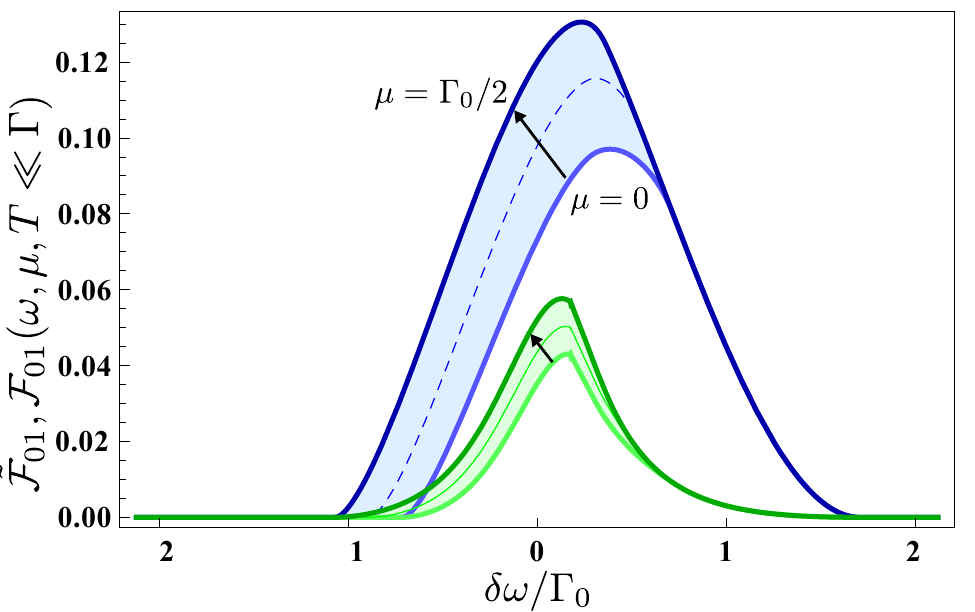}
	\caption{The evolution of the functions $\mathcal{F}_{01}(\omega,\mu,T)$ (top) from Eq.~\eqref{sec:app_Fintegrals:eq:fctCalFKM} and $\tilde{\mathcal{F}}_{01}(\omega,\mu,T)$ (bottom) from Eq.~\eqref{sec:app_Fintegrals:eq:fctTildecalF_vertex} for fixed $T\ll\Gamma$, both drawn for a chemical potential varying from $\mu=0$ to $\mu=\Gamma/2$.}
	\label{sec:app_Fintegrals:fig:F01}
\end{figure}

The functions entering Eq.~\eqref{sec:conductivity:eq:fctF_high_T_1} high $T\gg\Gamma$ are
\begin{equation}
	\begin{split}
		& F_{KM}\left(\frac{\delta\omega}{\Gamma}\right) = \Theta(\Gamma_K+\Gamma_M-|\delta\omega|) \frac{3\Gamma}{4\Gamma_K} \\
		& \times \int_{\text{max}\{-\Gamma_K,-\Gamma_M-\delta\omega\}/\Gamma}^{\text{min}\{\Gamma_K,\Gamma_M-\delta\omega\}/\Gamma}\dd\ep \: \sqrt{1-\frac{\ep^2}{c_K}} \\
		& \times \sqrt{1-\frac{(\ep+\delta\omega/\Gamma)^2}{c_M}} \: .
		\label{sec:app_Fintegrals:eq:fctFKM}
	\end{split}
\end{equation}
In the case $K=M$ (intra-LL transition) Eq.~\eqref{sec:app_Fintegrals:eq:fctFKM} turns into
\begin{equation}
	\begin{split}
		& F_{KK}\left(\frac{\omega}{\Gamma_K}\right) = \frac{3}{4}\Theta\left(2-\left|\frac{\omega}{\Gamma_K}\right|\right)\\
		& \times \int_{-1}^{1-\omega/\Gamma_K}\dd\ep \: \sqrt{1-\ep^2}\sqrt{1-\left(\ep+\frac{\omega}{\Gamma_K}\right)^2} \: ,
	\end{split}
	\label{sec:app_Fintegrals:eq:fctF}
\end{equation}
which can be expressed in terms of elliptic integrals, see e.g. Ref.~\onlinecite{FracMIRO07}.

\begin{figure}[h]
	\centering
	\includegraphics[width=0.45\textwidth]{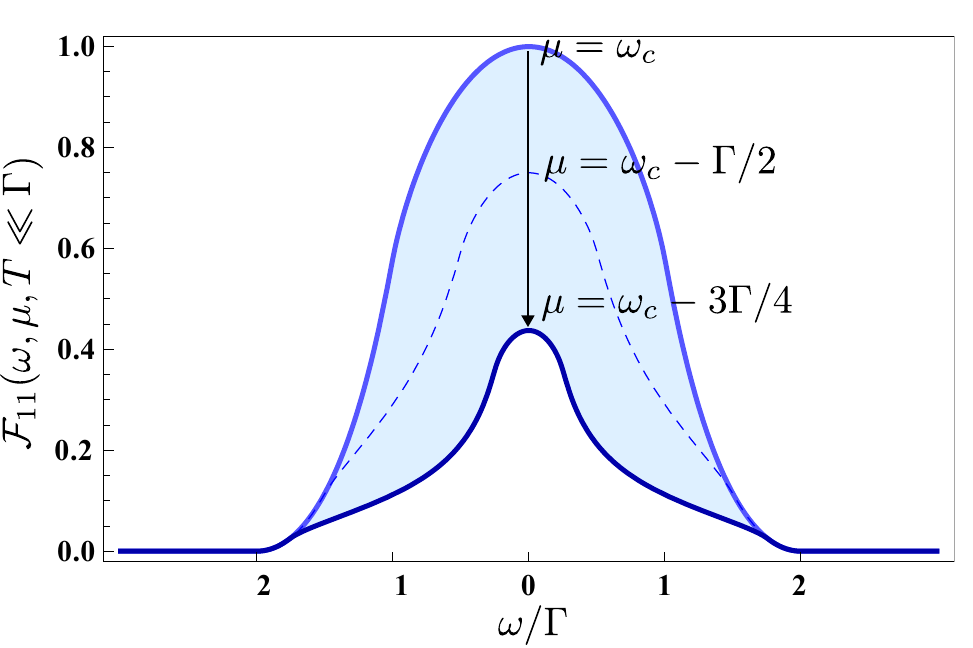}
	\caption{The evolution of the function $\mathcal{F}_{11}(\omega,\mu,T)$, Eq.~\eqref{sec:app_Fintegrals:eq:fctCalFKM}, for fixed $T\ll\Gamma$ and for $\mu$ varying from the value $\mu=\omega_c$ to $\mu=\omega_c-3\Gamma/4$ inside the first LL.}
	\label{sec:app_Fintegrals:fig:F11}
\end{figure}

At the cyclotron resonance $||K|-|M||=1$, the inclusion of the vertex corrections [Eq.~\eqref{sec:app_vertex_corrections:eq:sep_Pi_cycl_res_vertex}], in the case $T\ll\Gamma$, yields for Eq.~\eqref{sec:app_vertex_corrections:eq:def_F_tilde}
\begin{equation}
	\begin{split}
		& \tilde{\mathcal{F}}_{KM}(\omega,\mu) = \frac{\Gamma}{\omega} \: \Theta(\Gamma_K+\Gamma_M-|\delta\omega|) \\
		& \times \Theta(\delta\mu+\text{min}\{\Gamma_K,\delta\omega+\Gamma_M\})\\
		& \times \Theta(\text{min}\{\Gamma_K-\delta\omega,\Gamma_M\}-\delta\mu) \\
		& \times \int_{\text{max}\{-\Gamma_K,-\Gamma_M-\delta\omega,\delta\mu-\omega\}/\Gamma}^{\text{min}\{\Gamma_K,\Gamma_M-\delta\omega,\delta\mu\}/\Gamma}
		\frac{3\sqrt{1-\frac{\ep^2}{c_K}}\sqrt{1-\frac{(\ep+\delta\omega/\Gamma)^2}{c_M}}\dd\ep}{9+8[\ep+2\delta\omega/\Gamma)][\delta\omega/\Gamma-\ep]} \: ,
	\end{split}
	\label{sec:app_Fintegrals:eq:fctTildecalF_vertex}
\end{equation}
where an additional energy dependence enters in the denominator of the integrand in accord with Eq.~\eqref{sec:app_vertex_corrections:eq:sep_Pi_cycl_res_vertex}. This energy dependence also enters
the function

\begin{equation}
	\begin{split}
		& \tilde{F}_{KM}\left(\frac{\delta\omega}{\Gamma}\right) = \Theta(\Gamma_K+\Gamma_M-|\delta\omega|) \frac{3\Gamma}{4\Gamma_K} \\
 		& \times \int_{\text{max}\{-\Gamma_K,-\Gamma_M-\delta\omega\}/\Gamma}^{\text{min}\{\Gamma_K,\Gamma_M-\delta\omega\}/\Gamma}
		\frac{3\sqrt{1-\frac{\ep^2}{c_K}}\sqrt{1-\frac{(\ep+\delta\omega/\Gamma)^2}{c_M}}\dd\ep}{9+8[\ep+2\delta\omega/\Gamma][\delta\omega/\Gamma-\ep]} \: ,
		\label{sec:app_Fintegrals:eq:fctTildeF_vertex}
	\end{split}
\end{equation}

from Eq.~\eqref{sec:conductivity:eq:fctF_high_T_2} for $T\gg\Gamma$.

Figure.~\ref{sec:app_Fintegrals:fig:F01_2} illustrates the function $\mathcal{F}_{01}$, Eq.~\eqref{sec:app_Fintegrals:eq:fctCalFKM}, for $T\ll\Gamma$. The arrow indicates the change with increasing chemical potential from ${\mu=-\Gamma_0}$ to ${\mu=+\Gamma_0}$. The function $\mathcal{F}_{01}$ ($\tilde{\mathcal{F}}_{01}$) shows an overall increase from ${\mu=-\Gamma_0}$ to ${\mu=+\Gamma_0}$ followed by a decrease if the chemical potential jumps into the first LL and increases further, see Fig.~\ref{sec:app_Fintegrals:fig:F01_3}. Figure \ref{sec:app_Fintegrals:fig:F01} compares the function $\tilde{\mathcal{F}}_{01}$, Eq.~\eqref{sec:app_Fintegrals:eq:fctTildecalF_vertex}, to $\mathcal{F}_{01}$. We observe that the vertex corrections included in $\tilde{\mathcal{F}}$ reduce the strength and width of the cyclotron resonance similar to functions $F_{KM}$ and  $\tilde{F}_{KM}$ illustrated in Fig.~\ref{sec:conductivity:fig:functionF}. Figure \ref{sec:app_Fintegrals:fig:F11} shows the function $\mathcal{F}_{11}$, Eq.~\eqref{sec:app_Fintegrals:eq:fctF_intra}, for $T\ll\Gamma$ and different chemical potentials.

\begin{acknowledgments}
We are thankful to P. S. Alekseev, I. V. Gornyi, E. K\"onig, P. M. Ostrovsky and M. Sch\"utt for discussions. This work was supported by DFG-RFBR, and DFG-SPP 1285 and 1459.
\end{acknowledgments}

%\bibliography{paper_dt_01.bib}

\begin{thebibliography}{47}%
\makeatletter
\providecommand \@ifxundefined [1]{%
 \@ifx{#1\undefined}
}%
\providecommand \@ifnum [1]{%
 \ifnum #1\expandafter \@firstoftwo
 \else \expandafter \@secondoftwo
 \fi
}%
\providecommand \@ifx [1]{%
 \ifx #1\expandafter \@firstoftwo
 \else \expandafter \@secondoftwo
 \fi
}%
\providecommand \natexlab [1]{#1}%
\providecommand \enquote  [1]{``#1''}%
\providecommand \bibnamefont  [1]{#1}%
\providecommand \bibfnamefont [1]{#1}%
\providecommand \citenamefont [1]{#1}%
\providecommand \href@noop [0]{\@secondoftwo}%
\providecommand \href [0]{\begingroup \@sanitize@url \@href}%
\providecommand \@href[1]{\@@startlink{#1}\@@href}%
\providecommand \@@href[1]{\endgroup#1\@@endlink}%
\providecommand \@sanitize@url [0]{\catcode `\\12\catcode `\$12\catcode
  `\&12\catcode `\#12\catcode `\^12\catcode `\_12\catcode `\%12\relax}%
\providecommand \@@startlink[1]{}%
\providecommand \@@endlink[0]{}%
\providecommand \url  [0]{\begingroup\@sanitize@url \@url }%
\providecommand \@url [1]{\endgroup\@href {#1}{\urlprefix }}%
\providecommand \urlprefix  [0]{URL }%
\providecommand \Eprint [0]{\href }%
\providecommand \doibase [0]{http://dx.doi.org/}%
\providecommand \selectlanguage [0]{\@gobble}%
\providecommand \bibinfo  [0]{\@secondoftwo}%
\providecommand \bibfield  [0]{\@secondoftwo}%
\providecommand \translation [1]{[#1]}%
\providecommand \BibitemOpen [0]{}%
\providecommand \bibitemStop [0]{}%
\providecommand \bibitemNoStop [0]{.\EOS\space}%
\providecommand \EOS [0]{\spacefactor3000\relax}%
\providecommand \BibitemShut  [1]{\csname bibitem#1\endcsname}%
\let\auto@bib@innerbib\@empty
%</preamble>
\bibitem [{\citenamefont {Novoselov}\ \emph {et~al.}(2004)\citenamefont
  {Novoselov}, \citenamefont {Geim}, \citenamefont {Morozov}, \citenamefont
  {Jiang}, \citenamefont {Zhang}, \citenamefont {Dubonos}, \citenamefont
  {Grigorieva},\ and\ \citenamefont {Firsov}}]{Nov04Graphene}%
  \BibitemOpen
  \bibfield  {author} {\bibinfo {author} {\bibfnamefont {K.~S.}\ \bibnamefont
  {Novoselov}}, \bibinfo {author} {\bibfnamefont {A.~K.}\ \bibnamefont {Geim}},
  \bibinfo {author} {\bibfnamefont {S.~V.}\ \bibnamefont {Morozov}}, \bibinfo
  {author} {\bibfnamefont {D.}~\bibnamefont {Jiang}}, \bibinfo {author}
  {\bibfnamefont {Y.}~\bibnamefont {Zhang}}, \bibinfo {author} {\bibfnamefont
  {S.}~\bibnamefont {Dubonos}}, \bibinfo {author} {\bibfnamefont {I.~V.}\
  \bibnamefont {Grigorieva}}, \ and\ \bibinfo {author} {\bibfnamefont {A.~A.}\
  \bibnamefont {Firsov}},\ }\href@noop {} {\bibfield  {journal} {\bibinfo
  {journal} {Science}\ }\textbf {\bibinfo {volume} {306}},\ \bibinfo {pages}
  {666} (\bibinfo {year} {2004})}\BibitemShut {NoStop}%
\bibitem [{\citenamefont {Semenoff}(1984)}]{Sem84}%
  \BibitemOpen
  \bibfield  {author} {\bibinfo {author} {\bibfnamefont {G.~W.}\ \bibnamefont
  {Semenoff}},\ }\href@noop {} {\bibfield  {journal} {\bibinfo  {journal}
  {Phys. Rev. Lett.}\ }\textbf {\bibinfo {volume} {53}},\ \bibinfo {pages}
  {2449} (\bibinfo {year} {1984})}\BibitemShut {NoStop}%
\bibitem [{\citenamefont {Castro~Neto}\ \emph {et~al.}(2009)\citenamefont
  {Castro~Neto}, \citenamefont {Guinea}, \citenamefont {Peres}, \citenamefont
  {Novoselov},\ and\ \citenamefont {Geim}}]{CNeto09}%
  \BibitemOpen
  \bibfield  {author} {\bibinfo {author} {\bibfnamefont {A.~H.}\ \bibnamefont
  {Castro~Neto}}, \bibinfo {author} {\bibfnamefont {F.}~\bibnamefont {Guinea}},
  \bibinfo {author} {\bibfnamefont {N.~M.~R.}\ \bibnamefont {Peres}}, \bibinfo
  {author} {\bibfnamefont {K.~S.}\ \bibnamefont {Novoselov}}, \ and\ \bibinfo
  {author} {\bibfnamefont {A.~K.}\ \bibnamefont {Geim}},\ }\href@noop {}
  {\bibfield  {journal} {\bibinfo  {journal} {Rev. Mod. Phys.}\ }\textbf
  {\bibinfo {volume} {81}},\ \bibinfo {pages} {109} (\bibinfo {year}
  {2009})}\BibitemShut {NoStop}%
\bibitem [{\citenamefont {Nair}\ \emph {et~al.}(2008)\citenamefont {Nair},
  \citenamefont {Blake}, \citenamefont {Grigorenko}, \citenamefont {Novoselov},
  \citenamefont {Booth}, \citenamefont {Stauber}, \citenamefont {Peres},\ and\
  \citenamefont {Geim}}]{Nair08}%
  \BibitemOpen
  \bibfield  {author} {\bibinfo {author} {\bibfnamefont {R.~R.}\ \bibnamefont
  {Nair}}, \bibinfo {author} {\bibfnamefont {P.}~\bibnamefont {Blake}},
  \bibinfo {author} {\bibfnamefont {A.~N.}\ \bibnamefont {Grigorenko}},
  \bibinfo {author} {\bibfnamefont {K.~S.}\ \bibnamefont {Novoselov}}, \bibinfo
  {author} {\bibfnamefont {T.~J.}\ \bibnamefont {Booth}}, \bibinfo {author}
  {\bibfnamefont {T.}~\bibnamefont {Stauber}}, \bibinfo {author} {\bibfnamefont
  {N.~M.~R.}\ \bibnamefont {Peres}}, \ and\ \bibinfo {author} {\bibfnamefont
  {A.~K.}\ \bibnamefont {Geim}},\ }\href@noop {} {\bibfield  {journal}
  {\bibinfo  {journal} {Science}\ }\textbf {\bibinfo {volume} {320}},\ \bibinfo
  {pages} {1308} (\bibinfo {year} {2008})}\BibitemShut {NoStop}%
\bibitem [{\citenamefont {Ostrovsky}\ \emph {et~al.}(2008)\citenamefont
  {Ostrovsky}, \citenamefont {Gornyi},\ and\ \citenamefont
  {Mirlin}}]{Ostr08AnQH}%
  \BibitemOpen
  \bibfield  {author} {\bibinfo {author} {\bibfnamefont {P.~M.}\ \bibnamefont
  {Ostrovsky}}, \bibinfo {author} {\bibfnamefont {I.~V.}\ \bibnamefont
  {Gornyi}}, \ and\ \bibinfo {author} {\bibfnamefont {A.~D.}\ \bibnamefont
  {Mirlin}},\ }\href@noop {} {\bibfield  {journal} {\bibinfo  {journal} {Phys.
  Rev. B}\ }\textbf {\bibinfo {volume} {77}},\ \bibinfo {pages} {195430}
  (\bibinfo {year} {2008})}\BibitemShut {NoStop}%
\bibitem [{\citenamefont {Novoselov}\ \emph {et~al.}(2006)\citenamefont
  {Novoselov}, \citenamefont {McCann}, \citenamefont {Morozov}, \citenamefont
  {Fal'ko}, \citenamefont {Katsnelson}, \citenamefont {Zeitler}, \citenamefont
  {Jiang}, \citenamefont {Schedin},\ and\ \citenamefont
  {Geim}}]{novoselov:2006}%
  \BibitemOpen
  \bibfield  {author} {\bibinfo {author} {\bibfnamefont {K.~S.}\ \bibnamefont
  {Novoselov}}, \bibinfo {author} {\bibfnamefont {E.}~\bibnamefont {McCann}},
  \bibinfo {author} {\bibfnamefont {S.~V.}\ \bibnamefont {Morozov}}, \bibinfo
  {author} {\bibfnamefont {V.~I.}\ \bibnamefont {Fal'ko}}, \bibinfo {author}
  {\bibfnamefont {M.~I.}\ \bibnamefont {Katsnelson}}, \bibinfo {author}
  {\bibfnamefont {U.}~\bibnamefont {Zeitler}}, \bibinfo {author} {\bibfnamefont
  {D.}~\bibnamefont {Jiang}}, \bibinfo {author} {\bibfnamefont
  {F.}~\bibnamefont {Schedin}}, \ and\ \bibinfo {author} {\bibfnamefont
  {A.~K.}\ \bibnamefont {Geim}},\ }\href@noop {} {\bibfield  {journal}
  {\bibinfo  {journal} {Nat. Phys.}\ }\textbf {\bibinfo {volume} {2}},\
  \bibinfo {pages} {177} (\bibinfo {year} {2006})}\BibitemShut {NoStop}%
\bibitem [{\citenamefont {Novoselov}\ \emph {et~al.}(2007)\citenamefont
  {Novoselov}, \citenamefont {Jiang}, \citenamefont {Zhang}, \citenamefont
  {Morozov}, \citenamefont {Stormer}, \citenamefont {Zeitler}, \citenamefont
  {Maan}, \citenamefont {Boebinger}, \citenamefont {Kim},\ and\ \citenamefont
  {Geim}}]{Nov07QH}%
  \BibitemOpen
  \bibfield  {author} {\bibinfo {author} {\bibfnamefont {K.~S.}\ \bibnamefont
  {Novoselov}}, \bibinfo {author} {\bibfnamefont {Z.}~\bibnamefont {Jiang}},
  \bibinfo {author} {\bibfnamefont {Y.}~\bibnamefont {Zhang}}, \bibinfo
  {author} {\bibfnamefont {S.~V.}\ \bibnamefont {Morozov}}, \bibinfo {author}
  {\bibfnamefont {H.~L.}\ \bibnamefont {Stormer}}, \bibinfo {author}
  {\bibfnamefont {U.}~\bibnamefont {Zeitler}}, \bibinfo {author} {\bibfnamefont
  {J.~C.}\ \bibnamefont {Maan}}, \bibinfo {author} {\bibfnamefont {G.~S.}\
  \bibnamefont {Boebinger}}, \bibinfo {author} {\bibfnamefont {P.}~\bibnamefont
  {Kim}}, \ and\ \bibinfo {author} {\bibfnamefont {A.~K.}\ \bibnamefont
  {Geim}},\ }\href@noop {} {\bibfield  {journal} {\bibinfo  {journal}
  {Science}\ }\textbf {\bibinfo {volume} {315}},\ \bibinfo {pages} {1379}
  (\bibinfo {year} {2007})}\BibitemShut {NoStop}%
\bibitem [{\citenamefont {Bonaccorso}\ \emph {et~al.}(2010)\citenamefont
  {Bonaccorso}, \citenamefont {Sun}, \citenamefont {Hasan},\ and\ \citenamefont
  {Ferrari}}]{FerrariGrapheneOptoRev10}%
  \BibitemOpen
  \bibfield  {author} {\bibinfo {author} {\bibfnamefont {F.}~\bibnamefont
  {Bonaccorso}}, \bibinfo {author} {\bibfnamefont {Z.}~\bibnamefont {Sun}},
  \bibinfo {author} {\bibfnamefont {T.}~\bibnamefont {Hasan}}, \ and\ \bibinfo
  {author} {\bibfnamefont {A.~C.}\ \bibnamefont {Ferrari}},\ }\href@noop {}
  {\bibfield  {journal} {\bibinfo  {journal} {Nat. Photon.}\ }\textbf {\bibinfo
  {volume} {4}},\ \bibinfo {pages} {611} (\bibinfo {year} {2010})}\BibitemShut
  {NoStop}%
\bibitem [{\citenamefont {Avouris}(2010)}]{AvGraphenePhotRev10}%
  \BibitemOpen
  \bibfield  {author} {\bibinfo {author} {\bibfnamefont {P.}~\bibnamefont
  {Avouris}},\ }\href@noop {} {\bibfield  {journal} {\bibinfo  {journal} {Nano
  Lett.}\ }\textbf {\bibinfo {volume} {10}},\ \bibinfo {pages} {4285} (\bibinfo
  {year} {2010})}\BibitemShut {NoStop}%
\bibitem [{\citenamefont {Koppens}\ \emph {et~al.}(2011)\citenamefont
  {Koppens}, \citenamefont {Chang},\ and\ \citenamefont {Garcia~de
  Abajo}}]{Koppens2011}%
  \BibitemOpen
  \bibfield  {author} {\bibinfo {author} {\bibfnamefont {F.~H.~L.}\
  \bibnamefont {Koppens}}, \bibinfo {author} {\bibfnamefont {D.~E.}\
  \bibnamefont {Chang}}, \ and\ \bibinfo {author} {\bibfnamefont {F.~J.}\
  \bibnamefont {Garcia~de Abajo}},\ }\href@noop {} {\bibfield  {journal}
  {\bibinfo  {journal} {Nano Letters}\ }\textbf {\bibinfo {volume} {11}},\
  \bibinfo {pages} {3370} (\bibinfo {year} {2011})}\BibitemShut {NoStop}%
\bibitem [{\citenamefont {Grigorenko}\ \emph {et~al.}(2012)\citenamefont
  {Grigorenko}, \citenamefont {Polini},\ and\ \citenamefont
  {Novoselov}}]{Grigorenko2012}%
  \BibitemOpen
  \bibfield  {author} {\bibinfo {author} {\bibfnamefont {A.~N.}\ \bibnamefont
  {Grigorenko}}, \bibinfo {author} {\bibfnamefont {M.}~\bibnamefont {Polini}},
  \ and\ \bibinfo {author} {\bibfnamefont {K.~S.}\ \bibnamefont {Novoselov}},\
  }\href@noop {} {\bibfield  {journal} {\bibinfo  {journal} {Nat. Photon.}\
  }\textbf {\bibinfo {volume} {6}},\ \bibinfo {pages} {749} (\bibinfo {year}
  {2012})}\BibitemShut {NoStop}%
\bibitem [{\citenamefont {Engel}\ \emph {et~al.}(2012)\citenamefont {Engel},
  \citenamefont {Steiner}, \citenamefont {Lombardo}, \citenamefont {Ferrari},
  \citenamefont {von L\"ohneysen}, \citenamefont {Avouris},\ and\ \citenamefont
  {Krupke}}]{Engel2012}%
  \BibitemOpen
  \bibfield  {author} {\bibinfo {author} {\bibfnamefont {M.}~\bibnamefont
  {Engel}}, \bibinfo {author} {\bibfnamefont {M.}~\bibnamefont {Steiner}},
  \bibinfo {author} {\bibfnamefont {A.}~\bibnamefont {Lombardo}}, \bibinfo
  {author} {\bibfnamefont {A.~C.}\ \bibnamefont {Ferrari}}, \bibinfo {author}
  {\bibfnamefont {H.}~\bibnamefont {von L\"ohneysen}}, \bibinfo {author}
  {\bibfnamefont {P.}~\bibnamefont {Avouris}}, \ and\ \bibinfo {author}
  {\bibfnamefont {R.}~\bibnamefont {Krupke}},\ }\href@noop {} {\bibfield
  {journal} {\bibinfo  {journal} {Nat. Commun.}\ }\textbf {\bibinfo {volume}
  {3}},\ \bibinfo {pages} {906} (\bibinfo {year} {2012})}\BibitemShut {NoStop}%
\bibitem [{\citenamefont {Morimoto}\ \emph {et~al.}(2009)\citenamefont
  {Morimoto}, \citenamefont {Hatsugai},\ and\ \citenamefont
  {Aoki}}]{LLLaser09}%
  \BibitemOpen
  \bibfield  {author} {\bibinfo {author} {\bibfnamefont {T.}~\bibnamefont
  {Morimoto}}, \bibinfo {author} {\bibfnamefont {Y.}~\bibnamefont {Hatsugai}},
  \ and\ \bibinfo {author} {\bibfnamefont {H.}~\bibnamefont {Aoki}},\
  }\href@noop {} {\bibfield  {journal} {\bibinfo  {journal} {J. of Phys.: Conf.
  Ser.}\ }\textbf {\bibinfo {volume} {150}},\ \bibinfo {pages} {022059}
  (\bibinfo {year} {2009})}\BibitemShut {NoStop}%
\bibitem [{\citenamefont {Tan}\ \emph {et~al.}(2011)\citenamefont {Tan},
  \citenamefont {Tan}, \citenamefont {Ma}, \citenamefont {Liu}, \citenamefont
  {Lu},\ and\ \citenamefont {Yang}}]{Tan11SdH}%
  \BibitemOpen
  \bibfield  {author} {\bibinfo {author} {\bibfnamefont {Z.}~\bibnamefont
  {Tan}}, \bibinfo {author} {\bibfnamefont {C.}~\bibnamefont {Tan}}, \bibinfo
  {author} {\bibfnamefont {L.}~\bibnamefont {Ma}}, \bibinfo {author}
  {\bibfnamefont {G.~T.}\ \bibnamefont {Liu}}, \bibinfo {author} {\bibfnamefont
  {L.}~\bibnamefont {Lu}}, \ and\ \bibinfo {author} {\bibfnamefont {C.~L.}\
  \bibnamefont {Yang}},\ }\href@noop {} {\bibfield  {journal} {\bibinfo
  {journal} {Phys. Rev. B}\ }\textbf {\bibinfo {volume} {84}},\ \bibinfo
  {pages} {115429} (\bibinfo {year} {2011})}\BibitemShut {NoStop}%
\bibitem [{\citenamefont {Sadowski}\ \emph {et~al.}(2006)\citenamefont
  {Sadowski}, \citenamefont {Martinez}, \citenamefont {Potemski}, \citenamefont
  {Berger},\ and\ \citenamefont {de~Heer}}]{SadowskiGrLayers06}%
  \BibitemOpen
  \bibfield  {author} {\bibinfo {author} {\bibfnamefont {M.~L.}\ \bibnamefont
  {Sadowski}}, \bibinfo {author} {\bibfnamefont {G.}~\bibnamefont {Martinez}},
  \bibinfo {author} {\bibfnamefont {M.}~\bibnamefont {Potemski}}, \bibinfo
  {author} {\bibfnamefont {C.}~\bibnamefont {Berger}}, \ and\ \bibinfo {author}
  {\bibfnamefont {W.~A.}\ \bibnamefont {de~Heer}},\ }\href@noop {} {\bibfield
  {journal} {\bibinfo  {journal} {Phys. Rev. Lett.}\ }\textbf {\bibinfo
  {volume} {97}},\ \bibinfo {pages} {266405} (\bibinfo {year}
  {2006})}\BibitemShut {NoStop}%
\bibitem [{\citenamefont {Orlita}\ and\ \citenamefont
  {Potemski}(2010)}]{OrlitaPotMOptRev10}%
  \BibitemOpen
  \bibfield  {author} {\bibinfo {author} {\bibfnamefont {M.}~\bibnamefont
  {Orlita}}\ and\ \bibinfo {author} {\bibfnamefont {M.}~\bibnamefont
  {Potemski}},\ }\href@noop {} {\bibfield  {journal} {\bibinfo  {journal}
  {Semicond. Sci. Technol.}\ }\textbf {\bibinfo {volume} {25}},\ \bibinfo
  {pages} {063001} (\bibinfo {year} {2010})}\BibitemShut {NoStop}%
\bibitem [{\citenamefont {Orlita}\ \emph {et~al.}(2011)\citenamefont {Orlita},
  \citenamefont {Faugeras}, \citenamefont {Grill}, \citenamefont {Wysmolek},
  \citenamefont {Strupinski}, \citenamefont {Berger}, \citenamefont {de~Heer},
  \citenamefont {Martinez},\ and\ \citenamefont {Potemski}}]{Pot11Scattering}%
  \BibitemOpen
  \bibfield  {author} {\bibinfo {author} {\bibfnamefont {M.}~\bibnamefont
  {Orlita}}, \bibinfo {author} {\bibfnamefont {C.}~\bibnamefont {Faugeras}},
  \bibinfo {author} {\bibfnamefont {R.}~\bibnamefont {Grill}}, \bibinfo
  {author} {\bibfnamefont {A.}~\bibnamefont {Wysmolek}}, \bibinfo {author}
  {\bibfnamefont {W.}~\bibnamefont {Strupinski}}, \bibinfo {author}
  {\bibfnamefont {C.}~\bibnamefont {Berger}}, \bibinfo {author} {\bibfnamefont
  {W.~A.}\ \bibnamefont {de~Heer}}, \bibinfo {author} {\bibfnamefont
  {G.}~\bibnamefont {Martinez}}, \ and\ \bibinfo {author} {\bibfnamefont
  {M.}~\bibnamefont {Potemski}},\ }\href@noop {} {\bibfield  {journal}
  {\bibinfo  {journal} {Phys. Rev. Lett.}\ }\textbf {\bibinfo {volume} {107}},\
  \bibinfo {pages} {216603} (\bibinfo {year} {2011})}\BibitemShut {NoStop}%
\bibitem [{\citenamefont {Ando}\ \emph {et~al.}(1982)\citenamefont {Ando},
  \citenamefont {Fowler},\ and\ \citenamefont {Stern}}]{ando:1982a}%
  \BibitemOpen
  \bibfield  {author} {\bibinfo {author} {\bibfnamefont {T.}~\bibnamefont
  {Ando}}, \bibinfo {author} {\bibfnamefont {A.~B.}\ \bibnamefont {Fowler}}, \
  and\ \bibinfo {author} {\bibfnamefont {F.}~\bibnamefont {Stern}},\
  }\href@noop {} {\bibfield  {journal} {\bibinfo  {journal} {Rev. Mod. Phys.}\
  }\textbf {\bibinfo {volume} {54}},\ \bibinfo {pages} {437} (\bibinfo {year}
  {1982})}\BibitemShut {NoStop}%
\bibitem [{\citenamefont {Dmitriev}\ \emph {et~al.}(2012)\citenamefont
  {Dmitriev}, \citenamefont {Mirlin}, \citenamefont {Polyakov},\ and\
  \citenamefont {Zudov}}]{IvanReview11}%
  \BibitemOpen
  \bibfield  {author} {\bibinfo {author} {\bibfnamefont {I.~A.}\ \bibnamefont
  {Dmitriev}}, \bibinfo {author} {\bibfnamefont {A.~D.}\ \bibnamefont
  {Mirlin}}, \bibinfo {author} {\bibfnamefont {D.~G.}\ \bibnamefont
  {Polyakov}}, \ and\ \bibinfo {author} {\bibfnamefont {M.~A.}\ \bibnamefont
  {Zudov}},\ }\href@noop {} {\bibfield  {journal} {\bibinfo  {journal} {Rev.
  Mod. Phys.}\ }\textbf {\bibinfo {volume} {84}},\ \bibinfo {pages} {1709}
  (\bibinfo {year} {2012})}\BibitemShut {NoStop}%
\bibitem [{\citenamefont {Ando}(1974{\natexlab{a}})}]{Ando74IVOscCond}%
  \BibitemOpen
  \bibfield  {author} {\bibinfo {author} {\bibfnamefont {T.}~\bibnamefont
  {Ando}},\ }\href@noop {} {\bibfield  {journal} {\bibinfo  {journal} {J. Phys.
  Soc. Jpn.}\ }\textbf {\bibinfo {volume} {37}},\ \bibinfo {pages} {1233}
  (\bibinfo {year} {1974}{\natexlab{a}})}\BibitemShut {NoStop}%
\bibitem [{\citenamefont {Ando}(1975)}]{Ando75CRShape}%
  \BibitemOpen
  \bibfield  {author} {\bibinfo {author} {\bibfnamefont {T.}~\bibnamefont
  {Ando}},\ }\href@noop {} {\bibfield  {journal} {\bibinfo  {journal} {J. Phys.
  Soc. Jpn.}\ }\textbf {\bibinfo {volume} {38}},\ \bibinfo {pages} {989}
  (\bibinfo {year} {1975})}\BibitemShut {NoStop}%
\bibitem [{\citenamefont {Abstreiter}\ \emph {et~al.}(1976)\citenamefont
  {Abstreiter}, \citenamefont {Kotthaus}, \citenamefont {Koch},\ and\
  \citenamefont {Dorda}}]{Abstreiter76}%
  \BibitemOpen
  \bibfield  {author} {\bibinfo {author} {\bibfnamefont {G.}~\bibnamefont
  {Abstreiter}}, \bibinfo {author} {\bibfnamefont {J.~P.}\ \bibnamefont
  {Kotthaus}}, \bibinfo {author} {\bibfnamefont {J.~F.}\ \bibnamefont {Koch}},
  \ and\ \bibinfo {author} {\bibfnamefont {G.}~\bibnamefont {Dorda}},\ }\href
  {\doibase 10.1103/PhysRevB.14.2480} {\bibfield  {journal} {\bibinfo
  {journal} {Phys. Rev. B}\ }\textbf {\bibinfo {volume} {14}},\ \bibinfo
  {pages} {2480} (\bibinfo {year} {1976})}\BibitemShut {NoStop}%
\bibitem [{\citenamefont {Dmitriev}\ \emph {et~al.}(2003)\citenamefont
  {Dmitriev}, \citenamefont {Mirlin},\ and\ \citenamefont
  {Polyakov}}]{Ivan03CRHarmonics}%
  \BibitemOpen
  \bibfield  {author} {\bibinfo {author} {\bibfnamefont {I.~A.}\ \bibnamefont
  {Dmitriev}}, \bibinfo {author} {\bibfnamefont {A.~D.}\ \bibnamefont
  {Mirlin}}, \ and\ \bibinfo {author} {\bibfnamefont {D.~G.}\ \bibnamefont
  {Polyakov}},\ }\href@noop {} {\bibfield  {journal} {\bibinfo  {journal}
  {Phys. Rev. Lett.}\ }\textbf {\bibinfo {volume} {91}},\ \bibinfo {pages}
  {226802} (\bibinfo {year} {2003})}\BibitemShut {NoStop}%
\bibitem [{\citenamefont {Fedorych}\ \emph {et~al.}(2010)\citenamefont
  {Fedorych}, \citenamefont {Potemski}, \citenamefont {Studenikin},
  \citenamefont {Gupta}, \citenamefont {Wasilewski},\ and\ \citenamefont
  {Dmitriev}}]{Stud10QO2DEG}%
  \BibitemOpen
  \bibfield  {author} {\bibinfo {author} {\bibfnamefont {O.~M.}\ \bibnamefont
  {Fedorych}}, \bibinfo {author} {\bibfnamefont {M.}~\bibnamefont {Potemski}},
  \bibinfo {author} {\bibfnamefont {S.~A.}\ \bibnamefont {Studenikin}},
  \bibinfo {author} {\bibfnamefont {J.~A.}\ \bibnamefont {Gupta}}, \bibinfo
  {author} {\bibfnamefont {Z.~R.}\ \bibnamefont {Wasilewski}}, \ and\ \bibinfo
  {author} {\bibfnamefont {I.~A.}\ \bibnamefont {Dmitriev}},\ }\href@noop {}
  {\bibfield  {journal} {\bibinfo  {journal} {Phys. Rev. B}\ }\textbf {\bibinfo
  {volume} {81}},\ \bibinfo {pages} {201302} (\bibinfo {year}
  {2010})}\BibitemShut {NoStop}%
\bibitem [{\citenamefont {Gusynin}\ \emph
  {et~al.}(2007{\natexlab{a}})\citenamefont {Gusynin}, \citenamefont
  {Sharapov},\ and\ \citenamefont {Carbotte}}]{Gus07}%
  \BibitemOpen
  \bibfield  {author} {\bibinfo {author} {\bibfnamefont {V.~P.}\ \bibnamefont
  {Gusynin}}, \bibinfo {author} {\bibfnamefont {S.~G.}\ \bibnamefont
  {Sharapov}}, \ and\ \bibinfo {author} {\bibfnamefont {J.~P.}\ \bibnamefont
  {Carbotte}},\ }\href@noop {} {\bibfield  {journal} {\bibinfo  {journal} {Int.
  J. Mod. Phys. B}\ }\textbf {\bibinfo {volume} {21}},\ \bibinfo {pages} {4611}
  (\bibinfo {year} {2007}{\natexlab{a}})}\BibitemShut {NoStop}%
\bibitem [{\citenamefont {Gusynin}\ \emph
  {et~al.}(2007{\natexlab{b}})\citenamefont {Gusynin}, \citenamefont
  {Sharapov},\ and\ \citenamefont {Carbotte}}]{Gus07_2}%
  \BibitemOpen
  \bibfield  {author} {\bibinfo {author} {\bibfnamefont {V.~P.}\ \bibnamefont
  {Gusynin}}, \bibinfo {author} {\bibfnamefont {S.~G.}\ \bibnamefont
  {Sharapov}}, \ and\ \bibinfo {author} {\bibfnamefont {J.~P.}\ \bibnamefont
  {Carbotte}},\ }\href@noop {} {\bibfield  {journal} {\bibinfo  {journal} {J.
  Phys.: Condens. Matter}\ }\textbf {\bibinfo {volume} {19}},\ \bibinfo {pages}
  {026222} (\bibinfo {year} {2007}{\natexlab{b}})}\BibitemShut {NoStop}%
\bibitem [{\citenamefont {Gusynin}\ \emph
  {et~al.}(2007{\natexlab{c}})\citenamefont {Gusynin}, \citenamefont
  {Sharapov},\ and\ \citenamefont {Carbotte}}]{Gus07Anomalous}%
  \BibitemOpen
  \bibfield  {author} {\bibinfo {author} {\bibfnamefont {V.~P.}\ \bibnamefont
  {Gusynin}}, \bibinfo {author} {\bibfnamefont {S.~G.}\ \bibnamefont
  {Sharapov}}, \ and\ \bibinfo {author} {\bibfnamefont {J.~P.}\ \bibnamefont
  {Carbotte}},\ }\href@noop {} {\bibfield  {journal} {\bibinfo  {journal}
  {Phys. Rev. Lett.}\ }\textbf {\bibinfo {volume} {98}},\ \bibinfo {pages}
  {157402} (\bibinfo {year} {2007}{\natexlab{c}})}\BibitemShut {NoStop}%
\bibitem [{\citenamefont {Shon}\ and\ \citenamefont
  {Ando}(1998)}]{AndoJPSJ2DGraphite98}%
  \BibitemOpen
  \bibfield  {author} {\bibinfo {author} {\bibfnamefont {N.~H.}\ \bibnamefont
  {Shon}}\ and\ \bibinfo {author} {\bibfnamefont {T.}~\bibnamefont {Ando}},\
  }\href@noop {} {\bibfield  {journal} {\bibinfo  {journal} {J. Phys. Soc.
  Jpn.}\ }\textbf {\bibinfo {volume} {67}},\ \bibinfo {pages} {2421} (\bibinfo
  {year} {1998})}\BibitemShut {NoStop}%
\bibitem [{\citenamefont {Alekseev}\ \emph {et~al.}(2012)\citenamefont
  {Alekseev}, \citenamefont {Dmitriev}, \citenamefont {Gornyi},\ and\
  \citenamefont {Kachorovskii}}]{MagnetoRes2012Ioffe}%
  \BibitemOpen
  \bibfield  {author} {\bibinfo {author} {\bibfnamefont {P.~S.}\ \bibnamefont
  {Alekseev}}, \bibinfo {author} {\bibfnamefont {A.~P.}\ \bibnamefont
  {Dmitriev}}, \bibinfo {author} {\bibfnamefont {I.~V.}\ \bibnamefont
  {Gornyi}}, \ and\ \bibinfo {author} {\bibfnamefont {V.~Y.}\ \bibnamefont
  {Kachorovskii}},\ }\href@noop {} {\bibfield  {journal} {\bibinfo  {journal}
  {arXiv:1210.6081}\ } (\bibinfo {year} {2012})}\BibitemShut {NoStop}%
\bibitem [{\citenamefont {Peres}\ \emph {et~al.}(2006)\citenamefont {Peres},
  \citenamefont {Guinea},\ and\ \citenamefont {Castro~Neto}}]{PeresGuinea06}%
  \BibitemOpen
  \bibfield  {author} {\bibinfo {author} {\bibfnamefont {N.~M.~R.}\
  \bibnamefont {Peres}}, \bibinfo {author} {\bibfnamefont {F.}~\bibnamefont
  {Guinea}}, \ and\ \bibinfo {author} {\bibfnamefont {A.~H.}\ \bibnamefont
  {Castro~Neto}},\ }\href@noop {} {\bibfield  {journal} {\bibinfo  {journal}
  {Phys. Rev. B}\ }\textbf {\bibinfo {volume} {73}},\ \bibinfo {pages} {125411}
  (\bibinfo {year} {2006})}\BibitemShut {NoStop}%
\bibitem [{\citenamefont {Ando}(1974{\natexlab{b}})}]{Ando74ILevlBroad}%
  \BibitemOpen
  \bibfield  {author} {\bibinfo {author} {\bibfnamefont {T.}~\bibnamefont
  {Ando}},\ }\href@noop {} {\bibfield  {journal} {\bibinfo  {journal} {J. Phys.
  Soc. Jpn.}\ }\textbf {\bibinfo {volume} {36}},\ \bibinfo {pages} {959}
  (\bibinfo {year} {1974}{\natexlab{b}})}\BibitemShut {NoStop}%
\bibitem [{\citenamefont {Peres}\ \emph {et~al.}(2007)\citenamefont {Peres},
  \citenamefont {Lopes~dos Santos},\ and\ \citenamefont
  {Stauber}}]{Peres07Drude}%
  \BibitemOpen
  \bibfield  {author} {\bibinfo {author} {\bibfnamefont {N.~M.~R.}\
  \bibnamefont {Peres}}, \bibinfo {author} {\bibfnamefont {J.~M.~B.}\
  \bibnamefont {Lopes~dos Santos}}, \ and\ \bibinfo {author} {\bibfnamefont
  {T.}~\bibnamefont {Stauber}},\ }\href@noop {} {\bibfield  {journal} {\bibinfo
   {journal} {Phys. Rev. B}\ }\textbf {\bibinfo {volume} {76}},\ \bibinfo
  {pages} {073412} (\bibinfo {year} {2007})}\BibitemShut {NoStop}%
\bibitem [{\citenamefont {Gusynin}\ and\ \citenamefont
  {Sharapov}(2006)}]{Gus06HallOptCond}%
  \BibitemOpen
  \bibfield  {author} {\bibinfo {author} {\bibfnamefont {V.~P.}\ \bibnamefont
  {Gusynin}}\ and\ \bibinfo {author} {\bibfnamefont {S.~G.}\ \bibnamefont
  {Sharapov}},\ }\href@noop {} {\bibfield  {journal} {\bibinfo  {journal}
  {Phys. Rev. B}\ }\textbf {\bibinfo {volume} {73}},\ \bibinfo {pages} {245411}
  (\bibinfo {year} {2006})}\BibitemShut {NoStop}%
\bibitem [{\citenamefont {Horng}\ \emph {et~al.}(2011)\citenamefont {Horng},
  \citenamefont {Chen}, \citenamefont {Geng}, \citenamefont {Girit},
  \citenamefont {Zhang}, \citenamefont {Hao}, \citenamefont {Bechtel},
  \citenamefont {Martin}, \citenamefont {Zettl}, \citenamefont {Crommie},
  \citenamefont {Shen},\ and\ \citenamefont {Wang}}]{Horng11DrudeGraphene}%
  \BibitemOpen
  \bibfield  {author} {\bibinfo {author} {\bibfnamefont {J.}~\bibnamefont
  {Horng}}, \bibinfo {author} {\bibfnamefont {C.~F.}\ \bibnamefont {Chen}},
  \bibinfo {author} {\bibfnamefont {B.}~\bibnamefont {Geng}}, \bibinfo {author}
  {\bibfnamefont {C.}~\bibnamefont {Girit}}, \bibinfo {author} {\bibfnamefont
  {Y.}~\bibnamefont {Zhang}}, \bibinfo {author} {\bibfnamefont
  {Z.}~\bibnamefont {Hao}}, \bibinfo {author} {\bibfnamefont {H.~A.}\
  \bibnamefont {Bechtel}}, \bibinfo {author} {\bibfnamefont {M.}~\bibnamefont
  {Martin}}, \bibinfo {author} {\bibfnamefont {A.}~\bibnamefont {Zettl}},
  \bibinfo {author} {\bibfnamefont {M.~F.}\ \bibnamefont {Crommie}}, \bibinfo
  {author} {\bibfnamefont {Y.~R.}\ \bibnamefont {Shen}}, \ and\ \bibinfo
  {author} {\bibfnamefont {F.}~\bibnamefont {Wang}},\ }\href@noop {} {\bibfield
   {journal} {\bibinfo  {journal} {Phys. Rev. B}\ }\textbf {\bibinfo {volume}
  {83}},\ \bibinfo {pages} {165113} (\bibinfo {year} {2011})}\BibitemShut
  {NoStop}%
\bibitem [{Note1()}]{Note1}%
  \BibitemOpen
  \bibinfo {note} {\label {footnote_interactions} Interactions may contribute a
  third damping mechanism. For systems with a parabolic spectrum,
  electron-electron interactions are known to have no direct influence on the
  damping of SdH oscillations. Only if one considers the combined effect of
  disorder and interactions, the damping of SdH oscillations is influenced by a
  renormalization of the effective mass and scattering time.\cite
  {Martin03QMO_FL,AdamovGornyiMirlin06Interactions} By contrast, interactions
  directly influence\cite {dmitriev:2009b} the higher-order quantum corrections
  $\propto \lambda ^2$ similar to these in Eq.~(\ref
  {sec:conductivity:eq:highT_QO_1}).}\BibitemShut {Stop}%
\bibitem [{\citenamefont {M\"uller}\ \emph {et~al.}(2009)\citenamefont
  {M\"uller}, \citenamefont {Schmalian},\ and\ \citenamefont
  {Fritz}}]{MuellerSchmalian09}%
  \BibitemOpen
  \bibfield  {author} {\bibinfo {author} {\bibfnamefont {M.}~\bibnamefont
  {M\"uller}}, \bibinfo {author} {\bibfnamefont {J.}~\bibnamefont {Schmalian}},
  \ and\ \bibinfo {author} {\bibfnamefont {L.}~\bibnamefont {Fritz}},\
  }\href@noop {} {\bibfield  {journal} {\bibinfo  {journal} {Phys. Rev. Lett.}\
  }\textbf {\bibinfo {volume} {103}},\ \bibinfo {pages} {025301} (\bibinfo
  {year} {2009})}\BibitemShut {NoStop}%
\bibitem [{\citenamefont {Sch\"utt}\ \emph {et~al.}(2011)\citenamefont
  {Sch\"utt}, \citenamefont {Ostrovsky}, \citenamefont {Gornyi},\ and\
  \citenamefont {Mirlin}}]{MichaRates11}%
  \BibitemOpen
  \bibfield  {author} {\bibinfo {author} {\bibfnamefont {M.}~\bibnamefont
  {Sch\"utt}}, \bibinfo {author} {\bibfnamefont {P.~M.}\ \bibnamefont
  {Ostrovsky}}, \bibinfo {author} {\bibfnamefont {I.~V.}\ \bibnamefont
  {Gornyi}}, \ and\ \bibinfo {author} {\bibfnamefont {A.~D.}\ \bibnamefont
  {Mirlin}},\ }\href@noop {} {\bibfield  {journal} {\bibinfo  {journal} {Phys.
  Rev. B}\ }\textbf {\bibinfo {volume} {83}},\ \bibinfo {pages} {155441}
  (\bibinfo {year} {2011})}\BibitemShut {NoStop}%
\bibitem [{\citenamefont {Sch\"utt}\ \emph {et~al.}(2013)\citenamefont
  {Sch\"utt}, \citenamefont {Ostrovsky}, \citenamefont {Titov}, \citenamefont
  {Gornyi}, \citenamefont {Narozhny},\ and\ \citenamefont
  {Mirlin}}]{MichaelCoulDrag12}%
  \BibitemOpen
  \bibfield  {author} {\bibinfo {author} {\bibfnamefont {M.}~\bibnamefont
  {Sch\"utt}}, \bibinfo {author} {\bibfnamefont {P.~M.}\ \bibnamefont
  {Ostrovsky}}, \bibinfo {author} {\bibfnamefont {M.}~\bibnamefont {Titov}},
  \bibinfo {author} {\bibfnamefont {I.~V.}\ \bibnamefont {Gornyi}}, \bibinfo
  {author} {\bibfnamefont {B.~N.}\ \bibnamefont {Narozhny}}, \ and\ \bibinfo
  {author} {\bibfnamefont {A.~D.}\ \bibnamefont {Mirlin}},\ }\href@noop {}
  {\bibfield  {journal} {\bibinfo  {journal} {Phys. Rev. Lett.}\ }\textbf
  {\bibinfo {volume} {110}},\ \bibinfo {pages} {026601} (\bibinfo {year}
  {2013})}\BibitemShut {NoStop}%
\bibitem [{\citenamefont {Tielrooij}\ \emph {et~al.}(2012)\citenamefont
  {Tielrooij}, \citenamefont {Song}, \citenamefont {Jensen}, \citenamefont
  {Centeno}, \citenamefont {Pesquera}, \citenamefont {Zurutuza~Elorza},
  \citenamefont {Bonn}, \citenamefont {Levitov},\ and\ \citenamefont
  {Koppens}}]{KoppensCascade2012}%
  \BibitemOpen
  \bibfield  {author} {\bibinfo {author} {\bibfnamefont {K.~J.}\ \bibnamefont
  {Tielrooij}}, \bibinfo {author} {\bibfnamefont {J.~C.~W.}\ \bibnamefont
  {Song}}, \bibinfo {author} {\bibfnamefont {S.~A.}\ \bibnamefont {Jensen}},
  \bibinfo {author} {\bibfnamefont {A.}~\bibnamefont {Centeno}}, \bibinfo
  {author} {\bibfnamefont {A.}~\bibnamefont {Pesquera}}, \bibinfo {author}
  {\bibfnamefont {A.}~\bibnamefont {Zurutuza~Elorza}}, \bibinfo {author}
  {\bibfnamefont {M.}~\bibnamefont {Bonn}}, \bibinfo {author} {\bibfnamefont
  {L.~S.}\ \bibnamefont {Levitov}}, \ and\ \bibinfo {author} {\bibfnamefont
  {F.~H.~L.}\ \bibnamefont {Koppens}},\ }\href@noop {} {\bibfield  {journal}
  {\bibinfo  {journal} {arXiv:1210.1205}\ } (\bibinfo {year}
  {2012})}\BibitemShut {NoStop}%
\bibitem [{\citenamefont {Song}\ \emph {et~al.}(2012)\citenamefont {Song},
  \citenamefont {Tielrooij}, \citenamefont {Koppens},\ and\ \citenamefont
  {Levitov}}]{KoppensCascade2}%
  \BibitemOpen
  \bibfield  {author} {\bibinfo {author} {\bibfnamefont {J.~C.~W.}\
  \bibnamefont {Song}}, \bibinfo {author} {\bibfnamefont {K.~J.}\ \bibnamefont
  {Tielrooij}}, \bibinfo {author} {\bibfnamefont {F.~H.~L.}\ \bibnamefont
  {Koppens}}, \ and\ \bibinfo {author} {\bibfnamefont {L.~S.}\ \bibnamefont
  {Levitov}},\ }\href@noop {} {\bibfield  {journal} {\bibinfo  {journal}
  {arXiv:1209.4346}\ } (\bibinfo {year} {2012})}\BibitemShut {NoStop}%
\bibitem [{\citenamefont {Li}\ \emph {et~al.}(2012)\citenamefont {Li},
  \citenamefont {Luo}, \citenamefont {Hupalo}, \citenamefont {Zhang},
  \citenamefont {Tringides}, \citenamefont {Schmalian},\ and\ \citenamefont
  {Wang}}]{PopInv12}%
  \BibitemOpen
  \bibfield  {author} {\bibinfo {author} {\bibfnamefont {T.}~\bibnamefont
  {Li}}, \bibinfo {author} {\bibfnamefont {L.}~\bibnamefont {Luo}}, \bibinfo
  {author} {\bibfnamefont {M.}~\bibnamefont {Hupalo}}, \bibinfo {author}
  {\bibfnamefont {J.}~\bibnamefont {Zhang}}, \bibinfo {author} {\bibfnamefont
  {M.~C.}\ \bibnamefont {Tringides}}, \bibinfo {author} {\bibfnamefont
  {J.}~\bibnamefont {Schmalian}}, \ and\ \bibinfo {author} {\bibfnamefont
  {J.}~\bibnamefont {Wang}},\ }\href@noop {} {\bibfield  {journal} {\bibinfo
  {journal} {Phys. Rev. Lett.}\ }\textbf {\bibinfo {volume} {108}},\ \bibinfo
  {pages} {167401} (\bibinfo {year} {2012})}\BibitemShut {NoStop}%
\bibitem [{\citenamefont {Ryzhii}\ \emph {et~al.}(2007)\citenamefont {Ryzhii},
  \citenamefont {Ryzhii},\ and\ \citenamefont {Otsuji}}]{RyzhiiNegCond07}%
  \BibitemOpen
  \bibfield  {author} {\bibinfo {author} {\bibfnamefont {V.}~\bibnamefont
  {Ryzhii}}, \bibinfo {author} {\bibfnamefont {M.}~\bibnamefont {Ryzhii}}, \
  and\ \bibinfo {author} {\bibfnamefont {T.}~\bibnamefont {Otsuji}},\
  }\href@noop {} {\bibfield  {journal} {\bibinfo  {journal} {J. Appl. Phys.}\
  }\textbf {\bibinfo {volume} {101}},\ \bibinfo {pages} {083114} (\bibinfo
  {year} {2007})}\BibitemShut {NoStop}%
\bibitem [{\citenamefont {Dmitriev}\ \emph {et~al.}(2009)\citenamefont
  {Dmitriev}, \citenamefont {Khodas}, \citenamefont {Mirlin}, \citenamefont
  {Polyakov},\ and\ \citenamefont {Vavilov}}]{dmitriev:2009b}%
  \BibitemOpen
  \bibfield  {author} {\bibinfo {author} {\bibfnamefont {I.~A.}\ \bibnamefont
  {Dmitriev}}, \bibinfo {author} {\bibfnamefont {M.}~\bibnamefont {Khodas}},
  \bibinfo {author} {\bibfnamefont {A.~D.}\ \bibnamefont {Mirlin}}, \bibinfo
  {author} {\bibfnamefont {D.~G.}\ \bibnamefont {Polyakov}}, \ and\ \bibinfo
  {author} {\bibfnamefont {M.~G.}\ \bibnamefont {Vavilov}},\ }\href@noop {}
  {\bibfield  {journal} {\bibinfo  {journal} {Phys. Rev. B}\ }\textbf {\bibinfo
  {volume} {80}},\ \bibinfo {pages} {165327} (\bibinfo {year}
  {2009})}\BibitemShut {NoStop}%
\bibitem [{\citenamefont {Ostrovsky}\ \emph {et~al.}(2006)\citenamefont
  {Ostrovsky}, \citenamefont {Gornyi},\ and\ \citenamefont
  {Mirlin}}]{Ostr06Dis}%
  \BibitemOpen
  \bibfield  {author} {\bibinfo {author} {\bibfnamefont {P.~M.}\ \bibnamefont
  {Ostrovsky}}, \bibinfo {author} {\bibfnamefont {I.~V.}\ \bibnamefont
  {Gornyi}}, \ and\ \bibinfo {author} {\bibfnamefont {A.~D.}\ \bibnamefont
  {Mirlin}},\ }\href@noop {} {\bibfield  {journal} {\bibinfo  {journal} {Phys.
  Rev. B}\ }\textbf {\bibinfo {volume} {74}},\ \bibinfo {pages} {235443}
  (\bibinfo {year} {2006})}\BibitemShut {NoStop}%
\bibitem [{\citenamefont {Dmitriev}\ \emph {et~al.}(2007)\citenamefont
  {Dmitriev}, \citenamefont {Mirlin},\ and\ \citenamefont
  {Polyakov}}]{FracMIRO07}%
  \BibitemOpen
  \bibfield  {author} {\bibinfo {author} {\bibfnamefont {I.~A.}\ \bibnamefont
  {Dmitriev}}, \bibinfo {author} {\bibfnamefont {A.~D.}\ \bibnamefont
  {Mirlin}}, \ and\ \bibinfo {author} {\bibfnamefont {D.~G.}\ \bibnamefont
  {Polyakov}},\ }\href@noop {} {\bibfield  {journal} {\bibinfo  {journal}
  {Phys. Rev. Lett.}\ }\textbf {\bibinfo {volume} {99}},\ \bibinfo {pages}
  {206805} (\bibinfo {year} {2007})}\BibitemShut {NoStop}%
\bibitem [{\citenamefont {Martin}\ \emph {et~al.}(2003)\citenamefont {Martin},
  \citenamefont {Maslov},\ and\ \citenamefont {Reizer}}]{Martin03QMO_FL}%
  \BibitemOpen
  \bibfield  {author} {\bibinfo {author} {\bibfnamefont {G.~W.}\ \bibnamefont
  {Martin}}, \bibinfo {author} {\bibfnamefont {D.~L.}\ \bibnamefont {Maslov}},
  \ and\ \bibinfo {author} {\bibfnamefont {M.~Y.}\ \bibnamefont {Reizer}},\
  }\href@noop {} {\bibfield  {journal} {\bibinfo  {journal} {Phys. Rev. B}\
  }\textbf {\bibinfo {volume} {68}},\ \bibinfo {pages} {241309} (\bibinfo
  {year} {2003})}\BibitemShut {NoStop}%
\bibitem [{\citenamefont {Adamov}\ \emph {et~al.}(2006)\citenamefont {Adamov},
  \citenamefont {Gornyi},\ and\ \citenamefont
  {Mirlin}}]{AdamovGornyiMirlin06Interactions}%
  \BibitemOpen
  \bibfield  {author} {\bibinfo {author} {\bibfnamefont {Y.}~\bibnamefont
  {Adamov}}, \bibinfo {author} {\bibfnamefont {I.~V.}\ \bibnamefont {Gornyi}},
  \ and\ \bibinfo {author} {\bibfnamefont {A.~D.}\ \bibnamefont {Mirlin}},\
  }\href {\doibase 10.1103/PhysRevB.73.045426} {\bibfield  {journal} {\bibinfo
  {journal} {Phys. Rev. B}\ }\textbf {\bibinfo {volume} {73}},\ \bibinfo
  {pages} {045426} (\bibinfo {year} {2006})}\BibitemShut {NoStop}%
\end{thebibliography}

%

\end{document}